\documentclass[12pt,preprint]{aastex}

\usepackage{natbib, epsfig}

\shorttitle{The FIRST-2MASS Red Quasar Survey}

\newcommand{\opt}{\phm{O}\phn\phn\ }

\begin{document}
\title {\bf The FIRST-2MASS Red Quasar Survey}

\author{Eilat Glikman\altaffilmark{1}}
\email{eilatg@astro.caltech.edu}
\author{David J. Helfand\altaffilmark{2}}
\email{djh@astro.columbia.edu}
\author{Richard L. White\altaffilmark{3}}
\email{rlw@stsci.edu}
\author{Robert H. Becker\altaffilmark{4,5}}
\email{bob@igpp.ucllnl.org}
\author{Michael D. Gregg\altaffilmark{4,5}}
\email{gregg@igpp.ucllnl.org}
\and
\author{Mark Lacy\altaffilmark{6}}
\email{mlacy@ipac.caltech.edu}

\altaffiltext{1}{Astronomy Department, California Institute of Technology, Pasadena, CA, 91125}
\altaffiltext{2}{Columbia Univeristy Department of Astronomy, 550 West 120th Street, New York, NY 10027}
\altaffiltext{3}{Space Telescope Science Institute, 3700 San Martin Dr., Baltimore, MD 21218}
\altaffiltext{4}{Department of Physics, University of California at Davis, 1 Shields Avenue, Davis, CA 95616}
\altaffiltext{5}{Institute of Geophysics and Planetary Physics, Lawrence Livermore National Laboratory, Livermore, CA, 94551}
\altaffiltext{6}{SIRTF Science Center, 120 E. California Ave., Pasadena, CA, 91125}

\begin{abstract}
Combining radio observations with optical and infrared color selection 
--  demonstrated in our pilot study to be an efficient selection 
algorithm for finding red quasars -- we have obtained optical and 
infrared spectroscopy for 120 objects in a complete sample of 156 
candidates from a sky area of 2716 square degrees. Consistent with our 
initial results, we find our selection 
criteria -- $J-K>1.7, R-K>4.0$ -- yield a $\sim 50\%$ success rate for 
discovering quasars substantially redder than those found in optical 
surveys. Comparison with UVX- and optical color-selected samples shows 
that $\gtrsim 10\%$ of the quasars are missed in a magnitude-limited survey. 
Simultaneous two-frequency radio observations for part of 
the sample indicate that a synchrotron continuum component is ruled out as a 
significant contributor to reddening the quasars' spectra. We go on to 
estimate extinctions for our objects assuming their red colors are 
caused by dust. Continuum fits and Balmer decrements suggest  $E(B-V)$ 
values ranging from near zero to 2.5 magnitudes. Correcting the $K$-band 
magnitudes for these extinctions, we find that for $K \leq 14.0$, red 
quasars make up between 25\% and 60\% of the underlying quasar 
population; owing to the incompleteness of the 2MASS survey at fainter 
$K$-band magnitudes, we can only set a lower limit to the radio-detected 
red quasar population of $>20-30\%$.
\end{abstract}

\keywords{dust, extinction; quasars: general; surveys}

\section {Introduction}

The definition and physical understanding of quasars has evolved and expanded since the first quasars were found over 40 years ago.  Discovered in the radio, quasars were thought to be radio-loud objects until \citet{Sandage65} uncovered a population of objects with optical properties identical to quasars, but without radio emission.  It is now known that there are roughly ten times more radio-quiet than radio-loud quasars.  

The spectroscopic properties of the earliest known luminous quasars showed blue optical-to-ultraviolet colors and broad emission lines, as well as narrow forbidden lines of highly ionized species.  This, coupled with their compact morphologies and extreme luminosities, led to the commonly accepted model for the quasar as powered by accretion onto a supermassive black hole.  This picture was expanded by invoking orientation-based arguments to include narrow emission-line spectra of AGN (e.g., Seyfert2 and narrow-line radio galaxies) as well as highly-variable featureless spectra (e.g., BL Lacs).  

Such orientation models depict an axisymmetric geometry with an accretion disk at the center surrounded by broad emission-line clouds.  Dusty, molecular clouds, often modeled as a torus coplanar with the accretion disk, obscure the broad emission lines along certain lines of sight.  Beyond the obscuring clouds lies the narrow emission-line region \citep{Urry95}.  This model is successful at explaining the different varieties of AGN spectra: whether broad lines are seen depends on our viewing angle toward the AGN.  It has also been successful at predicting AGN number counts in the infrared as well as explaining the shape of the extragalactic x-ray background \citep[e.g.,][]{Madau94,Comastri95,Gilli01,Ueda03,Treister04,Treister05,Treister06}.  

Although the unification scheme is well-established for the low-luminosity AGN, it has been problematic for quasars.  Several individual cases of Type 2 quasars at high redshift ($z>3$) were found in the X-ray.  These sources are found to have high column densities ($N_H > 10^{23}$ cm$^{-2}$) and X-ray luminosities ($L_X > 10^{44}$ erg s$^{-1}$) as well as narrow permitted emission-lines in their optical spectra and extremely weak optical continua \citep{Norman02,stern02}.  \citet{Zakamska03} have presented low-redshift Type 2 quasar population by taking advantage of the large spectral database in the SDSS.  Selecting for objects with $0.3 < z < 0.83$, \citet{Zakamska03} estimate that $50\%$ of their sample ($\sim 150$ out of 291 Type 2 AGN) have quasar luminosities ($M_B<-23$).  Their multiwavelength and polarization properties also support their Type 2 classification and the unification scheme for luminous AGN out to the quasar regime \citep{Zakamska04,Zakamska05}. 

Several observational findings linking supermassive black holes and quasar activity to their host galaxies, however, are not accounted for by these orientation models and may provide clues about joint formation and evolution of quasars and their hosts.   Most galaxies harbor a quiescent supermassive black hole, suggesting that a quasar may be a short-lived phase in the life of a galaxy. Moreover, the black hole mass appears to correlate with the host galaxy's bulge mass and stellar velocity dispersion \citep{Magorrian98,Ferrarese00}.  This, combined with the remarkably similar evolution of quasar activity and star formation over the history of the universe, suggests that the quasar is an essential ingredient in galaxy formation.  Initial results from semianalytical models successfully reproduce some of these observational trends \citep{Kauffmann00} and more recent numerical simulations of joint black hole/galaxy formation have been able to account for many of these observations \citep{DiMatteo05,Hopkins06}.  In addition, these models predict an obscured phase for the nascent quasar as a galaxy merger triggers star formation and funnels gas into the nucleus.  Eventually, feedback from the quasar blows the dust and gas away, quenching the accretion and ending the black hole's active phase \citep[see also][]{Sanders88a,Yun04}.

Although reddened quasars are predicted in both the static, orientation-based model and the evolutionary scenarios, this population has only begun to be uncovered.  The majority of quasars described in the literature have been found from optically selected samples and have remarkably homogeneous spectra with blue optical-to-ultraviolet spectral energy distributions.  The recent discovery of a population of red quasars has challenged the view that all quasars are blue, UV-luminous sources and suggests that a large population of quasars with red colors exists \citep{Webster95,Cutri01,Gregg02,Richards03,White03b,Glikman04}.  The advent of large near-infrared detectors that perform at the level of optical CCDs has enabled surveys for quasars that are largely insensitive to dust extinction.  Since most near-infrared sources in the sky are Galactic stars, radio surveys offer an effective complement to near-infrared quasar searches by selecting predominantly extragalactic sources, as well as being completely insensitive to dust extinction.  

\defcitealias{Glikman04}{Paper~I}

In \citet[hereafter, Paper~I]{Glikman04}, we used this principle to search for red quasars by studying radio sources that are bright at 2 \micron\ ($K\leq 16$) but undetectable at $R \gtrsim 20$.  Spectroscopic followup of these sources revealed that objects with $J-K>1.7$ and $R-K>4$ had a $50\%$ chance of being highly reddened quasars.  Seventeen red quasars were detected in that survey, and we concluded that red quasars make up $\gtrsim 20\%$ of the total quasar population.

In this paper, we expand our pilot survey to produce a complete catalog of objects obeying our optical-to-near-infrared color criteria over the same 2716 deg$^2$ overlap region between the VLA FIRST and 2MASS surveys\footnote{In this work we use the 2000 July FIRST catalog to maintain continuity with \citetalias{Glikman04}}.  We review the input catalogs and the candidate selection process in Section 2.  We describe the spectroscopic followup of our candidate list as well as the classification of the spectra and the resultant catalog statistics in Section 3.  In Section 4, we present the observed surface density of red quasars on the sky and compare it to that for optically selected quasars.  In Section 5, we derive and compare extinctions from continuum fits to a reddened quasar template (\S 5.1) and extinctions derived from Balmer decrements (\S 5.2).  In Section 6 we examine the radio spectral indices of the red quasar population.  We estimate the percentage of red quasars in the overall quasar population in Section 7, while Section 8 discusses variable blue quasars found in our survey.  Section 9 places our sample in context with other radio-selected reddened quasars in the literature.  Section 10 summarizes our findings.

\section {The Selection Process}

Our red quasar survey relies upon three wavelength regions to construct our candidate list.  We require that our quasars be radio sources to avoid confusion with Galactic stars.  Additionally, since we are looking for red objects, we require that our quasars be bright at 2 \micron.  Finally, we use optical data to select objects with red optical-to-near-infrared colors.

The FIRST Survey \citep{Becker95} has mapped $\sim 9000$ deg$^2$ of the sky at 20 cm using the Very Large Array (VLA) in the B-configuration.  Coverage includes $\sim 8400$ deg$^2$ in the north Galactic cap and $\sim 600$ deg$^2$ in the south Galactic cap. With the array in this configuration, the survey is roughly analogous to all-sky optical surveys, with 5\arcsec\ resolution and 0\farcs5 positional accuracy.  The 3-minute snapshot integration time yields a typical rms of 0.15 mJy.  The July 2000 catalog generated from this survey has $\sim 722,354$ sources brighter than the survey's detection threshold of 1 mJy over 7988 deg$^2$.

The 2MASS survey \citep{Strutskie06} used two telescopes, one at Mt. Hopkins, Arizona and the other at Cerro Tololo, Chile, to image the sky in three near-infrared bands, $J$ (1.24 \micron), $H$ (1.66 \micron), and $K_s$ (2.16 \micron), between 1997 and 2001.  This four-year effort has produced a Point Source Catalog (PSC) with $\sim 4.70\times 10^8$ sources.  The survey is $99\%$ complete to a magnitude limit at $10\sigma$ of $J=15.8$, $H=15.1$ $K=14.3$.  Fainter sources are included in the catalog (down to $K\lesssim 16$) but with lower signal-to-noise and completeness.  The survey has a resolution of 4\arcsec\ and an astrometric accuracy better than 0\farcs5.  We utilize the 2nd Incremental Release of 2MASS, which includes observations through 1999, covers 19,681 deg$^2$, and contains $>1.6\times10^8$ point sources \footnote{Although the 2MASS All-Sky Catalog is described as being of superior quality by the 2MASS Explanatory Supplement (\url{http://www.ipac.caltech.edu/2mass/releases/allsky/doc/explsup.html, \S I.6.v) }, we discuss the subsample selected from the 2nd Incremental Release Point Source Catalog in this work for continuity with \citetalias{Glikman04} and because of our high level of spectroscopic completeness over this well-defined area.}.

The Guide Star Catalog II (GSC-II) is an all-sky optical catalog produced by scanning the second-generation Palomar Observatory (POSS-II) and UK Schmidt Sky Surveys at 1\arcsec\ resolution and generating a catalog with positions, magnitudes and morphological classifications.  We employed the GSC 2.2.1 which is a pre-release version not available outside the STScI that reaches the POSS-II plate limits of $F < 20.80$ and $J < 22.50$.  It is similar to the publicly released catalog, version 2.3.2.  The astrometric accuracy of GSC-II is better than 1\arcsec.  

\subsection{The FIRST-2MASS Sample}

We began our search for a population of optically obscured, red quasars by looking for objects detected in the FIRST and 2MASS catalogs but which lacked an optical counterpart in the Cambridge Automated Plate Measuring Machine (APM) scans of the first-generation Palomar Observatory Sky Survey (POSS-I) plates.  Spectroscopic followup of 50 of the resulting 69 candidates allowed us to define a region in color-color space which maximized our efficiency for finding red quasars.  We found that $\sim 50\%$ of the spectroscopically identified objects in our initial sample defined by $R-K$\footnote{$B$ and $R$ are used to refer to ``blue'' and ``red'' filters in either the POSS-I $E$ and $O$ or the POSS-II $F$ and $J$ bands as well as the SDSS $r$ band.}$\geq4$ and $J-K\geq1.7$ were heavily obscured quasars. The other $\sim 50\%$ were galaxies exhibiting some form of activity such as a starburst or narrow-line AGN activity. Objects with bluer $J-K$ colors tended to be late-type stars and objects with $R-K<4$ tended to be elliptical galaxies; these results were reported in \citetalias{Glikman04}.

Based on these findings we chose to impose these color cuts on FIRST-2MASS sources, employing optical data from GSC-II to generate a sample of candidate red quasars.  As was shown in \citetalias{Glikman04}, a 2\arcsec\ search radius between FIRST and 2MASS yields a $94\%$ probability that the two sources are physically associated.  A match between FIRST and 2MASS using this search radius yielded 12,435 matches, keeping only the closest match (a catalog we refer to henceforth as F2M).  We matched these sources, using their FIRST positions, to the GSC-II catalog with a search radius of 2\arcsec.  There were 10,903 matches within 2\arcsec\ of the FIRST positions in F2M. 

In order to apply the optical color cut to the optical matches we need to take into account sources that have been detected in one band but not in another.  We replaced the magnitudes for undetected objects with their respective plate limits, $B=22.50$ and $R=20.80$.  We then applied our $J-K>1.7$ and $R-K>4$ color cuts, and reduced the size of our candidate list to 146 sources.  In cases where bright objects were detected on plates observed through the Photographic $J$ bandpass but were undetected in Photographic $F$ (e.g., saturated stars), artificially red colors may have been assigned.  We examined finding charts of all 35 sources with $R>20.80$ and found that 5 were indeed too faint to be discerned on the Digitized Sky Survey images; we kept them in the candidate list.  We obtained magnitudes for the remaining 30 sources using auxiliary catalogs.  Twenty-six objects had detections in the SDSS-DR5 catalog which contributed $r$ magnitudes -- all but three were too blue to remain in our candidate list.  We obtained APM $E$ magnitudes for the remaining 4 sources and removed all but one object, F2M J171058.4+310958, which has $E=19.34$ and $E-K=4.38$. The final list of FIRST-2MASS sources with GSC II matches within 2\arcsec\ thus contains 120 objects with $R-K\geq4$ and $J-K\geq1.7$. 

In addition, fourteen objects too faint to be detected in the GSC-II catalog were previously reported in \citetalias{Glikman04} as having $J-K>1.7$. We include these sources in our candidate list, adopting the plate limits as lower limits to their $B$ and $R$ magnitudes.  

The 2MASS survey quotes upper limits for objects with a photometric signal-to-noise ratio smaller than $\sim3$.  Objects that are below the 2MASS $J$-band limit can have $J-K$ values that are too blue to be selected into our sample, though they {\em might} have $J-K>1.7$.  There are 718 F2M sources with a $J$-band signal-to-noise ratio less than 3 and $J-K\leq 1.7$.  618 of these have a GSC-II match within 2\arcsec\ and 90 have no GSC-II match . In addition, we checked the 2MASS All-Sky catalog whose photometry is based on a different thresholding algorithm and found four $J$-band detections.  One of these had J-K=1.52, and we remove it from the candidate list.  We examined the POSS-II images of the optically {\em unmatched} objects and removed 72 large galaxies and saturated stars which made it into this list as a result of bad photometry in at least one survey.  We applied the $R-K$ color cut to the objects with GSC-II matches and were left with 20 sources.  We obtained SDSS $r$ magnitudes for 15 of the 18 optically unmatched sources and APM photometry for the other 3.  Two sources survive the $R-K$ color cut.  Therefore, in order for our survey to be a $K$-band limited survey, rather than a $J$-band limited survey (which would be more susceptible to dust extinction) we append to our candidate list 22 objects with a $J$-band signal-to-noise ratio smaller than 3 that have nominal $J-K$ values less than $1.7$, but have limits consistent with $J-K>1.7$.  This may contaminate our candidate sample slightly, but assures completeness.

The final candidate list thus contains 156 sources, 140 of which have detections in the GSC-II catalog, and 16 of which are fainter than the sensitivity limit of GSC-II.   This list includes the 31 objects from \citetalias{Glikman04} which had no detection in the APM catalog and meet our color criteria.  We present these candidates in Tables \ref{table:candlist} and \ref{table:jlim_cands} (which we describe in \S 3.2).  Table \ref{table:candlist} lists the sources that have $J$-band detections in 2MASS and Table \ref{table:jlim_cands} lists the 2MASS objects whose $J$-band detection had a signal-to-noise ratio less than 3 and $J-K<1.7$.

The 2nd Incremental Release of 2MASS is made up of a noncontiguous set of tiles over the celestial sphere, while the FIRST covers a solid ``footprint'' over the North and South Galactic caps. We computed the area of the F2M survey in \citetalias{Glikman04} by counting the number of FIRST sources that fell within the four corners of a 2MASS tile.  Since FIRST sources are smoothly distributed over the sky, the fraction of FIRST sources inside the 2MASS area determined the fraction of the FIRST area covered by the two surveys.  This amounted to $34\%$, or 2716 deg$^2$.

\begin{figure}
\epsscale{0.75}
\plotone{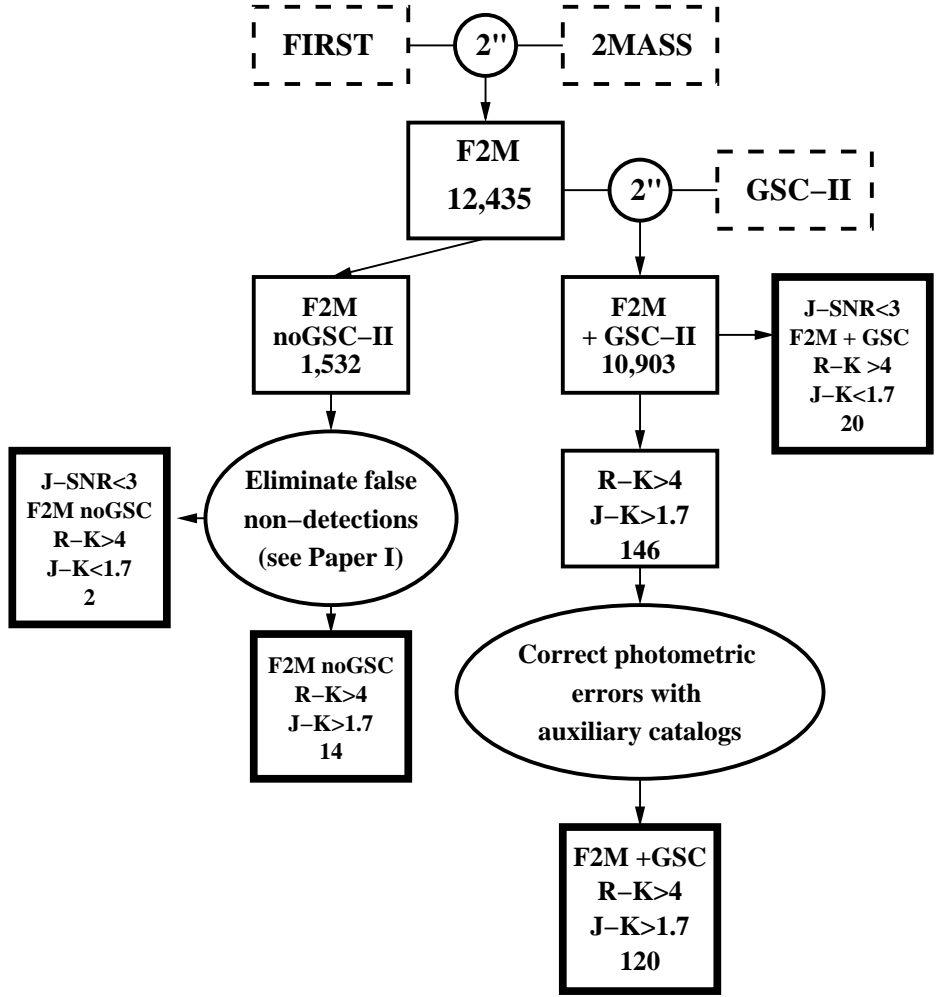}
\caption{Schematic diagram of our selection process.  Major catalogs are indicated by dotted-lined boxes and the matching radius is shown in the circle at the node of each match.  Sources in the final catalog are indicated by bold-lined boxes.}\label{fig:flowchart}
\end{figure}

Figure \ref{fig:flowchart} shows a schematic diagram of our selection process.  The boxes with dotted-line borders represent the input catalogs: FIRST, 2MASS and GSC II.  The matching radius is given in a circle at the node of each match.  The final catalog is indicated by four bold-lined boxes, one representing the 120 sources with FIRST, 2MASS and GSC II matches and the colors $R-K>4$ and $J-K>1.7$, a second for the 14 FIRST-2MASS sources with no GSC-II detection, and two final boxes indicating the 22 additional sources with $J$-band upper limits.

\section{Spectroscopic Followup}

\subsection{Observations}

Spectroscopic followup of our candidates was carried out over a five-year period at four observatories at both optical and near-infrared wavelengths.  A majority of the optical spectra were obtained at the 10 m W. M. Keck Observatory in Hawaii using the ESI and, to a lesser extent, the LRIS spectrographs.  A handful of spectra were obtained with the Kast spectrograph on the 3 m Shane telescope at Lick observatory.  We also obtained spectra from publicly available databases such as the FBQS II and III \citep{White00,Becker01} as well as the Sloan Digital Sky Survey Spectral database.  The LRIS spectra were reduced using standard IRAF procedures, while the ESI spectra were reduced using both IRAF tasks and software designed by the authors \citep[see][]{White03a}.

The near-infrared observations, spanning $0.8-2.5$ \micron, were conducted at the NASA Infrared Telescope Facility (IRTF) with the SpeX spectrograph \citep{Rayner03}.  We reduced these spectra using the Spextool software which is specially designed for SpeX data \citep{Cushing04}.  We corrected for telluric absorption using the spectrum of a nearby A0V star obtained immediately before or after each object spectrum using the technique outlined in \citet{Vacca03}.

Five near-infrared spectra, spanning $0.95-2.5$ \micron, were obtained at the MDM Observatory on Kitt Peak with the TIFKAM infrared camera in spectroscopic mode.  These spectra were reduced using standard IRAF procedures.  We corrected these spectra for telluric absorption using the spectrum of a nearby A0 V star as well as a G2 V star following the technique outlined in \citet{Hanson96}.

\subsection{Results}

We have obtained spectra for 120 of our 156 candidates from optical and/or near-infrared observations or from data found in the literature.  Our complete candidate lists are presented in Tables \ref{table:candlist} (for objects detected in the 2MASS $J$-band) and \ref{table:jlim_cands} (for objects with $J-K<1.7$ but with a $J$-band signal-to-noise ratio less than 3 which, as a result, {\em may} result in a true vale of $J-K>1.7$). The FIRST catalog J2000.0 Right Ascension and Declination is listed in columns (1) and (2) followed by the  FIRST peak and integrated 20cm flux densities in columns (3) and (4). We list the GSC-II $B$ and $R$\footnote{Photographic $F$ and $J$ magnitudes.} magnitudes  in columns (5) and (6) and the 2MASS $J$, $H$, and $K_s$ magnitudes in columns (7), (8), and (9), respectively.  We list the $J-K$ and $R-K$ colors used in the selection of this sample (columns 10 and 11).  When available, we present the object's redshift and classification (described below) in columns (12) and (13). In column (14) we list a reference for the optical and/or near-infrared spectrum used to classify our source.  Column (15) provides any references to the object in the literature.

\subsubsection{Object Classification and Narrow-Line AGN}

We classified our spectra using continuum and line analysis.  We cross-correlated our spectra with a set of galaxy templates from \citet{Kinney96}.  We confirmed the classification of spectra with no obvious emission lines that were well-correlated with an elliptical or S0 template by measuring the 4000\AA\ break at the same redshift.  In spectra with emission lines, we measured the redshifts and line profiles, classifying objects with line widths $\gtrsim$ 1000 km s$^{-1}$ as quasars.  We use the redshifts determined from the line-measurements to confirm the redshifts determined from the cross correlations. 

\begin{figure}
\epsscale{1}
\plotone{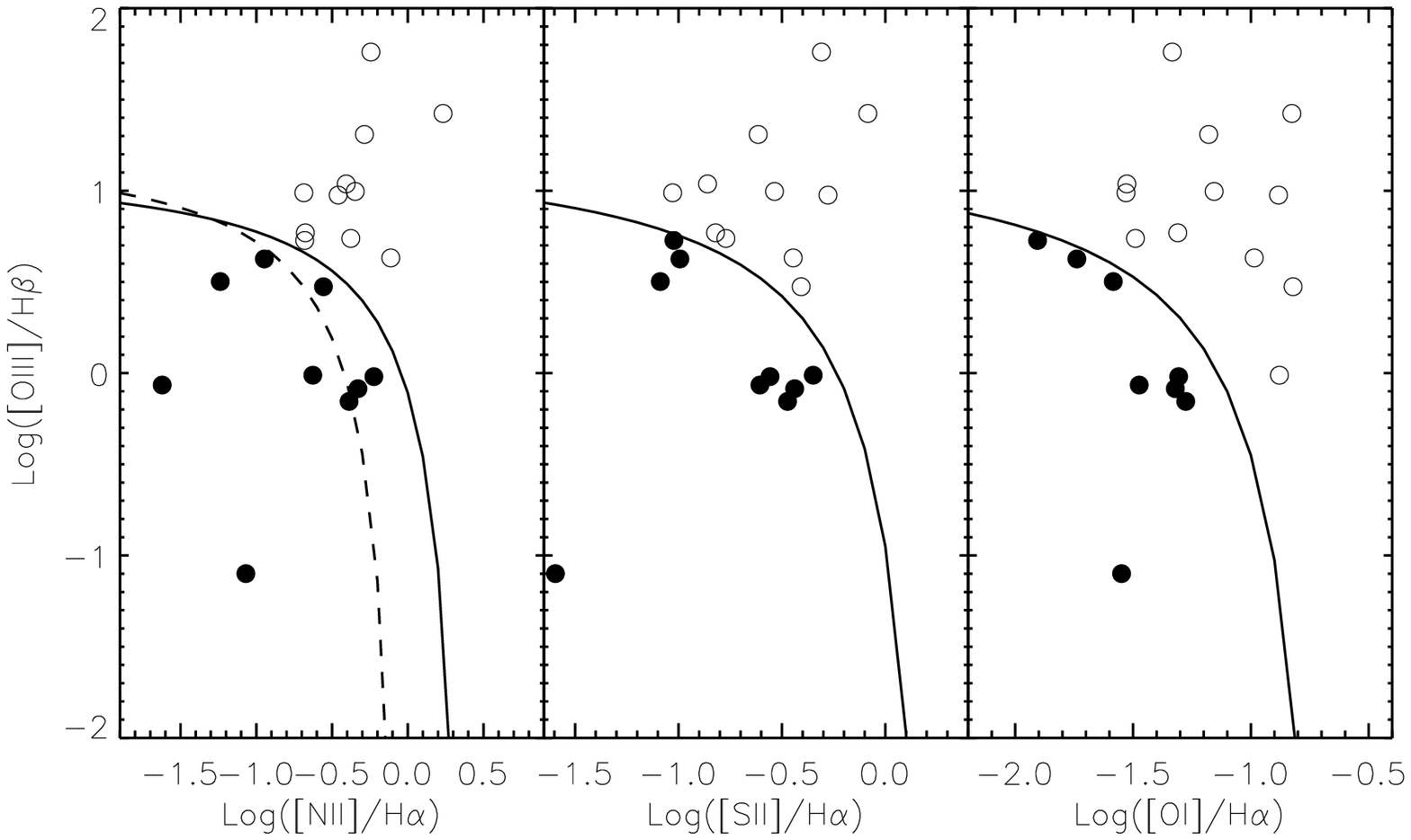}
\caption{Line diagnostics for our narrow-line spectra, plotted on diagrams from \citet{Kewley06}.  The left panel shows the [\ion{N}{2}]/H$\alpha$ versus [\ion{O}{3}]/H$\beta$ diagnostic.  In this panel, objects below and to the left of the dotted line are pure starbursts.  The middle panel shows the [\ion{S}{2}] /H$\alpha$ versus [\ion{O}{3}]/H$\beta$ diagnostic. The right panel shows the [\ion{O}{1}] /H$\alpha$ versus [\ion{O}{3}]/H$\beta$ diagnostic.  Objects to the left and below the solid line are starburst dominated (filled cirlcles), while objects above and to the right of the line are AGN dominated (open circles).}\label{fig:diagnostics}
\end{figure}

Twenty-five spectra had narrow, permitted emission lines. We classified twenty of them by fitting Gaussian profiles to the lines and performing line diagnostics following \citet{Kewley06}.  Figure \ref{fig:diagnostics} and Table \ref{table:diagnostics} list the results of these measurements.  Seven objects were classified as emission-line galaxies and ten objects were classified as AGN in all three diagrams. We assigned the classification that agreed in two diagrams for the remaining sources. The redshifts of two more objects were too high to measure H$\alpha$.  Their [\ion{O}{3}]/H$\beta$ ratios, however, are $\sim 1$ making them very likely AGN.  The three remaining narrow-line spectra have only a near-infrared spectrum revealing narrow Pa$\alpha$ and Pa$\beta$.  We label these as narrow-line emitters (``NLemit'') in Table \ref{table:candlist}.

The classification of the 120 spectra breaks down as follows: Thirty-seven galaxies (ten starburst galaxies and twenty-seven red elliptical galaxies);  fourteen narrow-line AGN (including one Seyfert 1.5 from \citealt{Smith02}) and three narrow-line emitting spectra that could not be further classified; one star;  five BL Lacs and 57 quasars (line widths $\gtrsim$ 1000 km s$^{-1}$) of which 52 are red quasars and 5 are blue quasars (discussed in more detail in \S 8). We also obtained two IRTF spectra and two optical spectra which we were unable to classify due to poor signal-to-noise and/or a lack of identifiable features.  We assume that these objects are not quasars in our subsequent analyses (although some of them could be). Our efficiency for finding broad-line quasars holds at $\sim50\%$ (57/120) as we predicted in \citetalias{Glikman04}.  

In this paper, we include only the (type I) quasars in our analysis.  Rest-frame visual extinction in narrow-line (type II) AGN can reach hundreds of magnitudes and the de-reddening techniques that we use (in \S5) cannot be applied.  Without a reddening estimate, we cannot compute the intrinsic luminosity of these sources for comparison with their unreddened counterparts.  Furthermore, the nuclear light in type II AGN is entirely obscured even in the near-infrared.  The near-infrared colors of type II AGN are consistent with coming from the host galaxy \citep{Zakamska04}.  Since our color-selection technique is designed to find lightly reddened quasars, and intentionally avoids galaxies \citepalias[see Figure 4 of][]{Glikman04}, our sampling of type II AGN is likely very incomplete.  Similar distinctions have been made by \citet{Lacy07} and \citet{Brown06} who estimate the fraction of reddened type I quasars separately from type II sources.

We note that as in \citetalias{Glikman04} several of our red quasars had all narrow-lines in their optical spectra.  It was the near-infrared spectrum that revealed broad Pa$\beta$ or H$\alpha$.  Without the near-infrared spectrum, these sources would have been classified as narrow-line AGN.  Fourteen of the sixteen objects classified as ``NLAGN'' or ``NLemit'' have only optical spectra.  It is possible that near-infrared spectroscopy would reveal broad emission lines for some of these objects. Therefore, for this and other reasons discussed below, our estimate for the fraction of missed red quasars is only a lower limit.

We lack a spectrum for one of our quasars, which we identified from the literature.  This is a $\sim 400$ mJy source and overlaps with the \citet{Webster95} red quasar sample \citep{Drinkwater97}.  The remaining 56 quasar spectra are presented in Figure \ref{fig:spectra}.  We have both an optical and near-infrared spectrum for 33 of our quasars. Since the overlapping regions between the optical and near-infrared spectra ($8000-11000$\AA) were often the noisiest portions of each spectrum, the size of the overlap region used in a given pair of spectra covered a large range ($100$\AA\ to over $2500$\AA). In addition, the optical and near-infrared observations were not contemporaneous and therefore the spliced spectra are susceptible to variability effects. We chose to scale the spectra individually using our own interactive software.  We overplotted the optical and near-infrared spectra and selected a region for each pair in the least noisy regions of overlap.  We then scaled the spectrum with the lower median flux in that region to that with the higher median.  Since our data were taken on different nights, seeing and transparency variations as well as slit losses resulting from different aperture sizes required scaling factors up to 12; most of the spectra (23/33) required scaling by less than a factor of 3. 

We note that we are $100\%$ spectroscopically complete to $K\leq14.5$; our completeness drops somewhat to $77\%$ for the entire sample, as a result of our $\sim 70\%$ completeness for $K>14.5$. Figure \ref{fig:khist} shows a histogram of the $K$ magnitudes binned by 0.25 magnitudes.  The unshaded histogram contains all 156 candidates, while the shaded histogram shows all identified objects (including the four unclassified spectra, which we include here as non-quasars).  We overplot the 57 quasars in black.

\section{The Observed Surface Density of Quasars on the Sky}

In order to conduct a meaningful comparison between our red quasar sample and the blue quasars found in optical surveys, we draw upon the FIRST Bright Quasar Survey \citep[FBQS;][]{Gregg96}.  The FBQS looked for quasars in matches between FIRST radio sources and sources in the APM scans of the first generation Palomar sky survey (POSS-I) plates (separations $<$ 1\farcs2) which were classified as point sources on either the photographic $E$ (red) or $O$ (blue) plates.  The initial catalog contains 636 quasars over 2682 deg$^2$ and imposed an $E=17.8$ magnitude limit on the APM sources as well as an $O-E \leq 2$ color criterion \citep[FBQS II;][]{White00}.  A deeper segment of the survey covered 589 deg$^2$ and found 321 quasars with the same color cut, but with a deeper magnitude limit of $E=18.9$ \citep[FBQS III;][]{Becker01}.  Since we are interested in objects detected in both FIRST and 2MASS, we matched the FBQS II and III catalogs to the 2MASS All-Sky Data Release with a search radius of 2\arcsec\ and found 503 and 134 matches to FBQS II and III, respectively. These quasars form an optically selected radio sample in the same $K$-band flux-limited survey and can be compared to our sample.

In Figure \ref{fig:bkhist} we plot the $B-K$ color distribution of FBQS II and III quasars along with the F2M red quasars as was done in \citetalias{Glikman04}.  This sample now fills in the $5\lesssim B-K\lesssim 6.5$ region of the plot which was previously unpopulated since our earlier optical selection criterion required objects to have $B\gtrsim 21.5$.

The redshift distribution is shown in Figure \ref{fig:z_dist}.  Our survey is able to reach a redshift range similar to the FBQS II and III samples, although the density of our sources decreases with redshift.  This is a consequence of the increased effect of dust on observed $K$-band light as a function of redshift.  At higher redshifts, the rest-frame optical light is shifted into the $K$-band.  This light is more susceptible to extinction resulting in an observed magnitude that falls below the 2MASS detection threshold for all but the most luminous objects.

We present the spatial density of red quasars compared with FBQS II, III and a UVX-selected sample in Table \ref{table:surfden}. The UVX-selected sample is described in \citetalias{Glikman04}.  Briefly, it was constructed by combining the Large Bright Quasar Survey (LBQS) described in \citet{Hewett95} and the Two-Degree Field QSO Redshift Survey (2QZ) described in \citet{Croom01}. We matched these sources to FIRST and 2MASS and created a UVX-selected sample that is comparable to the F2M red quasars.  This sample goes deeper in the optical than the FBQS samples.  The magnitude range for the LBQS and 2QZ quasars is $16.0<B_J<20.85$.

 We plot the densities as a function of $K$ magnitude in Figure \ref{fig:space_dist}.  Assuming our selection method finds all obscured quasars bright enough to be detected in the 2MASS and FIRST surveys (which is unlikely -- see discussion below), the total surface density of quasars ought simply to be the sum of the blue quasars plus the red quasars, normalized by their respective coverage areas.  Depending on the optical survey used for comparison, the last line of Table \ref{table:surfden} reveals that FBQS II misses 12.0$\pm$2.1$\%$, FBQS III misses 12.4$\pm$2.1$\%$, and the UVX selection method misses 8.2$\pm$1.4$\%$ of all FIRST-detected quasars visible in the 2MASS survey.  These, of course, are lower limits to the fraction of red quasars at a given magnitude.  In addition to our spectroscopic incompleteness, the observed $K$-band magnitudes of our quasars are still affected by dust and we have compared counts as a function of unreddened magnitudes with magnitudes that need to be corrected for absorption.  In the following section we determine the reddening in each quasar using its spectrum.  We can then correct the observed $K$-band magnitudes for absorption and compare their counts to the unobscured samples.

We note that although we have made an effort to compare samples with as similar properties as possible by comparing the F2M red quasars to objects detected in both FIRST and 2MASS, the fraction of missed quasars that we compute is a lower limit to the real fraction of missed quasars.  This is because there are quasars with FIRST and 2MASS detections that may not have been selected by {\em any} of these surveys.  The FBQS II survey has an APM $E=17.8$ magnitude limit.  Any object that is fainter than  $K=13.8$ (which is true of $\sim 80\%$ of the F2M quasars) and that is a slightly bluer than $R-K=4$ will not be in either the FBQS or the F2M red quasar sample. In the FBQS III case (which has an $E=18.9$ magnitude limit) this is a smaller effect -- only quasars with $K>14.9$ (amounting to about $25\%$ of the F2M quasars) can be missed. Overall, however, this is likely to be a small effect, since the space densities of FBQS II and III quasars with 2MASS matches are nearly identical (see Figure \ref{fig:space_dist}), meaning that the fainter optical magnitude limit of FBQS III does not recover significant numbers of slightly reddened quasars missed by FBQS II.

Taking into account the possible existence of additional missed quasars, the actual observed fraction of red quasars is no longer simply the number F2M red quasars divided by sum of the blue quasars plus the red quasars, normalized by their respective coverage areas.  The quasars missed by all these surveys enter in both the numerator and denominator of this equation and as a result our computed missed fraction above must be a lower limit to the true missed fraction of quasars.  

Higher fractions of missed reddened type I quasars have been found by \citet{Brown06} and \citet{Lacy07} by selecting quasar candidates in the dust-insensitive mid-infrared.  \citet{Brown06} find that reddened type I quasars make up $\sim 20\%$ of the type I quasar population, while \citet{Lacy07} find an even higher fraction of $30\%$, based on a sample of 11 reddened and 25 unreddened type I quasars. When we take reddening into account for our sample, we also find a higher missed fraction (see \S 6).

\section{Reddening Investigation}

\subsection{Continuum Fits to a Reddened Quasar Template}

In \citetalias{Glikman04} we assumed that the red colors of our quasars were due to extinction by dust and derived the color excess, $E(B-V)$, in the rest frame for each source.  Here, we fit to each object a reddened quasar composite from the FIRST Bright Quasar Survey \citep{Brotherton01} combined with a near-infrared quasar composite \citep{Glikman06}, resulting in wavelength coverage from 800 to 35,000 \AA.  We linearly fit the reddening law parameterized by \citet{Fitzpatrick99}, $k(\lambda)$, to the reddening of each individual quasar, $f(\lambda)$, with respect to the unreddened composite, $f_0(\lambda)$. Assuming standard extinction,
\begin{equation}
f(\lambda) = f_0(\lambda) e^{-\tau_\lambda},
\end{equation}
our linear fit has the form:
\begin{equation}
\log \left(\frac{f(\lambda)}{f_0(\lambda)} \right) = -\tau_\lambda = -\frac{k(\lambda)E(B-V)}{1.086}.
\end{equation}
We exclude strong emission lines and use a robust linear fit which ignores extreme outliers in very noisy spectra and returns $E(B-V)$ as the slope.  

Twenty-two of our 55 quasar spectra have only optical or near-infrared spectra and we adopt the amount of reddening that gave the best fit for each of these spectra.  Figure \ref{fig:single_fit} shows examples of these fits for two objects, one with an optical spectrum and one with a near-infrared spectrum.  

This method of determining $E(B-V)$ does have several drawbacks for determining the reddening in objects with combined optical and near-infrared spectral coverage.  Our spectra were not obtained simultaneously;  the interval between observations ranges from several weeks to a few years.  Aside from variability in brightness, which involves a simple scaling of one spectrum to the other, variability in the spectral shape is also likely and cannot be accounted for.  This is an especially important caveat in the case of dust-obscured quasars, as the obscuring material close to the nucleus may cross the line-of-sight over these timescales leading to different measurements of the source's extinction \citep[cf.][]{Risaliti05}.

In addition, our choice of scaling described in \S 3.2 may have a strong effect on the fit.  Since rest frame UV-optical light is more sensitive to extinction than the near-infrared, we expect that the optical spectrum, when available, will provide a more reliable measure of the extinction.  To determine the effect on our reddening measurements, we fit the reddening both to the combined, as well as to the individual optical spectra alone.  In most cases, fitting to the combined spectra gave satisfactory fits.  Figure \ref{fig:combined_fit} shows an example of two optical-to-near-infrared spectra, and the fitted composite.  However, in about a third of the cases, the combined fit was extremely poor, while fitting the optical spectrum alone yielded an excellent fit.  Figure \ref{fig:compare_fits} shows an example of such a case.

Another caveat to fitting a reddened quasar template to our spectra is the presence of galactic starlight.  Although this is not an issue for luminous, unreddened AGN, where the nuclear light outshines the galaxy, in our highly reddened objects the presence of a galactic component can have an effect on our fits.  This is especially a concern at low redshifts where we are sensitive to both intrinsically lower luminosity quasars, and more heavily obscured objects.  Spectropolarimetry of near-infrared selected quasars from 2MASS confirmed that host galaxy light can contribute significantly to the spectra of type II AGN, diluting the polarized flux from the AGN \citep{Smith03}.

We conducted a two-component fit to our spectra, using the same quasar template plus an elliptical galaxy template from \citet{Mannucci01} allowing the reddening and fraction of quasar and galaxy light to be fit simultaneously.  The fitting function has the form:

\begin{equation}
f(\lambda) = A f_{\rm gal}(\lambda) + B f_{\rm QSO}(\lambda) e^{-\tau_\lambda}.
\end{equation}

Since the strongest feature in an elliptical galaxy template is the 4000\AA\ break, which moves out of the optical at $z\sim 1.5$, and the scaling between the optical and near-infrared spectra could introduce additional uncertainty, we used only the optical spectrum in these fits. The fits converged and significant galactic components were found only for eight objects with $z\lesssim 0.5$.  We subtracted the galactic component from the optical spectrum, and re-scaled it to the near-infrared spectrum.  We then ran the linear fitting procedure (equation [2]) using only the quasar template on these objects.  We compared $E(B-V)$ derived from the three techniques: (1) the original linear fits to the combined spectrum, (2) the two-component fits to the optical spectrum and (3) the galaxy-subtracted, linear fits to the recombined spectrum.  The results generally found the highest derived extinctions using method (2) followed by method (3).   

The following examples illustrate these findings.  For F2M170802.4$+$222725, at $z=0.377$, we derive $E(B-V)=1.05$ using technique (1), $E(B-V)=1.57$ using technique (2), and $E(B-V)=1.10$ using technique (3).  F2M130700.6$+$233805 has only an optical spectrum, so scaling is not a concern.  We derive $E(B-V)=0.58$ using technique (1) for this object.  Technique (2) measures $E(B-V)=0.55$ and technique (3) yields $E(B-V)=0.56$.

These results underscore the difficulty of estimating reddening in quasars spanning such a large redshift range, whose data are highly heterogeneous.  The nature of our sample and of the data introduces unconstrained degrees of freedom into our fitting techniques, including the optical to near-infrared spectrum scaling, possibly variable reddening as well as other variability, and an unknown fraction of galactic starlight contamination.

Taking all of this into account, while striving for consistency, we disregard the galactic contribution and adopted the amount of reddening from either the combined optical and near-infrared spectra or just the optical spectrum.  Whichever gave the best fit upon examination.  The measured extinctions range from $E(B-V) = -0.05$ to $1.5$.  We deredden the spectra to compute their intrinsic $K$-band magnitudes. 

Table \ref{table:ebv} presents the reddening parameters derived for our spectra.  We list each object's J2000 coordinates in columns (1) and (2) followed by the redshift in column (3) and the $K$-band magnitude in column (4).  Column (5) lists the $K$-band extinction based on the $E(B-V)$ value reported in column (9).  Column (6) lists the object's $R-K$ color.  We compare reddenings derived from two different methods in the next three columns.  Columns (7) and (8) list the $E(B-V)$ values derived using Balmer decrements in the total hydrogen line profile and only the broad component of the line profile, respectively, for eligible spectra (see \S 5.2), while column (9) lists the reddening from our composite fitting technique described in this section.  We list in column (10) the spectrum used to measure the extinction.  

\subsection{Reddening Determined from Balmer Decrements}

The line ratios seen in our spectra can provide useful information about the reddening processes along the line of sight.  Many of our objects have been sufficiently extinguished such that the broad emission lines are no longer present in the rest frame optical; it is only from broad Paschen lines that we are able to classify them as quasars.  In some cases the Balmer line profiles show a weakened broad component with sharp narrow lines superimposed.  In these cases we can measure the extinction in the broad and narrow components of the Balmer lines separately to determine if the dust is located behind or in front of the narrow line region.  We can also correlate the extinctions measured directly from Balmer decrements with the extinctions derived from our continuum fits.

Forty-five quasars in our sample have spectra with both H$\alpha$ and H$\beta$ lines included in our wavelength coverage.  Of these, nineteen have both lines in a single optical or near-infrared spectrum.  To avoid uncertainty in the Balmer decrements from the scaling of the infrared spectrum to the optical, we measure Balmer decrements for only these nineteen quasars.  We follow the line-fitting technique outlined in \citet{Greene04} to model the line profiles of the broad and narrow components of H$\alpha$ and H$\beta$ and use these to measure the total line intensity as well as that of the broad components.  We begin by fitting the [\ion{S}{2}] $\lambda\lambda 6716, 6731$ profile, when available, fixing the line separations and limiting the line centers to within $\pm5$ \AA\ from the theoretical values.  We use up to three Gaussian components to model the line profile and allow the scaling between the two [\ion{S}{2}] lines to be a free parameter.  Using this as our narrow-line model, we then fit a narrow H$\alpha$+[\ion{N}{2}] $\lambda\lambda 6548,6583$ model fixing the profiles of the lines to those we derived from [\ion{S}{2}] and the relative strength of $\lambda 6583$ to $\lambda 6548$ to the theoretical value of 2.96. We simultaneously fit a broad H$\alpha$ line allowing up to three Gaussian components to accommodate the fit.  If [\ion{S}{2}] is unavailable, we model the narrow-line profile with [\ion{O}{3}] $\lambda$5007 when it has high signal-to-noise.  In about a third of the objects, no narrow-line profile was available.  These are all higher-redshift objects (all but one with $z>1$) with lower signal-to-noise spectra, especially at H$\beta$.  These objects are also among the highest luminosity sources (see Figure \ref{fig:eb_min_v_color}) and have very broad H$\alpha$ profiles.  In these sources, we fit up to three Gaussian components to H$\alpha$ ignoring any (relatively weak) narrow-line component.

To fit a line profile to the broad and narrow components of H$\beta$, we first fit and subtract the [\ion{O}{3}] $\lambda\lambda 4959, 5007$ doublet, fixing the line separation and allowing the widths and intensities to be fit.  We perform this subtraction on all our spectra, regardless of whether we used the [\ion{O}{3}] profile as a narrow-line model, so that even low signal-to-noise emission from the [\ion{O}{3}] doublet is removed before H$\beta$ is fit.  In the cases where we use a narrow-line model, we fit H$\beta$ with narrow- and broad-line components, fixing the widths of the broad-line components and the relative positions of the broad and narrow components based on the H$\alpha$ fits.   

We compute the line intensity for the broad-only as well as total line profile,  subtracting the continuum from the fit.  We divide the integrated intensity of H$\alpha$ by that of H$\beta$ to obtain the Balmer decrements.  

To obtain extinction values from our Balmer decrements, we follow \citet{Calzetti94} 
\begin{equation}
E(B-V) = \frac{1.086}{k(\mathrm{H}\beta)-k(\mathrm{H}\alpha)}\ln \left(\frac{[\mathrm{H}\alpha / \mathrm{H}\beta]_{measured}}{[\mathrm{H}\alpha / \mathrm{H}\beta]_{FBQS}}\right).
\end{equation}

In order to make a meaningful comparison between the continuum fit and Balmer decrement measurements of the reddening, we use the Balmer decrement from the FBQS composite spectrum as our intrinsic $\mathrm{H}\alpha / \mathrm{H}\beta$ ratio which we measure to be 4.526 for the broad-only components and 4.091 for the combined lines, rather than the theoretical value of 2.88 \citep{Osterbrock89}.  Presumably the FBQS composite already includes some reddening; adopting the composite H$\alpha$/H$\beta$ ratio allows a direct comparison with the $E(B-V)$ values derived from the continuum fits to the composite.  We also use the same SMC extinction law, $k(\lambda)$, that we used in the composite fitting procedure, which gives $k(\mathrm{H}\beta)-k(\mathrm{H}\alpha) = 1.259$ (using the extinction curve parameters from \citealt{Gordon98} and the reddening formula from \citealt{Fitzpatrick99}). 

Figure \ref{fig:bd_example} shows four examples of our Gaussian fitting for computing Balmer decrements.  The {\em solid line} shows the total profile fit.  In panels (c) and (d) where narrow-line models were used, the H$\alpha$+[\ion{N}{2}] profile and the narrow H$\beta$ profile are overplotted with a {\em dotted line}.  The broad-line is overplotted with a {\em dashed line}.  When computing H$\alpha$/H$\beta$ we use both the total hydrogen line profile as well as only the broad components.  We present the measurements of $E(B-V)$ based on Balmer decrements from the total line profile listed in column (7) of Table \ref{table:ebv}.

\subsection{Discussion of Reddening}

In figure \ref{fig:eb_min_v_color} we plot dereddened absolute $K$-band magnitude versus redshift. We color the points according to their measured extinctions derived from the continuum fits (from unreddened in yellow to heavily obscured in red) and overplot them on top of the FBQS II and III quasars which we assume to be largely unabsorbed.  The dashed lines indicate the detection limit of a flux-limited survey with $K<15.5$ for different amounts of extinction. 

The F2M red quasars are among the more luminous objects in this plot especially at higher redshifts.  Still, even the high-redshift quasars are not luminous enough to allow the detection of quasars reddened more than $E(B-V)\sim 0.5$.  At redshifts below $z\sim 0.5$ we find quasars with extinctions of up to $E(B-V)\sim 1.5$.  This trend was noted in \citet{White03b} and in \citetalias{Glikman04}, and demonstrates an inherent bias in flux-limited surveys against finding highly extinguished objects. Even at near-infrared wavelengths, the radiation in the rest frame is optical light beyond $z \sim 1$ and is quite sensitive to extinction.  Ongoing quasar surveys using the {\it Spitzer Space Telescope} should avoid this problem and yield quasar samples unbiased by extinction \citep[cf.][]{Lacy04,Stern04}.

Since almost all the data points in a given spectrum (excluding strong emission lines and regions of strong atmospheric absorption) are used in the composite fitting, this method in effect determines the reddening of the continuum.  \citet{Richards03} found that intrinsically blue quasars selected from the Sloan Digital Sky Survey may have spectral slopes that vary between $-0.25<\alpha_{\nu}<-0.76$. They conclude that these variations are intrinsic and not a result of dust extinction.  However, our extinction measurements are not sensitive to this distinction and may fit an intrinsically  ``optically steep'' quasar with too large an extinction value.  Therefore, it is useful to compare our derived $E(B-V)$ values from continuum fits with  $E(B-V)$ derived from Balmer decrements to study possible discrepancies between continuum shape and line ratios.

To determine if the intrinsic slope of a given quasar is responsible for the discrepancy between the extinction derived from Balmer decrements and the composite fitting, we varied the slope of our composite and recomputed the extinctions.  It is well-known that quasar spectra, though well-approximated by a power-law over a limited wavelength range, change shape in different wavelength regimes and are best described by several broken power-laws \citep{Elvis94,Zheng97,VandenBerk01,Glikman06}.  The UV-optical part of the spectrum covering the wavelength range $\sim 1300$ to $\sim 5000$ \AA\ is best described by $f_\nu \propto \nu^{-0.46}$ on average \citep{Brotherton01,VandenBerk01}.  Therefore we vary the spectral index of our quasar template only over this wavelength range, leaving intact the original shape of the rest of the template.  To shift the slope to a new spectral index, $\alpha_\nu$, we must multiply the spectrum by a power-law with an index which is the difference between the two, $\Delta \alpha_\nu = \alpha_\nu - \alpha_{\nu 0}$.  The new template becomes:

\begin{equation}
F_\lambda = F_{\lambda 0} \left(\frac{\lambda}{\lambda_0}\right)^{-\Delta \alpha_\nu} \qquad \mathrm{for} \ \lambda < \lambda_0 = 5500 \ \mathrm{\AA}
\end{equation}

We tested the extreme values found by \citet{Richards03} ($\alpha_\nu=-0.25$ and $-0.76$) on all our spectra. From these results we compared the 11 objects for which the Balmer decrement fits and continuum fits are discrepant by $E(B-V)\geq 0.2$ and whose individual fits are very good, since for such cases the discrepancies cannot be attributed to poor data and might be explained by an extreme intrinsic slope of the quasar.   We use three power-law shapes for our template fitting \citep[see Figure 7 in][]{Richards03}.  The default spectrum is the mean spectrum derived in \citet{Brotherton01} which we have used for fitting all the red quasar spectra to derive the $E(B-V)$ values listed in Table \ref{table:ebv}.  We also modify our template to have the steepest (in wavelength space) spectral shape found by \citet{Richards03} with $\alpha_\nu=-0.25$.  It is clear from this curve that in cases where the Balmer decrement measures a higher reddening than the continuum fit, we expect that a template with $\alpha_\nu=-0.25$ will provide better agreement since more extinction is required to redden this continuum.  When the Balmer decrement underestimates the reddening compared to the continuum fitting, it may be because the quasar has a flatter (in wavelength space) intrinsic spectrum requiring less extinction to produce the observed spectral shape.  

Four of these objects have $E(B-V) > 1$ measured from their Balmer decrements.  Although using the template with $\alpha_\nu=-0.25$ does yield higher extinctions, the increase is typically $< 0.1$ mag, not enough to match the color-excess derived from the Balmer decrements.  Similarly, we compared continuum fits using a template with $\alpha_\nu=-0.76$ to four quasars whose Balmer decrements measure little or no reddening ($E(B-V) < 0.1$).  Again, though the derived $E(B-V)$ values decreased, it was only by $< 0.1$ mag.  It is possible that these objects require even more extreme slopes in their intrinsic spectra; however it is unlikely that such a large percentage (8/19) of the objects in our sample with both continuum fits and Balmer decrements all have extreme intrinsic continua.

Figure \ref{fig:ebv_compare} compares the results of the two techniques for measuring extinction, using the same extinction law and with respect to the same intrinsic quasar template.  The color-excess derived using Balmer decrements of the total hydrogen line profile are plotted with square symbols, while the color-excess computed from Balmer decrements of the broad component of the lines are plotted with stars.  The horizontal error-bars indicate the range of $E(B-V)$ values obtained when using templates with the $\alpha_\nu=-0.25$ and $-0.76$ spectral slopes.  The error bars for the $E(B-V)$ from the broad component are much larger than from the total line flux, and increase with reddening.   A linear fit correlating the two techniques reveals that the total line profile (solid line, slope=$1.62\pm0.39$, intercept=$-0.35\pm0.23$) is closer to a one-to-one correlation (plotted with a dotted line for comparison) than the broad component (dashed line, slope=$2.99\pm0.57$, intercept=$-0.61\pm0.34$).  However, even when using the total line flux to compute extinctions, this plot shows that although the two techniques yield similar reddening measurements {\it on average}, the scatter in the fit is $\sim 0.5$ magnitudes and cannot be attributed to a steeper or flatter intrinsic spectrum.  Furthermore, Balmer decrements can yield drastically higher extinction values by $E(B-V)\sim 1-2$ in noisy spectra, since H$\beta$ is a weaker line, especially when reddened, and has large errors in the fit.  

The better correlation between the color-excess derived from Balmer decrements from the total line profile, including both the narrow and broad components, with the color-excess derived from reddening in the continuum suggests that the dust in these systems is not necessarily confined to the ``dusty torus'' invoked in quasar unification models \citep{Urry95}.  AGN unification {\em is} effective at explaining the Seyfert1 to Seyfert2 class continuum, where, in this toy-model picture, broad lines seen in the former diminish in the latter as the viewing angle approaches 90 degrees from the jet-axis.  The narrow-line region is farther from the nucleus than the broad line region and is the only contributor of lines in Seyfert2 spectra.  Since the Balmer decrements derived from the broad components of the hydrogen lines in the F2M spectra correlate poorly with the reddening derived from the continuum, it is more likely that the dust in these objects resides outside the nuclear AGN, possibly in starforming regions in the host galaxy.  {\em Hubble Space Telescope} images of 13 F2M red quasars show and unsusually high fraction of these objects are mergin and/or interacting systems.  These images suggest that the obscuring dust is broadly distributed in these systems \citep{Urrutia05b}.  This picture is consistent with the starburst-AGN connection suggested by \citet{Sanders88a} and recently modeled by \citet{Hopkins06}.

In Figure \ref{fig:z_ebv} we show the reddening as a function of redshift (for both estimators) using all 55 extinction measurements from the continuum reddening ({\em open circles}) and the 19 extinction measurements from the Balmer decrements determined from the total line fluxes ({\em filled circles}). The error bars on the open circles indicate the range of $E(B-V)$ values obtained from fitting the templates with extreme steep and flat spectral slopes.  We connect the points from the same object with a line.  The discrepancies between the two techniques do not appear to depend on redshift. However, both techniques do show the bias toward redder objects at lower redshifts noted above.

\subsection{Are There Variations in the Dust Law?}

In addition to the variations in the quasar template, variations in the dust law could also affect our color-excess measurements and the quality of our fits.  Figure \ref{fig:spectra} shows that the F2M red quasar sample includes a highly non-uniform set of spectra.  In addition to intrinsic variations in the spectral shapes of these quasars, our wavelength sampling and coverage as well as signal-to-noise are also highly diverse.  In cases where the continuum fits have not been improved by changing the slope of the quasar template, it is possible that the dust law itself is different enough to affect the fit.

The extinction law for the Milky Way Galaxy has been studied extensively using UV-though-infrared observations of stars along many lines of sight.  The extinction as a function of wavelength is typically parameterized by the ratio of total to selective extinction in the $V$-band, $R_V\equiv A(V)/E(B-V)$ \citep{Cardelli89}.  Galactic values range from 2.2 to 5.5 with a mean value of $\sim 3.1$.  Our extinction law is that derived by \citet{Gordon98} for the Small Magellanic Cloud and is shown in Figure \ref{fig:extinction}.  The solid line represents $R_V=3.1$, while the more extreme values are overplotted.  The only difference between the dust laws occurs in the near-infrared. Dust laws for dust composed of larger grains are characterized by larger values of $R_V$ and have a weaker effect in the near-infrared causing less extinction.  Shortward of $\sim 7000\AA$, the dust laws behave the same, and therefore we do not expect differences in the dust law to affect the extinctions that we derive from the Balmer decrements.

We computed $E(B-V)$ for all 55 of our quasar spectra using three extinction curves, with $R_V=$ 2.2, 3.1, and 5.5, and compared the results.  The quality of the fits were mostly indistinguishable. In a few cases where the combined optical plus near infrared spectra were well-fit by the original $R_V=3.1$ dust law, the extreme values produced slightly poorer fits.  The fits to the $R_V=2.2$ dust law yielded higher $E(B-V)$ values while the $R_V=5.5$ law yielded lower $E(B-V)$ values.  Figure \ref{fig:rv_compare} plots the color excess as a function of redshift for $R_V=3.1$ in filled circles.  The error bars represent the extreme extinctions derived from the $R_V=2.2$ and 5.5 dust laws. The results varied most in spectra for lower redshift objects where we only had a near infrared spectrum.  In these cases the varying sensitivities of the dust laws produced strongly differing results, since the $R_V=2.2$ dust law is more sensitive in the infrared and therefore a higher $E(B-V)$ value is computed.  When only an optical spectrum was available, or at higher redshifts where rest frame optical light is in the observed near-infrared spectrum, the derived extinctions were the same regardless of the dust law.  Combined spectra were mildly affected, and only at low redshifts.  Therefore the apparent trend with redshift is more a result of the heterogeneous nature of our spectra, than of any physical effect of dust law with redshift.  To better determine the nature of the dust in our spectra, longer wavelength data are needed (e.g., {\em Spitzer} IRS spectra).

\section{Radio Spectral Indices}

We have so far explored the red colors of our F2M quasars by assuming that they are a result of dust reddening.  It is possible that there is also a contribution to the $K$-band flux from nonthermal synchrotron emission from jets.  \citet{Serjeant96} argued that enhanced synchrotron emission from relativistic jets lying in the line of sight contributes to the near-infrared flux and is responsible for both the flat radio spectrum and the red optical to near-infrared colors in the red quasars found by \citet{Webster95}.  \citet{Whiting01} looked for a synchrotron signature in the spectral energy distributions of the Parkes red quasars and found evidence for a red synchrotron component in the near-infrared.

The Parkes quasars were drawn from the Parkes Half-Jansky Flat-spectrum Sample \citep[PHFS;][]{Drinkwater97} which included only sources brighter than 0.5 Jy at a radio frequency of 2.7 GHz.  It is possible for these synchrotron spectra to extend toward higher frequencies and contribute enough near-infrared flux to account for the large $R-K$ colors.  On the other hand, the FIRST survey is sensitive down to 1 mJy and 43/56 of the F2M quasars are fainter than 30 mJy at 20 cm (1.4 GHz).  We can put limits on the $K$-band light expected from synchrotron radiation by extrapolating the radio spectrum into the near infrared.  Figure \ref{fig:synch_rad} shows the magnitude of a series of 1 mJy FIRST sources (y-axis on the right) with flat to inverted spectral indices ($\alpha_\nu$ ranging from $-0.5$ to 0.2) at 2.2 \micron\ ($K$ band, y-axis on the left).  This computation makes the simple (and probably naive) assumption that there are no breaks in the radio spectrum.  Although this is likely to be unrealistic, it gives us the ``worst case scenario'' of how much synchrotron flux could be contaminating the $K$ band part of the spectrum.  Under these assumptions, a radio source with a flux density of 1 mJy at 20cm with a radio spectral index of $\gtrsim 0.0$ could contribute $\lesssim 14.6$ magnitudes in $K$ band, which would be detectable in 2MASS.  A 10 mJy source at 20 cm would contribute 14.6 magnitudes in $K$ band even with a spectral index of $\alpha_\nu = -0.2$ and 12 magnitudes with $\alpha_\nu = 0.0$.

Since the faintness of our radio sources does not rule out the possibility that a flat-spectrum source can contaminate our $K$ band light and redden otherwise blue, unobscured quasars, it is necessary to check whether the F2M quasars (selected to be red) have a higher fraction of flat-spectrum sources (accounting for their color) compared with optically-selected radio-detected quasars, such as FBQS.

The quasars in our study reside in a portion of color-color space largely unexplored by optical surveys.  Of the more than 14,000 quasars in the \citet{Veron-Cetty01} catalog, there are only 21 with 2MASS matches that fall outside our survey area, but that reside in our prescribed region of color-color space ($R-K>4$ and $J-K>1.7$).  Nineteen of the 21 are radio-selected objects.  {\it All} of these have flat spectral indices ($\alpha>-0.5$) when radio fluxes at two or more frequencies are available.

These findings are suggestive but are not conclusive. Furthermore, they do not probe sources of intermediate radio loudness, with 20 cm flux densities in the 1-30 mJy range, such as the majority of the objects in the F2M survey. Since the fraction of quasars that are radio sources rises steeply with a decreasing flux density threshold \citep{White00}, it is important to see if a large number of missing (red) quasars continues to appear as we go to fainter fluxes.

\citet{Webster95} and others have shown that quasars {\it selected} by their flat radio spectra have redder colors than optically selected quasars. In order to conclude that the two phenomena are linked, optically selected red quasars ought to have flat radio spectra as well.  To test this, we obtained simultaneous 3.6 cm and 20 cm flux density measurements for 44 of our 57 F2M quasars using the VLA in the CnD configuration in order to determine their radio spectral indices and to test whether the F2M sample exhibits enhanced synchrotron emission.  Data were reduced in the standard manner using the AIPS package.  

Flux densities were derived by fitting two-dimensional elliptical Gaussians to the images; for the weakest sources, the Gaussian peaks were fixed at the optical positions and the source sizes were fixed to the size of the synthesized beam. Peak and integrated flux densities were derived; since we are interested in the contribution any core component makes to the IR emission from the nucleus, we use the peak flux densities to calculate the spectral indices (all but 2 of the 42 objects detected at 3.6~cm are point-like).

The results are presented in Table \ref{table:vla}.  We list the quasar's name in column (1), followed by the FIRST peak and integrated flux densities in columns (2) and (3), respectively.  Our newly measured 20 cm peak and integrated flux densities in the VLA CnD configuration are listed in columns (4) and (5), respectively.  In column (6) we list the NVSS 20 cm flux density in the VLA D configuration as measured by \citet{Condon98}.  Columns (7) and (8) list our newly measured 3.6 cm peak and integrated flux densities, again in the CnD configuration.  Column (9) lists the radio spectral-index, $\alpha_{1.4~\mathrm{GHz}/{8.3~\mathrm(GHz)}}$, computed from the peak flux densities in columns (4) and (7).

We compare these measurements with contemporaneous 20 cm and 3.6 cm VLA measurements for 214 FBQS quasars \citep{Lacy01}.  We divide the samples into a radio-bright and radio-faint sample by their 20 cm flux density to avoid any biases in the FBQS subsample which is biased toward brighter radio sources and thus has a higher proportion of flat-spectrum objects.  Figure \ref{fig:vla} presents spectral index histograms for the F2M and FBQS quasars, normalized by the number of quasars in each sample for ease of comparison. The bright samples (left panel) have indistinguishable distributions.  There is a large fraction of flat-spectrum objects ($\alpha_{1.4 \mathrm{GHz}/8.3 \mathrm{GHz}} > -0.5$) in this subset and enhanced synchrotron emission should be considered as a possible source for the red colors in these quasars.  The faint sample (right panel) shows an excess of {\em steep} spectrum quasars in the F2M sample.  This may be due to the slightly lower median 20 cm flux density (2.68 mJy for F2M versus 3.1 mJy for FBQS).  In these objects dust is the likely cause for the reddening.

In particular, in only eight of the 44 objects observed would a smooth extrapolation from 3.6~cm to $2\mu$m significantly boost the K-band magnitude. Three of these eight are blue quasars (see \S 6), and so are not considered part of the red sample. The other five (representing less than 10\% of our red quasar sample) have $E(B-V)$ values of $\sim 0.2 -- 0.7$, intermediate values for this population. Thus, we conclude that synchrotron contamination does not have a significant impact on any of our conclusions.

\section{The Intrinsic Surface Density of Quasars on the Sky}
 
Having measured the extinctions for our red quasars, we have corrected their observed $K$-band magnitudes to their unobscured values.  The amount of $K$-band absorption ranges from $A_K \sim 0 - 1.2$ and once corrected, the intrinsic magnitude-limit of our survey becomes $K<15.0$.  Figure \ref{fig:acor_sd} shows the density of quasars as a function of dereddened $K$-magnitude.  The number of red quasars begins to decline at $K>14.5$.  This is a result of the 2MASS limit of $K\simeq 15.5$ corrected for a median $\langle A_K\rangle \sim 0.5$.  We ignore the 36 candidates without spectra, among which are the 19 objects that are undetected in the 2MASS $J$-band with $J-K<1.7$ (half of which have $J-K<1.4$).  We also omit F2MJ021542.0$-$022257, the $z=1.187$ quasar from \citet{Drinkwater97}, since we do not have a spectrum from which to measure $E(B-V)$ and determine its intrinsic $K$-band magnitude.  These objects were included in Figure \ref{fig:space_dist}.

We can assume our quasar detection efficiency of $50\%$ applies to the 17 objects with $J$-band detections and expect $\sim 8$ additional red quasars whose observed magnitude range lies between $14.6 < K < 15.5$ (the fainter half of the $K$-magnitude distribution for the whole sample, see Figure \ref{fig:khist}).  Since it is possibile that many of the objects lacking a $J$-band detection actually have $J-K<1.7$ and are therefore likely to be stars, the probability that these objects are quasars is likely less than $50\%$. They fall into the same  $14.6 < K < 15.5$ magnitude range as the former group.

For all the unclassified candidates, we estimate that there are fewer than $18$ quasars.  Assuming the absorption for these objects is the median $A_K$ of the rest of the quasar sample puts them in the dereddened range of $14.1<K<15$.  These objects may increase the density of objects with dereddened $K > 14.0$, where our sample appears to be significantly incomplete.  Of course, their extinctions may be higher than the median, adding counts to the density measurements at $K<14.0$.

Below $K=14.0$ we are able to make a comparison between the dereddened F2M quasars and FBQS II and III (with poorer statistics).  The density of F2M quasars with dereddened $K < 14.0$ is $0.011 \pm 0.002$ deg$^{-2}$.  At the same limit, FBQS II has a density of $0.028\pm0.003$ deg$^{-2}$ (summing the first six rows in column (1) of Table \ref{table:surfden}), implying a missed red quasar fraction of $28\pm6\%$, using FBQS II.  The FBQS III survey covers a smaller area and therefore has fewer objects brighter than $K=14$.  The density of FBQS III objects at this limit is $0.008\pm0.004$ deg$^{-2}$, implying a missed red quasar fraction of $56\pm16\%$.   
 
To a dereddened limit of $K<14.5$ the spatial density of F2M quasars is $0.016 \pm 0.002$ deg$^{-2}$ which implies that red quasars make up $22 \pm 4\%$ of the total quasar population, using the FBQS II sample, and $30 \pm 5\%$ using FBQS III.  These values are lower limits, since the 17 unobserved quasars may contribute counts to the F2M sample at these magnitudes; in addition, quasars with $\langle A_K\rangle > 1.2$ would fall below the 2MASS threshold. 

Again, these are likely lower limits, based on the arguments put forth in Section 4.  In addition, there may be redshift-dependent magnitude biases that exist for reddened objects (as we discussed in Section 5.3) which may impact our estimates of the fraction of red quasars.  Our plot of dereddened absolute $K$-magnitude versus redshift (Figure \ref{fig:eb_min_v_color}) shows regions where the red quasars appear to make up more than $25-50\%$.  For example, the $M_K < -30$,  $0.5<z<1$ region appears to have about 7 red quasars and 6 unreddened quasars, before making any correction for red objects that can't be detected.  On the other hand, the number of red quasars at the fainter, lower redshift end, where the selection effects against finding red quasars are weaker, appears significantly smaller than $50\%$.  To untangle the combined effects of luminosity, reddening and redshift on a flux-limited survey, even in the $K$ band, requires a larger sample reaching fainter flux-limits. 

\section{Blue Quasars in our Red Quasar Survey} 

Though our survey is highly efficient at finding red quasars by selecting for red colors in two sets of filters, we find a handful of blue quasars contaminating our survey.  In \citetalias{Glikman04} we found one blue quasar, F2M J090651.5+495235, at a redshift of 1.635.  We recover this source, along with four others in this survey.  Why do we find blue quasars in a survey targeted for red objects? 

We define ``blue'' quasars as those with $E(B-V)\leq 0.1$ (in column 8 of Table \ref{table:ebv}). In Tables \ref{table:blue_optical} and \ref{table:blue_radio} we present data we have collected from the literature on these objects.  We present optical photometry from three different epochs, GSC II, APM and SDSS to look for evidence of variability to explain why blue quasars end up in our survey; e.g., objects observed at a bright phase with 2MASS but a faint phase in the POSS-II observations.  We also present the FIRST and NVSS radio flux densities to search for variability in the radio which may further suggest variability in the optical. Four of the five ($80\%$) blue quasars have 20 cm flux densities above 100 mJy, compared with 6 of the 52 ($12\%$) red quasars in the whole F2M sample\footnote{We compare radio brightness, rather than radio-loudness (defined in Table \ref{table:blue_radio}) because the large extinction in the optical will produce an artificially large radio-loudness.  To compute the true value of radio-to-optical ratio we must de-redden the $B$ magnitudes of our quasars by $A_B$, which, as we have shown in \S 5, can be highly uncertain.}; since radio-loud quasars are more highly variable than radio-intermediate and radio-quiet populations, this discrepancy is consistent with variability being the explanation for at least some of the blue quasars in our sample.

Tables \ref{table:blue_optical} and \ref{table:blue_radio} summarize the optical and radio properties of these objects, respectively.  Columns (1) and (2) of Table \ref{table:blue_optical} list the J2000.0 Right Ascension and Declination, followed by the redshift in column (3).  The optical photometry is presented in the following nine columns.  Columns (4), (5), and (6) present the GSC-II catalog's photographic $F$ (red) and $J$ (blue) magnitudes, and $J-F$ color, respectively.  Columns (7), (8), and (9) list the APM photographic $E$ (red) and $O$ (blue) magnitudes, and $O-E$ color, respectively. Columns (10), (11), and (12) list the SDSS $r$ and $g$ magnitudes, and $g-r$ color, respectively.  Column (13) lists the 2MASS $K$-band magnitude and column (14) lists the color excess, $E(B-V)$, derived from Table \ref{table:ebv}.  Table 6 presents the J2000.0 coordinates of the five blue quasars in columns (1) and (2) followed by the FIRST peak and integrated radio flux densities in columns (3) and (4).  We list the NVSS 20 cm flux density in column (5) and the 6 cm flux density, mainly derived from \citet{Becker91}, in column (6).  We present the radio spectral index, $\alpha_{20 \mathrm{cm}/6 \mathrm{cm}}$, using the 20 cm flux density from the FIRST (integrated) and the NVSS flux density in columns (7) and (8), respectively.  Column (9) lists the radio-loudness parameters.  Column (10) indicates whether the object is a well-known radio source, based on references listed in the NASA/IPAC Extragalactic Database (NED) and column (11) lists any comments on the source based on observations at other wavelengths or from the literature.

All of these objects have flat or inverted radio spectra ($\alpha > -0.5$, $S_\nu \propto \nu^\alpha$) and have been studied at multiple wavelengths.  F2M115931.8$+$291443 and F2M164258.8$+$394837 are well-known Blazars with well-studied variability properties from $\gamma$-rays to radio wavelengths.  F2M115931.8$+$291443 (4C $+$29.45) has shown optical variability of $\Delta V > 5$ magnitudes \citep{Wills83,Ghosh00}.  F2M164258.8$+$394837 (3C 345) underwent an outburst in 1991 of $\Delta B \sim 2.5$ magnitudes \citep{Webb94,Raiteri98}.  A similar change of $\sim 2$ magnitudes is seen between the APM and GSC-II photometric data presented in Table \ref{table:blue_optical}.  

Figure \ref{fig:blue} shows our three other blue quasars overplotted by their SDSS $g$, $r$, $i$ and 2MASS $J$, $H$, $K_s$ photometry.  The objects in the top two panels show a strong discontinuity between the optical photometric points from SDSS and the 2MASS near-infrared photometry.  These objects also have inverted radio spectral indices consistent with radio and possibly optical variability.  The spectrum in the bottom panel shows a discontinuity between the optical and near-infrared photometry, but also indicates a flattening in the optical spectral shape compared with the spectrophotometric points.

\citet{Giommi04} reveal that the spatial density of Blazars (also referred to as Flat-Spectrum Radio Quasars; FSRQ's) brighter than 0.1 Jy at 5 GHz is 0.18 deg$^{-2}$.  The spatial density of featureless BL Lac objects brighter than 0.1 Jy at 5 GHz is 0.025 deg$^{-2}$.

Our survey is $77\%$ complete over 2716 deg$^2$ suggesting 376 (2716 deg$^2 \times 0.77 \times 0.18$ deg$^{-2}$) expected Blazars and 52 (2716 deg$^2 \times 0.77 \times 0.025$ deg$^{-2}$) BL Lacs.  All five sources with computed spectral indices in Table \ref{table:blue_radio} are have $S_{5\mathrm{GHz}} \gtrsim 0.1$ Jy. Even if we assume that all five of these quasars are Blazars (two are confirmed as such in the literature and the other three have strong circumstantial evidence that they belong to this class) and add the five BL Lacs from the main sample, we are finding less than 10\% of the numbers expected, demonstrating that our survey for red objects strongly selects against these sources.

\section{Red Quasars in Radio Surveys}

In Section 6 we showed that the red colors of the F2M quasars are unlikely to be caused by synchrotron radiation and are most probably a result of dust extinction.  Although it is hard to conceive of a causal connection between the radio properties of quasars and their extinction, there are far more radio-emitting red quasars known than radio-quiet red quasars.

The best explanation for this is that reddened quasars elude optically-selected samples both because they fall out of flux-limited samples and because their colors begin to resemble those of low mass stars, a huge source of contamination.  Samples selected in the radio are insensitive to this bias. Although only $10-20\%$ of quasars are radio-emitters, stars are effectively radio-silent.  Radio selected quasar samples are therefore more effective at finding red quasars.  And quasar composites made from radio-loud and radio-intermediate quasars tend to have redder colors than radio-quiet quasars for this reason \citep[cf.][]{Brotherton01}.

There are several examples of obscuration in very radio-bright samples, such  the Third Cambridge (3C) catalog of radio sources (whose members are brighter than 10 Jy).  \citet{Hill96} looked for broad Pa$\alpha$ in low-redshift narrow-line radio galaxies (NLRGs) from the 3C radio catalog.  They reclassify three of the NLRGs as broad-line radio galaxies after detecting broad Pa$\alpha$ in their near-infrared spectra, as was the case with some of the F2M red quasars.  \citet{Rawlings95} found a broad H$\alpha$ line redshifted into the near-infrared in 3C22, whose observed optical spectrum shows only narrow lines.  Similar to many of the F2M red quasars,  3C68.1 shows broad lines in the optical atop a reddened, highly polarized continuum \citep{Brotherton98}.  Without their strong radio emission, none of these objects would have been selected as quasars in optical surveys.  

\citet{Hill96} were able to explain the extinction in their radio galaxies in terms of orientation.  They found that the broad lines were more reddened than the narrow lines in their spectra, placing the absorbing dust between the broad- and narrow-line regions.  We were not able to confirm this finding for the F2M quasars.  In fact, as we showed in Section 5.3 and Figure \ref{fig:ebv_compare}, the reddening that we measured from the continuum correlated more strongly with the reddening derived from Balmer decrements that were measured from the {\em total} (broad plus narrow) line profiles, rather than the broad-only components.  This suggests that the dust in the F2M quasars is probably dispersed throughout the host galaxy.  The obscuration seen in our sources is likely not an orientation effect, but rather an evolutionary one.  

Although the F2M red quasars are radio-emitters, by definition, they are not necessarily radio-loud.  It is misleading to apply the traditional definition of radio-loudness, $R=f(5 {\rm GHz})/f(B)$ \citep{Stocke92}, to the F2M quasars without correcting for extinction, as most ($70\%$) would fall into the radio-loud regime.  However, to compute the unreddened $B$-band flux-density we must correct for the extinction, which, as we have argued in Section 5.1, is highly uncertain.  \citetalias{Glikman04} computed $R$ for the optically-faintest subset of the F2M sample and found that two-thirds of the quasars fell into the radio-intermediate regime ($3\lesssim R\lesssim 100$), similar to the FBQS quasars.  

\section{Summary and Conclusions}

We have presented the FIRST-2MASS red quasar survey based on sources matched between the FIRST and 2MASS catalogs with optical-to-near-infrared colors obeying $J-K>1.7$ and $R-K>4$.  Our catalog consists of 156 quasar candidates of which 120 have been spectroscopically identified.  We have found 57 quasars, $\sim 50$ of which are true red quasars, most likely obscured by dust.  

We have applied two techniques for determining the extinction in these sources and have compared the results. By applying a Small Magellanic Could extinction law to a quasar template, we fitted the spectral shape of our reddened quasars and derived $E(B-V)$.   In objects where H$\alpha$ and H$\beta$ appear in the same (optical or near-infrared) spectrum we compute extinctions based on the Balmer decrement for both the broad-only component of the lines as well as the total H$\alpha$ and H$\beta$ line profiles. We find better agreement between the two techniques when comparing to the Balmer decrements derived from the total line profile, though there is a large scatter ($\sim 0.5$ magnitudes) between the extinction values derived from the two techniques.  We investigate the effect of changing the optical slope of the quasar template and find that this has a small impact on the values derived.  We also explore varying the grain size in our dust law by changing the $R_V$ parameter.  The results from these tests were inconclusive due to the highly non-uniform nature of our spectra.  Since the SMC dust law varies only at long wavelengths (see Figure \ref{fig:rv_compare}), {\em Spitzer} IRS spectra would cover the ideal wavelength regime to determine the dust law for these quasars.

We note a bias in our sample based on the flux-limited nature of our survey (and lower spectroscopic completeness at fainter magnitudes) which prevents our detection of highly reddened quasars at high redshift.  Our sensitivity to reddened quasars as a function of redshift declines and we can only detect the most luminous sources with moderate extinction at $z>1.5$.  Deeper $K$-band surveys such as the UKIDSS survey (sensitive to $K=18.4$) will provide an opportunity to detect highly obscured quasars at higher redshifts. We can then begin to look for dependence of extinction on redshift and intrinsic luminosity.

We have estimated the percentage of missed red quasars in flux-limited optical surveys that would be recovered in flux-limited near-infrared surveys.  We find that optically selected flux-limited quasar surveys (such as the FBQS II and III) miss $\sim 12\%$ of quasars as a consequence of extinction pushing the observed magnitudes below the surveys' optical flux limit.  We also estimate the true fraction of red quasars after correcting for extinction effects.  We find that for quasars brighter than $K=14.0$ red quasars make up between $\sim 25\%$ and $60\%$ of the total quasar population (depending on the optically selected comparison sample).  Below $K=14.5$ we find that red quasars make up $> 20\%$ to $30\%$ of the total quasar population; this represents a lower limit due to our survey's incompleteness at fainter $K$-band magnitudes.

Finally, we looked at the handful of blue quasars that we find in this survey, concluding that most are selected as a result of their optical variability. We present radio spectral indices for a subset of our red quasars based on contemporaneous 20 cm and 3.6 cm measurements with the VLA.  We conclude that synchrotron radiation may only contribute to the quasars' red colors in the most radio-bright sources ($S_{20~\mathrm{cm}} > 10$ mJy), but the distribution of radio-spectral indices of our sample suggests that reddening as a result of dust obscuration cannot be explained by orientation models for the F2M quasars with flat radio spectra.

We thank Sally Laurent-Muehleisen for assisting with the VLA observations and Johnny Metelsky for reducing and analyzing them.  

This publication makes use of data products from the Two Micron All Sky Survey, which is a joint project of the University of Massachusetts and the Infrared Processing and Analysis Center/California Institute of Technology, funded by the National Aeronautics and Space Administration and the National Science Foundation.

Funding for the creation and distribution of the SDSS Archive has been provided by the Alfred P. Sloan Foundation, the Participating Institutions, the National Aeronautics and Space Administration, the National Science Foundation, the U.S. Department of Energy, the Japanese Monbukagakusho, and the Max Planck Society. The SDSS Web site is http://www.sdss.org/.

The SDSS is managed by the Astrophysical Research Consortium (ARC) for the Participating Institutions. The Participating Institutions are The University of Chicago, Fermilab, the Institute for Advanced Study, the Japan Participation Group, The Johns Hopkins University, the Korean Scientist Group, Los Alamos National Laboratory, the Max-Planck-Institute for Astronomy (MPIA), the Max-Planck-Institute for Astrophysics (MPA), New Mexico State University, University of Pittsburgh, University of Portsmouth, Princeton University, the United States Naval Observatory, and the University of Washington.

The Guide Star CatalogueÐII is a joint project of the Space Telescope Science Institute and the Osservatorio Astronomico di Torino. Space Telescope Science Institute is operated by the Association of Universities for Research in Astronomy, for the National Aeronautics and Space Administration under contract NAS5-26555. The participation of the Osservatorio Astronomico di Torino is supported by the Italian Council for Research in Astronomy. Additional support is provided by  European Southern Observatory, Space Telescope European Coordinating Facility, the International GEMINI project and the European Space Agency Astrophysics Division.

TIFKAM was funded by the Ohio State University, the MDM consortium, MIT, and NSF grant AST-9605012. NOAO and USNO paid for the development of the ALADDIN arrays and contributed the array currently in use in TIFKAM. 

\pagebreak


\begin{deluxetable}{ccccccccccccclc}

\rotate

\tabletypesize{\scriptsize}

\tablewidth{0pt}

\tablecaption{F2M Red Quasar Candidates \label{table:candlist}}


\tablehead{\colhead{R.A.} & \colhead{Dec } & \colhead{$F_{pk}$} & \colhead{$F_{int}$ } & \colhead{$B$} & \colhead{$R$} & \colhead{$J$} & \colhead{$H$} & \colhead{$K_s$} & \colhead{$J-K$} & \colhead{$R-K$} & \colhead{Type} & \colhead{Redshift}  & \colhead{Observations} & \colhead{Ref.} \\ 
\colhead{(J2000)} & \colhead{(J2000)} & \colhead{(mJy)} & \colhead{(mJy)} & \colhead{(mag)} & \colhead{(mag)} & \colhead{(mag)} & \colhead{(mag)} & \colhead{(mag)} & \colhead{(mag)} & \colhead{(mag)} & \colhead{} & \colhead{} & \colhead{} & \colhead{} \\
\colhead{(1)} & \colhead{(2)} & \colhead{(3)} & \colhead{(4)} & \colhead{(5)} & \colhead{(6)} & \colhead{(7)} & \colhead{(8)} & \colhead{(9)} & \colhead{(10)} & \colhead{(11)} &\colhead{(12)} & \colhead{(13)} & \colhead{(14)} & \colhead{(15)}
} 

\startdata
00 23 47.96  &$-$01 45 42.7 &     1.09 &     0.79 &   21.44  &   19.71  & 17.199  & 16.040  & 15.106  &  2.09  &   4.60  & Galaxy    & 0.342  & O10     \\ 
00 36 59.85  &$-$01 13 32.3 &     1.92 &     0.82 &   99.90  &   19.53  & 16.532  & 15.106  & 13.645  &  2.89  &   5.89  & QSO       & 0.294  & O5      \\ 
00 44 02.81  &$-$10 54 18.9 &    38.60 &    42.57 &   22.34  &   19.10  & 17.765  & 16.099  & 15.058  &  2.71  &   4.04  & Galaxy    & 0.431  & O7      & Gl04 \\ 
01 17 12.76  &$-$01 27 55.4 &     8.34 &     8.66 &$>$22.50  &$>$20.80  & 17.278  & 16.361  & 15.333  &  1.94  &   5.47  & Galaxy    & 0.864  & O10     & Gl04 \\ 
01 34 35.68  &$-$09 31 03.0 &   900.38 &   920.52 &$>$22.50  &$>$20.80  & 16.171  & 14.748  & 13.546  &  2.62  &$>$7.25  & QSO       & 2.220  & O20     & Gr02 \\ 
01 58 59.96  &$-$01 28 01.7 &     1.72 &     1.15 &   21.95  &   19.51  & 17.009  & 15.922  & 15.065  &  1.94  &   4.45  & \nodata   & \nodata&         \\ 
02 29 50.66  &$-$08 42 35.5 &     1.52 &     1.72 &   20.28  &   19.18  & 17.264  & 16.005  & 14.843  &  2.42  &   4.34  & QSO       & 0.440  & O10 IR2 \\ 
02 47 59.32  &$-$08 34 08.9 &    33.04 &    34.81 &   20.33  &   19.10  & 17.337  & 16.602  & 15.069  &  2.27  &   4.03  & \nodata   & \nodata&         \\ 
07 01 10.72  &  +52 30 12.0 &     1.64 &     1.36 &   21.35  &   19.30  & 17.104  & 16.118  & 15.266  &  1.84  &   4.03  & SbGal     & 0.363  &         \\ 
07 24 23.98  &  +26 26 44.4 &     1.71 &     1.90 &   21.27  &$>$20.80  & 17.272  & 16.099  & 15.222  &  2.05  &$>$5.58  & Galaxy    & 0.308  & O6      & Gl04 \\ 
07 29 10.35  &  +33 36 34.0 &     3.26 &     3.17 &   22.12  &   20.16  & 16.267  & 15.389  & 14.516  &  1.75  &   5.64  & QSO       & 0.957  & O4      \\ 
07 30 51.03  &  +25 38 59.0 &     1.23 &     1.14 &   22.03  &   19.56  & 17.011  & 15.384  & 14.190  &  2.82  &   5.37  & SbGal     & 0.290  & O5      \\ 
07 33 39.64  &  +45 25 21.5 &     1.69 &     2.28 &$>$22.50  &   19.09  & 16.864  & 16.062  & 15.042  &  1.82  &   4.05  & Galaxy    & 0.453  & O7     & Gl04 \\ 
07 38 06.06  &  +21 41 03.1 &     2.28 &     1.76 &$>$22.50  &$>$20.80  & 17.127  & 16.227  & 15.025  &  2.10  &$>$5.77  & NLAGN     & 0.275  & O3     & Gl04 \\ 
07 38 20.10  &  +27 50 45.5 &     2.64 &     2.33 &$>$22.50  &$>$20.80  & 17.053  & 16.174  & 15.257  &  1.80  &   5.54  & QSO       & 1.985  & O1\phn\ IR4 & Gr02 \\ 
07 38 49.75  &  +31 56 11.8 &    28.17 &    29.18 &   21.00  &   19.19  & 16.973  & 16.121  & 15.025  &  1.95  &   4.16  & NLAGN     & 0.295  & O10 IR2 \\ 
07 53 41.11  &  +25 06 39.4 &     1.51 &     1.83 &   21.62  &   19.42  & 17.254  & 16.209  & 15.217  &  2.04  &   4.20  & QSO       & 0.292  & O10     \\ 
08 10 29.65  &  +16 03 09.3 &     1.03 &     1.07 &   21.55  &   19.15  & 16.920  & 16.949  & 14.922  &  2.00  &   4.23  & SbGal     & 0.287  & O10     \\ 
08 10 58.98  &  +41 34 02.6 &   169.98 &   189.34 &   18.27  &   18.33  & 15.966  & 15.003  & 14.219  &  1.75  &   4.11  & QSO       & 0.506  & O10 IR2 \\ 
08 17 05.51  &  +19 58 43.0 &   271.16 &   280.57 &   19.88  &   19.07  & 16.811  & 15.868  & 14.934  &  1.88  &   4.14  & SbGal     & 0.138  & O4      \\ 
08 17 39.57  &  +43 54 20.0 &     3.10 &     2.80 &   20.95  &   18.66  & 16.570  & 15.375  & 14.092  &  2.48  &   4.57  & QSO       & 0.186  & O10 IR2 \\ 
08 18 16.01  &  +42 22 45.6 &   944.26 &  1004.62 &   18.83  &   18.49  & 15.888  & 14.879  & 13.975  &  1.91  &   4.51  & BLLac     & 0.245  & O20     & Law96 \\ 
08 18 30.91  &  +13 05 27.5 &     1.02 &     1.19 &   21.80  &   19.27  & 17.333  & 16.051  & 15.252  &  2.08  &   4.02  & NLAGN     & 0.310  & O10     \\ 
08 25 02.05  &  +47 16 52.0 &    61.12 &    63.24 &   22.36  &   20.45  & 17.011  & 15.786  & 14.111  &  2.90  &   6.34  & QSO       & 0.803  & O4      & Gl04 \\ 
08 30 11.13  &  +37 59 51.9 &     6.42 &     6.50 &   20.07  &   18.80  & 16.860  & 15.778  & 14.577  &  2.28  &   4.22  & QSO       & 0.414  & O5      \\ 
08 34 07.01  &  +35 06 01.8 &     1.22 &     0.55 &   21.51  &   18.96  & 16.545  & 15.718  & 14.652  &  1.89  &   4.31  & QSO       & 0.470  & O10 IR8 & Gl04 \\ 
08 41 04.98  &  +36 04 50.1 &     6.49 &     6.58 &$>$22.50  &$>$20.80  & 17.534  & 16.305  & 14.916  &  2.62  &$>$5.88  & QSO       & 0.553  & O6\phn\ IR1 & Gl04 \\ 
08 43 32.28  &  +49 21 16.2 &     3.80 &     3.82 &   20.34  &   19.50  & 17.417  & 15.912  & 15.128  &  2.29  &   4.37  & QSO       & 0.420  & O11     \\ 
08 49 38.56  &  +14 39 10.8 &     1.10 &     0.70 &   20.72  &   19.38  & 17.280  & 16.168  & 15.359  &  1.92  &   4.02  & SbGal     & 0.218  &         \\ 
09 04 50.51  &$-$01 45 24.5 &     2.23 &     2.75 &   21.26  &   20.36  & 16.725  & 15.964  & 14.904  &  1.82  &   5.46  & QSO       & 1.005  & O10 IR2 \\ 
09 06 51.52  &  +49 52 36.0 &    67.87 &    75.11 &   20.82  &$>$20.80  & 17.057  & 15.831  & 15.144  &  1.91  &$>$5.66  & QSO       & 1.635  & O1      & Gl04 \\ 
09 13 03.81  &  +19 42 34.7 &     1.57 &     2.31 &$>$22.50  &   19.96  & 17.466  & 16.556  & 15.508  &  1.96  &   4.45  & Galaxy?   & \nodata& O11     \\ 
09 15 01.71  &  +24 18 12.2 &     1.00 &     1.45 &   20.92  &$>$20.80  & 16.513  & 15.253  & 13.790  &  2.72  &$>$7.01  & QSO       & 0.842  & O4\phn\ IR2 \\ 
09 15 09.55  &  +18 19 24.6 &     9.80 &    10.12 &   20.89  &   19.42  & 17.237  & 16.065  & 15.257  &  1.98  &   4.16  & SbGal     & 0.290  & O11     \\ 
09 15 43.73  &  +28 50 34.7 &     1.48 &     1.14 &   20.38  &   19.35  & 16.944  & 15.977  & 15.024  &  1.92  &   4.33  & NLAGN     & 0.270  & O11     \\ 
09 16 08.57  &$-$03 19 39.9 &   117.25 &   125.77 &   21.54  &   19.54  & 16.674  & 15.282  & 14.816  &  1.86  &   4.72  & QSO       & 1.560  & O3\phn\ IR1 \\ 
09 16 48.91  &  +38 54 28.3 &   967.97 &  1063.49 &   19.34  &   18.77  & 16.628  & 15.965  & 14.753  &  1.88  &   4.02  & QSO       & 1.267  & O20     \\ 
09 19 20.64  &  +15 31 51.5 &     1.53 &     1.45 &   20.81  &   18.87  & 16.610  & 15.475  & 14.836  &  1.77  &   4.03  & SbGal     & 0.175  & O12     \\ 
09 21 22.13  &  +30 29 09.9 &     1.25 &     2.15 &   21.59  &   20.38  & 17.582  & 16.192  & 14.993  &  2.59  &   5.39  & Galaxy    & 0.423  & O10     \\ 
09 21 45.69  &  +19 18 12.6 &     4.40 &     4.72 &   21.32  &   20.12  & 16.749  & 15.429  & 14.548  &  2.20  &   5.57  & QSO       & 1.791  & O9\phn\ IR8 & S02\\ 
09 22 04.17  &  +13 21 53.5 &     4.73 &     4.63 &   20.52  &   18.72  & 16.386  & 15.465  & 14.681  &  1.70  &   4.04  & \nodata   & \nodata&         \\ 
09 27 44.38  &  +39 30 38.4 &     1.32 &     1.53 &   20.74  &   18.70  & 16.698  & 15.745  & 14.591  &  2.11  &   4.11  & QSO       & 1.170  & O12     \\ 
09 38 05.72  &  +17 55 50.4 &   106.63 &   122.73 &   20.05  &   18.78  & 16.441  & 15.527  & 14.725  &  1.72  &   4.06  & \nodata   & \nodata&         \\ 
09 40 14.71  &  +26 03 29.9 &   380.89 &   455.66 &   19.53  &   18.73  & 16.643  & 15.427  & 14.729  &  1.91  &   4.00  & BLLac     & 0.498? &        & P98 \\ 
09 43 20.51  &  +36 54 45.5 &     1.33 &     1.19 &   21.69  &   20.06  & 17.098  & 15.749  & 15.107  &  1.99  &   4.95  & Galaxy    & 0.555  & O11     \\ 
09 44 51.93  &  +31 13 52.8 &     1.52 &     1.33 &$>$22.50  &   19.65  & 17.005  & 16.227  & 15.283  &  1.72  &   4.37  & Galaxy    & 0.512  &        & Gl04 \\ 
09 44 54.18  &  +41 56 59.0 &     1.71 &     1.59 &   19.06  &$>$20.80  & 16.963  & 15.830  & 15.091  &  1.87  &$>$5.71  & \nodata   & \nodata&         \\ 
09 46 42.55  &  +18 40 34.0 &     3.69 &     5.00 &$>$22.50  &$>$20.80  & 19.032  & 17.953  & 15.464  &  3.57  &$>$5.34  & SbGal     & 0.403  &        & Gl04 \\ 
09 49 47.37  &  +23 25 22.0 &     1.05 &     1.64 &   21.18  &   19.27  & 17.206  & 15.774  & 15.253  &  1.95  &   4.02  & NLAGN     & 0.379  & O11     \\ 
09 50 32.44  &  +18 52 22.8 &     1.51 &     1.40 &$>$22.50  &   19.64  & 17.336  & 16.151  & 15.434  &  1.90  &   4.21  & Galaxy    & 0.480  & O15    & Gl04 \\ 
09 58 53.63  &  +12 38 23.0 &     3.31 &     3.25 &   20.74  &   19.34  & 16.761  & 15.798  & 14.739  &  2.02  &   4.60  & \nodata   & \nodata&         \\ 
10 01 25.31  &  +44 45 59.8 &     2.55 &     3.17 &   22.21  &   19.28  & 17.169  & 16.107  & 15.073  &  2.10  &   4.21  & Galaxy    & 0.466  & O11     \\ 
10 04 11.78  &  +29 43 28.6 &     1.11 &     2.01 &   21.94  &$>$20.80  & 17.030  & 16.933  & 15.047  &  1.98  &$>$5.75  & NLAGN     & 0.383  & O11     \\ 
10 04 24.87  &  +12 29 22.4 &    11.42 &    12.32 &$>$22.50  &$>$20.80  & 16.550  & 15.504  & 14.539  &  2.01  &$>$6.26  & QSO       & 2.658  & O4\phn\ IR1 & Lac02 \\ 
10 10 57.09  &  +30 00 20.4 &     1.65 &     1.47 &$>$22.50  &   19.40  & 17.266  & 15.945  & 14.952  &  2.31  &   4.45  & Galaxy    & 0.513  & O7      \\ 
10 12 30.49  &  +28 25 27.2 &     9.23 &     8.76 &$>$22.50  &$>$20.80  & 17.325  & 16.921  & 15.270  &  2.06  &$>$5.53  & QSO       & 0.937  & O1      & Gl04 \\ 
10 13 48.29  &  +17 20 39.1 &    60.84 &    61.39 &   20.72  &   19.78  & 16.832  & 15.864  & 15.105  &  1.73  &   4.68  & \nodata   & \nodata&         \\ 
10 15 28.64  &  +12 07 51.9 &    10.60 &    10.98 &   22.15  &   19.56  & 17.323  & 16.475  & 15.298  &  2.02  &   4.26  & NLAGN     & 0.426  & O4      & Gl04 \\ 
10 18 13.17  &  +19 03 04.1 &    60.65 &    66.85 &   20.70  &   19.95  & 17.811  & 16.157  & 15.173  &  2.64  &   4.78  & Galaxy?   & \nodata& O11     \\ 
10 18 29.72  &  +21 35 29.6 &     1.57 &     1.00 &   19.68  &   18.33  & 16.211  & 15.124  & 14.088  &  2.12  &   4.24  & SbGal     & 0.243  & O11     \\ 
10 22 29.39  &  +19 29 38.9 &     2.09 &     2.41 &   20.94  &   19.20  & 16.906  & 16.082  & 15.121  &  1.78  &   4.08  & NLAGN     & 0.406  &        & Gl04  \\ 
10 22 57.45  &  +16 28 40.5 &    22.22 &    22.37 &   21.19  &   20.42  & 17.728  & 16.152  & 15.357  &  2.37  &   5.06  & NLAGN     & 0.577  & O15     \\ 
10 36 33.54  &  +28 28 21.6 &     4.25 &     4.37 &   21.28  &   20.01  & 16.977  & 16.786  & 15.271  &  1.71  &   4.74  & QSO       & 1.762  & O5\phn\ IR1 \\ 
10 38 19.65  &  +22 18 47.4 &     1.37 &     1.52 &   18.98  &   19.64  & 16.900  & 15.888  & 14.925  &  1.97  &   4.05  & \nodata   & \nodata&         \\ 
10 43 35.72  &  +19 28 21.5 &     1.16 &     1.44 &   21.94  &   19.62  & 17.630  & 16.502  & 15.552  &  2.08  &   4.07  & Galaxy    & 0.570  & O11     \\ 
10 43 55.37  &  +21 57 17.6 &     1.74 &     1.78 &   21.32  &   19.65  & 17.713  & 16.590  & 15.488  &  2.22  &   4.16  & Galaxy    & 0.399  &  O7     \\ 
10 46 42.55  &  +16 04 17.4 &     1.49 &     1.30 &   22.20  &   19.94  & 17.432  & 16.002  & 15.466  &  1.97  &   4.47  & QSO       & 1.440  &         & X94 \\ 
10 49 02.95  &  +40 10 31.6 &     1.96 &     2.32 &   22.18  &   19.66  & 17.296  & 15.944  & 14.927  &  2.37  &   4.73  & SbGal     & 0.561  & O7      \\ 
             &              &          &          &          &          &         &         &         &        &         & NLemit    & 0.715  & O7      \\
10 49 18.24  &  +15 44 58.1 &     5.33 &     5.42 &$>$22.50  &$>$20.80  & 17.857  & 16.762  & 15.362  &  2.49  &$>$5.44  & Galaxy    & 0.677  & O15     & Gl04 \\ 
11 06 07.26  &  +28 12 47.0 &   211.47 &   214.79 &   19.90  &   20.02  & 16.556  & 15.544  & 14.629  &  1.93  &   5.39  & QSO       & 0.842  &         \\ 
11 13 54.68  &  +12 44 38.9 &     2.99 &     4.22 &   20.87  &   19.00  & 16.212  & 15.031  & 13.665  &  2.55  &   5.33  & QSO       & 0.681  & O5\phn\ IR1 \\ 
11 17 25.05  &  +31 55 30.4 &     5.51 &     6.57 &   21.51  &   19.24  & 17.234  & 15.717  & 15.238  &  2.00  &   4.00  & Galaxy    & 0.433  &         \\ 
11 17 33.83  &$-$02 36 00.3 &   608.04 &   961.26 &   20.27  &   18.75  & 15.952  & 14.980  & 14.047  &  1.90  &   4.70  & QSO       & 0.463  & O4      \\ 
11 18 11.06  &$-$00 33 41.9 &     1.30 &     1.81 &$>$22.50  &   19.87  & 17.043  & 15.637  & 14.584  &  2.46  &   5.29  & QSO       & 0.686  & O5\phn\ IR8 & Gl04 \\ 
11 39 19.45  &  +45 55 31.0 &     1.18 &     0.53 &   20.65  &   18.97  & 16.687  & 15.934  & 14.958  &  1.73  &   4.01  & \nodata   & \nodata&         \\ 
11 39 26.18  &  +27 43 56.6 &     2.63 &     2.43 &   20.72  &   19.12  & 17.051  & 15.088  & 13.063  &  3.99  &   6.06  & NLAGN     & 0.394  & O6      \\ 
11 46 58.31  &  +39 58 34.2 &   335.60 &   338.53 &   18.47  &   18.17  & 15.365  & 14.486  & 13.535  &  1.83  &   4.64  & QSO       & 1.088  & O11     \\ 
11 51 24.07  &  +53 59 57.4 &     3.52 &     3.60 &   21.92  &$>$20.80  & 17.105  & 16.277  & 15.102  &  2.00  &$>$5.70  & QSO       & 0.780  & O5\phn\ IR1 & Gl04 \\ 
11 53 44.48  &  +29 05 23.8 &     4.66 &     4.81 &   21.68  &   19.25  & 16.903  & 16.248  & 15.119  &  1.78  &   4.13  & \nodata   & \nodata&         \\ 
11 59 31.84  &  +29 14 43.9 &  1855.80 &  1952.59 &   17.67  &   17.59  & 13.180  & 12.304  & 11.477  &  1.70  &   6.11  & QSO       & 0.730  &         & Gr96 \\ 
12 01 11.14  &$-$03 32 19.7 &    55.27 &    56.56 &$>$22.50  &   19.40  & 16.680  & 15.849  & 14.921  &  1.76  &   4.48  & Galaxy    & 0.755  & O13     \\ 
12 02 55.33  &  +26 15 18.8 &    43.01 &    44.25 &   20.69  &   19.34  & 17.276  & 16.424  & 15.194  &  2.08  &   4.15  & NLAGN     & 0.566  & O5      & Gl04 \\ 
12 06 18.46  &  +28 57 19.7 &     5.24 &     5.59 &   20.76  &   19.30  & 17.388  & 16.061  & 15.269  &  2.12  &   4.03  & Galaxy    & 0.166  & O8      \\ 
12 09 21.17  &$-$01 07 17.0 &     1.42 &     1.25 &   20.35  &   17.86  & 15.917  & 14.607  & 13.718  &  2.20  &   4.14  & QSO       & 0.363  &\opt IR5 \\ 
12 25 11.36  &  +27 57 27.2 &     2.12 &     1.70 &   21.38  &   19.59  & 17.407  & 16.272  & 15.391  &  2.02  &   4.20  & \nodata   & \nodata& O21     \\ 
12 27 03.21  &  +50 53 56.3 &     3.87 &     4.27 &   20.05  &   18.77  & 16.324  & 15.524  & 14.613  &  1.71  &   4.16  & QSO       & 0.768  & O20 IR6 \\ 
13 03 03.23  &  +24 33 55.8 &   106.84 &   110.98 &   19.30  &   18.22  & 15.562  & 14.727  & 13.830  &  1.73  &   4.39  & BLLac     & 0.993  &         \\ 
13 04 04.07  &  +46 12 53.6 &     2.14 &     2.63 &   19.86  &   18.32  & 16.758  & 15.378  & 14.320  &  2.44  &   4.00  & QSO       & 0.315  &\opt IR5 \\ 
13 07 00.62  &  +23 38 05.3 &     2.99 &     2.63 &   20.79  &   19.37  & 16.788  & 15.202  & 13.446  &  3.34  &   5.92  & QSO       & 0.275  & O5      \\ 
13 21 15.23  &  +41 05 16.0 &    11.93 &    12.78 &   20.59  &   18.94  & 16.517  & 15.637  & 14.320  &  2.20  &   4.62  & NLemit    & 0.210  &\opt IR3 \\ 
13 41 08.11  &  +33 01 10.3 &    69.41 &    69.81 &$>$22.50  &$>$20.80  & 16.926  & 15.545  & 14.847  &  2.08  &$>$5.95  & QSO       & 1.720  & O11 IR3 & Gl04\\ 
13 44 08.31  &  +28 39 32.0 &     9.76 &    10.41 &   21.64  &$>$20.80  & 16.577  & 16.430  & 14.784  &  1.79  &$>$6.02  & QSO       & 1.770  & O11 IR3 \\ 
13 53 08.65  &  +36 57 51.2 &     3.71 &     3.32 &$>$22.50  &$>$20.80  & 17.410  & 15.582  & 14.264  &  3.15  &$>$6.54  & QSO       & 1.311  & O9\phn\ IR8 & Gl04 \\ 
13 57 32.14  &  +32 45 29.8 &     1.28 &     3.48 &   21.73  &   19.36  & 17.174  & 16.688  & 15.138  &  2.04  &   4.22  & \nodata   & \nodata&         \\ 
13 59 41.18  &  +31 57 40.5 &     1.33 &     1.28 &$>$22.50  &$>$20.80  & 16.927  & 15.835  & 14.788  & 2.14   &$>$6.01  & QSO       & 1.724  & \opt IR1    & Gl04\\
14 08 11.61  &  +32 43 50.2 &     1.09 &     2.26 &$>$22.50  &   19.59  & 17.152  & 16.168  & 15.353  &  1.80  &   4.24  & QSO       & 0.338  & O12     \\ 
14 12 14.30  &  +29 12 08.3 &    14.34 &    14.22 &   20.87  &   19.42  & 17.287  & 15.499  & 14.890  &  2.40  &   4.53  & \nodata   & \nodata&         \\ 
14 15 22.83  &  +33 33 06.5 &     5.21 &     6.21 &   20.58  &   18.92  & 16.716  & 15.372  & 14.304  &  2.41  &   4.62  & QSO       & 0.416  &\opt IR5 \\ 
14 22 46.15  &  +24 04 11.4 &     1.87 &     1.35 &$>$22.50  &$>$20.80  & 17.525  & 16.168  & 15.325  &  2.20  &$>$5.47  & Mstar     & \nodata& O1      & Gl04 \\ 
14 25 49.04  &  +14 24 56.7 &   509.19 &   523.79 &   19.28  &   19.19  & 16.636  & 15.941  & 14.843  &  1.79  &   4.44  & \nodata   & \nodata&         \\ 
14 27 44.34  &  +37 23 37.4 &     1.71 &     1.52 &   21.50  &   19.59  & 17.110  & 15.780  & 15.071  &  2.04  &   4.52  & QSO       & 2.168  & O11     \\ 
14 46 59.92  &$-$00 50 14.9 &    30.10 &    35.69 &   21.52  &   18.94  & 16.745  & 16.080  & 14.801  &  1.94  &   4.14  & \nodata   & \nodata&\opt IR6 \\ 
14 48 12.15  &  +30 56 14.5 &     1.02 &     0.94 &   22.02  &   19.39  & 16.672  & 16.197  & 14.960  &  1.71  &   4.43  & Galaxy    & 0.271  & O11 IR1 & Gl04 \\ 
14 54 06.68  &  +19 50 28.0 &     1.34 &     0.86 &   20.74  &   18.89  & 17.079  & 15.915  & 14.710  &  2.37  &   4.18  & Sey1.5    & 0.260  &         & S02 \\ 
14 58 43.41  &  +35 42 57.5 &   688.50 &   693.13 &   20.63  &   18.96  & 16.795  & 15.888  & 14.951  &  1.84  &   4.01  & \nodata   & \nodata& O19     \\ 
14 58 44.82  &  +37 20 21.6 &   266.39 &   269.98 &   18.63  &$>$18.50  & 16.128  & 15.141  & 14.269  &  1.86  &$>$4.23  & Galaxy?   & 0.333  &         & V96 \\ 
15 31 50.48  &  +24 23 17.6 &     2.24 &     2.00 &   20.52  &   19.17  & 16.667  & 15.783  & 14.693  &  1.97  &   4.48  & QSO       & 2.287  &\opt IR6 \\ 
15 32 33.19  &  +24 15 26.8 &     7.43 &     9.37 &   20.57  &   20.09  & 16.831  & 15.658  & 14.956  &  1.87  &   5.13  & QSO       & 0.562  &         \\ 
15 45 16.90  &  +38 15 39.6 &     1.86 &     1.33 &   19.96  &   18.20  & 16.612  & 15.314  & 14.069  &  2.54  &   4.13  & NLemit    & 0.290  &\opt IR5 \\ 
15 49 38.80  &  +12 45 07.9 &     1.80 &     1.79 &   18.92  &   17.61  & 15.824  & 14.565  & 13.524  &  2.30  &   4.09  & QSO       & 2.373  &\opt IR5 \\ 
15 55 23.72  &  +34 19 39.1 &     1.49 &     1.27 &   20.84  &   19.21  & 17.136  & 16.151  & 15.136  &  2.00  &   4.07  & \nodata   & \nodata&         \\ 
16 00 34.56  &  +35 22 27.0 &     3.17 &     3.30 &   20.52  &   18.59  & 16.412  & 15.400  & 14.170  &  2.24  &   4.42  & QSO       & 0.707  &\opt IR4 \\ 
16 12 53.62  &  +33 20 10.3 &     2.91 &     2.32 &   21.08  &   19.39  & 17.417  & 16.130  & 15.347  &  2.07  &   4.04  & \nodata   & \nodata&         \\ 
16 14 54.10  &  +35 14 13.2 &     1.87 &     1.43 &   22.12  &   19.32  & 17.167  & 16.136  & 15.306  &  1.86  &   4.01  & \nodata   & \nodata&         \\ 
16 18 09.72  &  +35 02 08.5 &    13.73 &    16.22 &$>$19.50  &   18.25  & 16.785  & 15.408  & 14.054  &  2.73  &   4.20  & QSO       & 0.447  &\opt IR5 \\ 
16 42 58.80  &  +39 48 37.2 &  6050.06 &  6598.61 &   17.42  &   16.68  & 14.260  & 13.289  & 12.360  &  1.90  &   4.32  & QSO       & 0.593  & O20 IR4 \\
16 54 07.69  &  +34 35 27.4 &     1.85 &     1.72 &   19.97  &   18.27  & 16.548  & 15.386  & 14.247  &  2.30  &   4.02  & NLAGN     & 0.215  &         \\ 
16 56 47.11  &  +38 21 36.7 &     4.12 &     3.89 &$>$22.50  &$>$20.80  & 17.299  & 17.718  & 15.087  &  2.21  &$>$5.71  & QSO       & 0.732  & O2      & Gl04 \\ 
17 04 25.31  &  +30 37 08.7 &     1.33 &     1.22 &   20.52  &$>$20.80  & 17.311  & 16.263  & 14.917  &  2.39  &   5.88  & QSO       & 0.311  &\opt IR4 \\  
17 08 02.53  &  +22 27 25.7 &     1.31 &     1.74 &   22.32  &   19.35  & 17.036  & 15.959  & 14.575  &  2.46  &   4.77  & QSO       & 0.377  &O3\phn\ IR6 \\ 
17 10 58.40  &  +31 09 58.1 &     6.09 &     6.16 &   16.32  &$>$20.80  & 16.742  & 15.818  & 14.853  &  1.89  &$>$5.95  & \nodata   & \nodata&\opt IR7 \\ 
17 24 05.43  &  +40 04 35.9 &   380.70 &   456.24 &   21.90  &   19.75  & 16.130  & 15.001  & 14.074  &  2.06  &   5.68  & Galaxy?   & 1.49   & IR3 IR5 & V96 \\ 
17 24 28.03  &  +33 31 19.9 &     1.58 &     1.35 &   20.88  &   18.48  & 16.927  & 15.734  & 14.461  &  2.47  &   4.02  & QSO       & 0.370  &         \\ 
17 33 53.18  &  +41 24 55.5 &     1.63 &     0.96 &   21.03  &   19.04  & 17.226  & 15.632  & 15.003  &  2.22  &   4.04  & \nodata   & \nodata&         \\ 
22 16 33.73  &$-$00 54 51.2 &     1.31 &     0.80 &   19.92  &   17.91  & 15.946  & 14.942  & 13.777  &  2.17  &   4.13  & QSO       & 0.200  & O18 IR2 \\ 
22 22 52.78  &$-$02 02 57.4 &     2.43 &     2.66 &   21.37  &   20.04  & 17.436  & 16.247  & 15.141  &  2.30  &   4.90  & QSO       & 2.252  & O10 IR2 \\ 
22 24 38.35  &$-$00 07 50.9 &     1.41 &     1.29 &$>$22.50  &   19.78  & 16.810  & 15.703  & 15.020  &  1.79  &   4.76  & Galaxy    & 0.452  & O10     & Gl04 \\ 
22 47 30.20  &  +00 00 06.7 &   183.71 &   190.56 &   17.99  &   18.17  & 15.649  & 14.820  & 13.860  &  1.79  &   4.31  & BLLac     & \nodata& O19 IR2 & B01\\ 
23 11 04.66  &$-$02 14 58.2 &     4.41 &     7.05 &   22.05  &   19.51  & 17.281  & 15.588  & 15.055  &  2.23  &   4.45  & Galaxy    & 0.499  & O7      \\ 
23 25 49.88  &$-$10 52 43.4 &     1.46 &     1.10 &   20.71  &   19.70  & 17.382  & 16.311  & 15.238  &  2.14  &   4.46  & QSO       & 0.564  & O10 IR2 \\ 
23 28 21.74  &$-$11 07 03.0 &     6.13 &     5.82 &   19.25  &   17.96  & 16.051  & 15.028  & 13.912  &  2.14  &   4.05  & NLAGN     & 0.232  & O14     \\ 
23 39 03.83  &$-$09 12 21.0 &     4.34 &     4.06 &   20.95  &   18.63  & 15.805  & 15.052  & 14.088  &  1.72  &   4.54  & QSO       & 0.660  &         \\ 
23 45 54.91  &$-$10 03 29.1 &     2.39 &     2.09 &   21.18  &   18.48  & 16.893  & 15.754  & 14.187  &  2.71  &   4.29  & Galaxy    & 0.263  & O6\phn\ IR2 \\ 
23 50 40.11  &$-$10 43 59.2 &     1.35 &     0.96 &   21.27  &   19.24  & 17.096  & 15.918  & 15.043  &  2.05  &   4.20  & Galaxy    & 0.253  &\opt IR4 \\ 
\enddata


\tablecomments{Optical spectra: (O1) ESI 2000 July; (O2) ESI 2000 August; (O3) ESI 2001 March; (O4) ESI 2001 May; (O5) ESI 2001 June; (O6) ESI 2002 January; (O7) ESI 2002 February; (O8) ESI 2002 March; (O9) ESI 2002 April; (O10) ESI 2002 December; (O11) ESI 2003 February; (O12) ESI 2004 March; (O13) ESI 2005 January; (O14) LRIS 1999 August; (O15) LRIS 2000 February; (O16) LRIS 2003 June; (O17) Kast 1998 December; (O18) Kast 2000 September; (O19) Kast 2001 June; (O20) SDSS Spectrum; (O21) Double Spectrograph 2006 April \\
Near-infrared spectra: (IR1) SpeX 2002 May; (IR2) Spex 2002 October; (IR3) Spex 2003 April; (IR4) Spex 2003 September; (IR5) Spex 2004 May; (IR6) Spex 2004 June; (IR7) Spex 2004 September; (IR8) TIFKAM 2002 April}

\tablerefs{ (B01) \citet{Becker01}; (Gr96) \citet{Gregg96}; (Gr02) \citet{Gregg02}; (Gl04) \citet{Glikman04}; (Lac02) \citet{Lacy02}; (Law96) \citet{Lawrence96}; (P98)\citet{Perlman98}; (S02) \citet{Smith02}; (V96) \citet{Vermeulen96}; (X94) \citet{Xu94} }

\end{deluxetable}


\begin{deluxetable}{ccccccccccccccc}

\rotate

\tabletypesize{\scriptsize}

\tablewidth{0pt}

\tablecaption{ F2M Candidates with $J_{\mathrm{SNR}} < 3$ and $J-K<1.7$  \label{table:jlim_cands}}


\tablehead{\colhead{R.A.} & \colhead{Dec } & \colhead{$F_{pk}$} & \colhead{$F_{int}$} & \colhead{$B$} & \colhead{$R$} & \colhead{$J$} & \colhead{$H$} & \colhead{$K_s$} & \colhead{$J-K$} & \colhead{$R-K$} & \colhead{Type} & \colhead{Redshift} & \colhead{Observations} &\colhead{Ref} \\ 
\colhead{(J2000)} & \colhead{(J2000)} & \colhead{(mJy)} & \colhead{(mJy)} & \colhead{(mag)} & \colhead{(mag)} & \colhead{(mag)} & \colhead{(mag)} & \colhead{(mag)} & \colhead{(mag)} & \colhead{(mag)} & \colhead{} & \colhead{} & \colhead{} & \colhead{}\\
\colhead{(1)} & \colhead{(2)} & \colhead{(3)} & \colhead{(4)} & \colhead{(5)} & \colhead{(6)} & \colhead{(7)} & \colhead{(8)} & \colhead{(9)} & \colhead{(10)} & \colhead{(11)} &\colhead{(12)} & \colhead{(13)} & \colhead{(14)} & \colhead{(15)}
} 

\startdata
02 15 42.02 &  $-$02 22 57.1 &   359.46 &   403.64 &     21.83                  &  20.38                  &  $>$16.42                     &  16.225                  &   15.059                  &   $>$1.36 &  5.32 & QSO    & 1.178  &  & D97\\ 
07 55 27.61 &  $+$24 46 44.5 &     1.13 &     1.25 &     20.33                  &  19.36                  &  $\phn$17.33\tablenotemark{a} &  15.885\tablenotemark{a} &   15.285\tablenotemark{a} &$\phn$2.05 &  4.08 &  &  & \\ 
08 20 25.97 &  $+$47 28 03.1 &     1.24 &     1.29 &     21.09                  &  19.36                  &  $>$16.74                     &  16.141                  &   15.117                  &   $>$1.62 &  4.24 &  &  & \\ 
08 23 24.76 &  $+$22 23 03.3 &  1919.32 &  2163.90 &     21.07                  &  19.73                  &  $>$16.98                     &  16.408                  &   15.418                  &   $>$1.56 &  4.31 & BLLac  & 0.951  & O20 &  S93 \\ 
08 46 22.29 &  $+$18 22 50.0 &    13.21 &    13.67 &  21.18\tablenotemark{b}    &  19.42\tablenotemark{b} &  $>$16.25                     &  15.986                  &   14.871                  &   $>$1.38 &  4.55 &  &  & \\ 
09 03 55.86 &  $+$20 20 19.0 &     2.26 &     5.21 &     22.50                  &  19.48                  &  $>$16.85                     &  16.415                  &   15.361                  &   $>$1.49 &  4.12 &  &  & \\ 
09 19 46.61 &  $+$22 58 22.1 &     2.27 &     2.56 &     22.00                  &  20.24                  &  $>$16.63                     &  15.458                  &   15.436                  &   $>$1.19 &  4.80 &  &  & \\ 
09 23 57.10 &  $+$15 38 29.5 &     1.20 &     0.72 &  $>$22.50                  &  19.20                  &  $>$16.64                     &  15.803                  &   15.022                  &   $>$1.62 &  4.18 &  &  & \\ 
09 24 22.34 &  $+$32 23 55.1 &     5.93 &     5.63 &     21.58                  &  19.48                  &  $>$16.47                     &  16.392                  &   15.309                  &   $>$1.16 &  4.17 &  &  & \\ 
09 52 13.18 &  $+$25 00 59.0 &    11.71 &    12.61 &     21.14                  &  18.92                  &  $>$16.51                     &  15.777                  &   14.895                  &   $>$1.61 &  4.02 &  &  & \\ 
10 00 06.47 &  $+$45 49 50.0 &     1.99 &     1.52 &  21.13\tablenotemark{b}    &  19.82\tablenotemark{b} &  $>$16.52                     &  16.111                  &   15.182                  &   $>$1.34 &  4.64 &  &  & \\ 
10 14 31.77 &  $-$02 20 53.4 &    18.02 &    27.44 &     21.08                  &  19.22                  &  $\phn$17.27\tablenotemark{a} &  16.079\tablenotemark{a} &   14.983\tablenotemark{a} &$\phn$2.29 &  4.24 &  &  & \\ 
10 20 54.07 &  $+$16 27 02.7 &     2.14 &     5.46 &     22.12                  &  19.43                  &  $>$16.86                     &  15.897                  &   15.176                  &   $>$1.68 &  4.25 &  &  & \\ 
11 06 22.48 &  $+$20 21 35.6 &     2.34 &     2.25 &     21.62                  &  19.67                  &  $>$16.73                     &  15.839                  &   15.306                  &   $>$1.42 &  4.36 &  &  & \\ 
11 40 32.29 &  $+$24 01 18.0 &     3.85 &     3.69 &     21.23                  &  19.33                  &  $>$16.74                     &  15.996                  &   15.047                  &   $>$1.69 &  4.28 &  &  & \\ 
11 42 07.75 &  $+$15 47 54.1 &   270.14 &   283.71 &     20.06                  &  19.48                  &  $>$16.36                     &  15.498                  &   15.305                  &   $>$1.06 &  4.17 &  &  & \\ 
12 08 27.76 &  $+$17 08 14.1 &     2.26 &     5.21 &     22.11                  &  19.40                  &  $>$16.62                     &  15.263                  &   15.322                  &   $>$1.30 &  4.08 & Galaxy  & 0.549 & O15 & Gl04 \\ 
12 29 24.80 &  $+$25 09 21.1 &     2.07 &     1.80 &     22.50                  &  20.21                  &  $>$16.78                     &  16.338                  &   15.434                  &   $>$1.35 &  4.78 &\nodata&\nodata&\nodata& Gl04 \\ 
12 40 23.71 &  $+$43 53 18.4 &     1.06 &     1.66 &     20.43                  &  19.01                  &  $>$16.68                     &  16.191                  &   15.008                  &   $>$1.67 &  4.00 &  &  & \\ 
14 21 12.88 &  $+$36 08 51.0 &     1.29 &     1.27 &     20.09                  &  19.85\tablenotemark{b} &  $>$15.95                     &  16.446                  &   15.002                  &   $>$0.95 &  4.85 &  &  & \\ 
14 34 34.95 &  $+$29 37 05.1 &     2.40 &     1.98 &     21.78                  &  19.39                  &  $>$16.81                     &  16.189                  &   15.142                  &   $>$1.67 &  4.25 &  &  & \\ 
16 07 43.42 &  $+$14 30 37.6 &     1.80 &     2.18 &     21.44                  &  19.57                  &  $>$16.85                     &  15.562                  &   15.293                  &   $>$1.56 &  4.28 &  &  & \\ 
\enddata

\tablenotetext{a}{$J$, $H$, and $K_s$ magnitudes are from the 2MASS All-Sky data Release catalog.}
\tablenotetext{b}{$B$ and $R$ magnitudes were undetected in GSC-II.  We replace these with $r$ and $g$ magnitudes from SDSS.}

\tablecomments{Reference codes for the observations are the same as in Table \ref{table:candlist}}

\tablerefs{(D97) \citet{Drinkwater97};  (Gl04) \citet{Glikman04}; (S93) \citet{Stickel93} }

\end{deluxetable}


\begin{deluxetable}{cccccccc}

\rotate
\tabletypesize{\footnotesize}

\tablewidth{0pt}

\tablecaption{Line Diagnostics of Narrow-Line Spectra\label{table:diagnostics}}


\tablehead{\colhead{Object} & \colhead{Log([\ion{O}{3}]/H$\beta$)} & \colhead{Log([\ion{N}{2}]/H$\alpha$)} & \colhead{Log([\ion{S}{2}]/H$\alpha$)} & \colhead{Log([\ion{O}{1}]/H$\alpha$)} & \colhead{Type 1} & \colhead{Type 2} & \colhead{Type 3} \\ 
\colhead{(1)} & \colhead{(2)} & \colhead{(3)} & \colhead{(4)} & \colhead{(5)} & \colhead{(6)} & \colhead{(7)} & \colhead{(8)} } 

\startdata
0701p5230.ms.dat &$-$0.01 &$-$0.63 &$-$0.35 &$-$0.88 &   starb &   starb &     agn\\ 
0730p2538.ms.dat &   0.73 &$-$0.68 &$-$1.02 &$-$1.91 &     agn &   starb &   starb\\ 
0738p2141.ms.dat &   1.31 &$-$0.29 &$-$0.61 &$-$1.18 &     agn &     agn &     agn\\ 
0738p3156.ms.dat &   1.04 &$-$0.41 &$-$0.86 &$-$1.53 &     agn &     agn &     agn\\ 
0810p1603.ms.dat &$-$0.02 &$-$0.22 &$-$0.56 &$-$1.31 &   starb &   starb &   starb\\ 
0817p1958.ms.dat &$-$1.10 &$-$1.07 &$-$1.60 &$-$1.55 &   starb &   starb &   starb\\ 
0818p1305.ms.dat &   1.00 &$-$0.35 &$-$0.54 &$-$1.16 &     agn &     agn &     agn\\ 
0849p1439.ms.dat &$-$0.07 &$-$1.62 &$-$0.61 &$-$1.47 &   starb &   starb &   starb\\ 
0915p1819.ms.dat &$-$0.16 &$-$0.39 &$-$0.47 &$-$1.28 &   starb &   starb &   starb\\ 
0915p2850.ms.dat &   1.42 &   0.23 &$-$0.08 &$-$0.83 &     agn &     agn &     agn\\ 
0919p1531.ms.dat &$-$0.09 &$-$0.33 &$-$0.44 &$-$1.32 &   starb &   starb &   starb\\ 
0946p1840.ms.dat &   0.63 &$-$0.95 &$-$0.99 &$-$1.74 &   starb &   starb &   starb\\ 
0949p2325.ms.dat &   1.76 &$-$0.24 &$-$0.31 &$-$1.33 &     agn &     agn &     agn\\ 
1004p2943.ms.dat &   0.74 &$-$0.38 &$-$0.77 &$-$1.49 &     agn &     agn &     agn\\ 
1015p1207.ms.dat &   0.98 &$-$0.46 &$-$0.28 &$-$0.88 &     agn &     agn &     agn\\ 
1018p2135.ms.dat &   0.50 &$-$1.24 &$-$1.09 &$-$1.58 &   starb &   starb &   starb\\ 
1022p1628.1d.dat &   0.98 & \nodata& \nodata& \nodata&agn\tablenotemark{a}&agn\tablenotemark{a}&agn\tablenotemark{a}\\
1022p1929.ms.dat &   0.77 &$-$0.68 &$-$0.82 &$-$1.31 &     agn &     agn &     agn\\ 
1139p2743.ms.dat &   0.47 &$-$0.56 &$-$0.41 &$-$0.82 &   starb &     agn &     agn\\ 
1202p2615.ms.dat &   1.10 & \nodata& \nodata& \nodata&agn\tablenotemark{a}&agn\tablenotemark{a}&agn\tablenotemark{a}\\
1654p3435.ms.dat &   0.99 &$-$0.69 &$-$1.03 &$-$1.53 &     agn &     agn &     agn\\ 
2328m1107.1d.dat &   0.63 &$-$0.11 &$-$0.45 &$-$0.99 &     agn &     agn &     agn\\ 
\enddata 
\tablenotetext{a}{These objects were at redshifts higher than 0.55 and H$\alpha$ was outside our spectroscopic range.  Nevertheless, we classify these sources as AGN since their [\ion{O}{3}]/H$\beta$ ratios were above the extreme starburst dividing line shown in Figure \ref{fig:diagnostics}}
\tablecomments{Type 1 is the classification determined from the [\ion{N}{2}]/H$\alpha$ versus [\ion{O}{3}]/H$\beta$ diagnostic.\\
Type 2 is the classification determined from the  [\ion{S}{2}] /H$\alpha$ versus [\ion{O}{3}]/H$\beta$ diagnostic.
Type 3 is the classification determined from the  [\ion{O}{1}] /H$\alpha$ versus [\ion{O}{3}]/H$\beta$ diagnostic.}


\end{deluxetable}


\begin{deluxetable}{cccccccccl}

\rotate

\tabletypesize{\small}

\tablewidth{0pt}

\tablecaption{Extinction Parameters for F2M Quasars\label{table:ebv}}


\tablehead{
\colhead{} & \colhead{} & \colhead{} & \colhead{} & \colhead{} & \colhead{} & \multicolumn{2}{c}{$E(B-V)$: Balmer Decrement} & \colhead{$E(B-V)$} & \colhead{}  \\

\colhead{R.A.} & \colhead{Dec } & \colhead{z} & \colhead{$K_s$ } & \colhead{$A_K$} & \colhead{$R-K$} & \colhead{Total line} & \colhead{Broad line} & \colhead{Cont. Fit} & \colhead{Spectrum Used} \\ 
\colhead{(J2000)} & \colhead{(J2000)} & \colhead{} & \colhead{(mag)} & \colhead{(mag)} & \colhead{(mag)} & \colhead{(mag)} & \colhead{(mag)} & \colhead{(mag)} & \colhead{}\\
\colhead{(1)} & \colhead{(2)} & \colhead{(3)} & \colhead{(4)} & \colhead{(5)} & \colhead{(6)} & \colhead{(7)} &\colhead{(8)} & \colhead{(9)} & \colhead{(10)}
 } 

\startdata
10 04 24.87  &    +12 29 22.4 &   2.658 &   14.54 & \phs1.09 &   6.26 &    \nodata          &    \nodata       & \phs0.48 &  Comb. \\ 
15 49 38.80  &    +12 45 07.9 &   2.373 &   13.52 & \phs0.65 &   4.09 & \phs0.188$\pm$0.012 &  0.141$\pm$0.013 & \phs0.26 &  Single:IR\\ 
15 31 50.48  &    +24 23 17.6 &   2.287 &   14.69 & \phs0.38 &   4.48 &  $-$0.005$\pm$0.046 &  0.170$\pm$0.073 & \phs0.16 &  Single:IR\\ 
22 22 52.78  &  $-$02 02 57.4 &   2.252 &   15.14 & \phs0.90 &   4.90 &  $-$0.632$\pm$0.038 &    \nodata       & \phs0.39 &  Comb. \\ 
01 34 35.68  &  $-$09 31 03.0 &   2.220 &   13.55 & \phs1.08 &   7.25 & \phs0.203$\pm$0.005 &    \nodata       & \phs0.58 &  Comb. \\ 
14 27 44.34  &    +37 23 37.4 &   2.168 &   15.07 & \phs0.60 &   4.52 & \phs0.072$\pm$0.024 &    \nodata       & \phs0.27 &  Comb.\\ 
07 38 20.10  &    +27 50 45.5 &   1.985 &   15.26 & \phs0.91 &   5.54 &    \nodata          &    \nodata       & \phs0.50 &  Comb. \\ 
09 21 45.69  &    +19 18 12.6 &   1.791 &   14.55 & \phs1.20 &   5.57 &    \nodata          &    \nodata       & \phs0.65 &  Opt.  \\ 
13 44 08.31  &    +28 39 32.0 &   1.770 &   14.78 & \phs0.96 &   6.02 &    \nodata          &    \nodata       & \phs0.53 &  Opt. \\ 
10 36 33.54  &    +28 28 21.6 &   1.762 &   15.27 & \phs0.85 &   4.74 &    \nodata          &    \nodata       & \phs0.47 &  Opt.  \\ 
13 59 41.18  &    +31 57 40.5 &   1.724 &   14.79 & \phs0.82 &   6.01 &    \nodata          &    \nodata       & \phs0.47 & Single:IR \\
13 41 08.11  &    +33 01 10.3 &   1.720 &   14.85 & \phs0.98 &   5.95 &    \nodata          &    \nodata       & \phs0.56 &  Comb.\\ 
09 06 51.52  &    +49 52 36.0 &   1.635 &   15.14 & \phs0.16 &   5.66 &    \nodata          &    \nodata       & \phs0.10 &  Single:Opt.\\ 
09 16 08.57  &  $-$03 19 39.9 &   1.560 &   14.82 & \phs0.52 &   4.72 & \phs0.917$\pm$0.075 &    \nodata       & \phs0.33 &  Comb.\\ 
10 46 42.55  &    +16 04 17.4 &   1.440 &   15.47 & \phs0.60 &   4.47 &    \nodata          &    \nodata       & \phs0.41 &  Single:Opt.\\ 
13 53 08.65  &    +36 57 51.2 &   1.311 &   14.26 & \phs1.20 &   6.54 &    \nodata          &    \nodata       & \phs0.90 &  Comb.\\ 
09 16 48.91  &    +38 54 28.3 &   1.267 &   14.75 & \phs0.17 &   4.02 &    \nodata          &    \nodata       & \phs0.13 &  Single:Opt.\\ 
09 27 44.38  &    +39 30 38.4 &   1.170 &   14.59 & \phs0.77 &   4.11 &    \nodata          &    \nodata       & \phs0.65 &  Single:Opt.\\ 
11 46 58.31  &    +39 58 34.2 &   1.088 &   13.53 &  $-$0.04 &   4.64 &    \nodata          &    \nodata       &  $-$0.03 &  Single:Opt.\\ 
09 04 50.51  &  $-$01 45 24.5 &   1.005 &   14.90 & \phs1.10 &   5.46 & \phs0.404$\pm$0.110 &    \nodata       & \phs1.06 &  Comb.\\ 
07 29 10.35  &    +33 36 34.0 &   0.957 &   14.52 & \phs0.83 &   5.64 &    \nodata          &    \nodata       & \phs0.84 &  Opt.\\ 
10 12 30.49  &    +28 25 27.2 &   0.937 &   15.27 & \phs0.62 &   5.53 &    \nodata          &    \nodata       & \phs0.46 &  Comb. \\ 
11 06 07.26  &    +28 12 47.0 &   0.842 &   14.63 & \phs0.17 &   5.39 &    \nodata          &    \nodata       & \phs0.19 &  Single:Opt.\\ 
09 15 01.71  &    +24 18 12.2 &   0.842 &   13.79 & \phs0.60 &   7.01 &    \nodata          &    \nodata       & \phs0.68 &  Comb.\\ 
08 25 02.05  &    +47 16 52.0 &   0.803 &   14.11 & \phs0.61 &   6.34 &    \nodata          &    \nodata       & \phs0.71 &  Comb.\\ 
11 51 24.07  &    +53 59 57.4 &   0.780 &   15.10 & \phs0.64 &   5.70 &    \nodata          &    \nodata       & \phs0.76 &  Comb.\\ 
12 27 03.21  &    +50 53 56.3 &   0.768 &   14.61 & \phs0.32 &   4.16 &  $-$0.226$\pm$0.065 &    \nodata       & \phs0.38 &  Comb.\\ 
16 56 47.11  &    +38 21 36.7 &   0.732 &   15.09 & \phs0.49 &   5.71 &    \nodata          &    \nodata       & \phs0.61 &  Opt.  \\ 
11 59 31.84  &    +29 14 43.9 &   0.730 &   11.48 &  $-$0.04 &   6.11 &    \nodata          &    \nodata       &  $-$0.05 &  Single:Opt.\\ 
16 00 34.56  &    +35 22 27.0 &   0.707 &   14.17 & \phs0.18 &   4.42 &    \nodata          &    \nodata       & \phs0.22 &  Single:IR\\ 
11 18 11.06  &  $-$00 33 41.9 &   0.686 &   14.58 & \phs0.61 &   5.29 &    \nodata          &    \nodata       & \phs0.79 &  Opt.\\ 
11 13 54.68  &    +12 44 38.9 &   0.681 &   13.66 & \phs0.88 &   5.33 &    \nodata          &    \nodata       & \phs1.15 &  Opt.  \\ 
23 39 03.83  &  $-$09 12 21.0 &   0.660 &   14.09 & \phs0.80 &   4.54 &    \nodata          &    \nodata       & \phs1.06 &  Opt.\\ 
16 42 58.80  &    +39 48 37.2 &   0.593 &   12.36 & \phs0.02 &   4.32 &    \nodata          &    \nodata       & \phs0.03 &  Comb.\\ 
23 25 49.88  &  $-$10 52 43.4 &   0.564 &   15.24 & \phs0.34 &   4.46 & \phs0.408$\pm$0.005 &  0.556$\pm$0.010 & \phs0.50 &  Comb.\\ 
15 32 33.19  &    +24 15 26.8 &   0.562 &   14.96 & \phs0.48 &   5.13 &    \nodata          &    \nodata       & \phs0.70 &  Comb.\\ 
08 41 04.98  &    +36 04 50.1 &   0.553 &   14.92 & \phs0.71 &   5.88 &    \nodata          &    \nodata       & \phs1.05 &  Comb.\\ 
08 10 58.98  &    +41 34 02.6 &   0.506 &   14.22 & \phs0.02 &   4.11 &  $-$0.159$\pm$0.008 &  0.245$\pm$0.009 & \phs0.04 &  Comb.\\ 
08 34 07.01  &    +35 06 01.8 &   0.470 &   14.65 & \phs0.46 &   4.31 &    \nodata          &    \nodata       & \phs0.74 &  Opt.\\ 
11 17 33.83  &  $-$02 36 00.3 &   0.463 &   14.05 & \phs0.17 &   4.70 & \phs0.121$\pm$0.013 &  0.032$\pm$0.015 & \phs0.29 &  Single:Opt.\\ 
16 18 09.72  &    +35 02 08.5 &   0.447 &   14.05 & \phs0.65 &   4.20 &    \nodata          &    \nodata       & \phs1.08 &  Single:IR\\ 
02 29 50.66  &  $-$08 42 35.5 &   0.440 &   14.84 & \phs0.09 &   4.34 & \phs0.233$\pm$0.015 &  0.362$\pm$0.028 & \phs0.15 &  Opt.\\ 
08 43 32.28  &    +49 21 16.2 &   0.420 &   15.13 & \phs0.22 &   4.37 &  $-$0.106$\pm$0.018 &  0.176$\pm$0.021 & \phs0.38 &  Single:Opt.\\ 
14 15 22.83  &    +33 33 06.5 &   0.416 &   14.30 & \phs0.35 &   4.62 &    \nodata          &    \nodata       & \phs0.60 &  Single:IR\\ 
08 30 11.13  &    +37 59 51.9 &   0.414 &   14.58 & \phs0.43 &   4.22 & \phs0.650$\pm$0.008 &  0.661$\pm$0.013 & \phs0.74 &  Single:Opt.\\ 
17 08 02.53  &    +22 27 25.7 &   0.377 &   14.57 & \phs0.59 &   4.77 & \phs2.408$\pm$0.130 &  3.480$\pm$0.566 & \phs1.05 &  Comb.\\ 
17 24 28.03  &    +33 31 19.9 &   0.370 &   14.46 & \phs0.51 &   4.02 &    \nodata          &    \nodata       & \phs0.91 &  Opt.\\ 
12 09 21.17  &  $-$01 07 17.0 &   0.363 &   13.72 & \phs0.49 &   4.14 &    \nodata          &    \nodata       & \phs0.89 &  Single:IR\\ 
14 08 11.61  &    +32 43 50.2 &   0.338 &   15.35 & \phs0.54 &   4.24 & \phs1.294$\pm$0.035 &  1.855$\pm$0.156 & \phs1.01 &  Single:Opt.\\ 
13 04 04.07  &    +46 12 53.6 &   0.315 &   14.32 & \phs0.77 &   4.00 &    \nodata          &    \nodata       & \phs1.49 &  Single:IR\\ 
17 04 25.31  &    +30 37 08.7 &   0.311 &   14.92 & \phs0.28 &   5.88 &    \nodata          &    \nodata       & \phs0.54 &  Single:IR\\ 
00 36 59.85  &  $-$01 13 32.3 &   0.294 &   13.65 & \phs0.40 &   5.89 & \phs1.313$\pm$0.027 &  2.544$\pm$0.256 & \phs0.79 &  Single:Opt.\\ 
07 53 41.11  &    +25 06 39.4 &   0.292 &   15.22 & \phs0.51 &   4.20 &    \nodata          &    \nodata       & \phs1.02 &  Single:Opt.\\ 
13 07 00.62  &    +23 38 05.3 &   0.275 &   13.45 & \phs0.28 &   5.92 & \phs1.106$\pm$0.031 &  1.736$\pm$0.101 & \phs0.58 &  Single:Opt.\\ 
22 16 33.73  &  $-$00 54 51.2 &   0.200 &   13.78 & \phs0.25 &   4.13 & \phs0.416$\pm$0.040 &  0.406$\pm$0.054 & \phs0.55 &  Comb.\\ 
08 17 39.57  &    +43 54 20.0 &   0.186 &   14.09 & \phs0.32 &   4.57 &    \nodata          &    \nodata       & \phs0.73 &  Opt.\\ 
\enddata




\end{deluxetable}

\begin{deluxetable}{ccccc}
\tablecolumns{5}
\tablewidth{0pt}
\tablecaption{QSO Counts and Surface Densities in $K$-Band \label{table:surfden}}
\tablehead{ 
\colhead{K Magnitude Range} & \colhead{FBQS II} & 
\colhead{FBQS III} & \colhead{UVX\tablenotemark{a}} & 
\colhead{F2M} \\
\colhead{$K$} & \multicolumn{4}{c}{(deg$^{-2}$ 0.5 mag$^{-1}$)} 
}
\startdata
11.0 - 11.5 & 0.0011$\pm$0.0006 & \nodata         & \nodata        & 0.0004$\pm$0.0004 \\
11.5 - 12.0 & 0.0004$\pm$0.0004 & 0.002$\pm$0.002 & \nodata        & \nodata \\
12.0 - 12.5 & 0.003$\pm$0.001   & \nodata         & \nodata        & 0.0004$\pm$0.0004 \\
12.5 - 13.0 & 0.004$\pm$0.001   & \nodata         & \nodata        & \nodata \\
13.0 - 13.5 & 0.006$\pm$0.002   & \nodata         & \nodata        & 0.0004$\pm$0.0004 \\
13.5 - 14.0 & 0.014$\pm$0.002   & 0.007$\pm$0.003 & 0.007$\pm$0.005& 0.0029$\pm$0.0011 \\
14.0 - 14.5 & 0.030$\pm$0.003   & 0.031$\pm$0.007 & 0.04$\pm$0.01  & 0.0041$\pm$0.0012 \\
14.5 - 15.0 & 0.044$\pm$0.004   & 0.042$\pm$0.009 & 0.07$\pm$0.02  & 0.0077$^{+0.0057}_{-0.0017}$\\
15.0 - 15.5 & 0.052$\pm$0.004   & 0.07$\pm$0.01   & 0.12$\pm$0.04  & 0.0052$^{+0.0119}_{-0.0014}$\\
15.5 - 16.0 & 0.030$\pm$0.003   & 0.07$\pm$0.01   & 0.04$\pm$0.02  & \nodata \\
16.0 - 16.5 & 0.002$\pm$0.001   & 0.010$\pm$0.004 & \nodata        & \nodata \\
\hline
\hline
                  & (deg$^{-2}$)    & (deg$^{-2}$)  & (deg$^{-2}$)  & (deg$^{-2}$)   \\
\phm{0.00} - 15.5 & 0.154$\pm$0.008 & 0.15$\pm$0.02 & 0.23$\pm$0.05 & 0.021$^{+0.004}_{-0.003}$\\
\enddata
\tablenotetext{a}{This sample is made up of the LBQS and 2QZ samples combined as described in \S 5.1 of \citetalias{Glikman04}}
\end{deluxetable}
 
\begin{deluxetable}{lcccccccc}
\tabletypesize{\footnotesize}
\tablecolumns{8}
\tablewidth{0pt}
\tablehead{ 
\colhead{} & \multicolumn{2}{c}{FIRST} &
\multicolumn{2}{c}{VLA 20 cm\tablenotemark{a}} & \colhead{NVSS} &
\multicolumn{2}{c}{VLA 3.6 cm\tablenotemark{a}} & \colhead{}\\
\colhead{Object} & \colhead{F$_{pk}$} & \colhead{F$_{int}$} & 
\colhead{F$_{pk}$} & \colhead{F$_{int}$} & \colhead{} &
\colhead{F$_{pk}$} & \colhead{F$_{int}$} & 
\colhead{$\alpha_{1.4 \mathrm{GHz}/8.3 \mathrm{GHz}}$} \\
\colhead{} & \colhead{(mJy)} & \colhead{(mJy)} & 
\colhead{(mJy)} & \colhead{(mJy)} & \colhead{(mJy)} &
\colhead{(mJy)} & \colhead{(mJy)} & \colhead{}\\
\colhead{(1)} & \colhead{(2)} & \colhead{(3)} & \colhead{(4)} & 
\colhead{(5)} & \colhead{(6)} & \colhead{(7)} &
\colhead{(8)} & \colhead{(9)}
}
\tablecaption{3.6 cm and 20 cm VLA Observations \label{table:vla}}
\startdata
F2MJ072910.3$+$333634                  &   3.26&   3.17&   2.31& \nodata&   2.5 &    0.56&  0.55 & $-$0.79 \\
F2MJ073820.1$+$275045                  &   2.64&   2.33&   1.40& \nodata&   3.7 &    0.47&  0.36 & $-$0.61 \\
F2MJ081058.9$+$413402                  & 169.98& 189.34& 218.42& 218.24 & 219.7 &  205.06&209.51 & $-$0.04 \\
F2MJ081739.5$+$435420                  &   3.10&   2.80&   2.52&   2.58 &   2.1 &    0.49&  0.62 & $-$0.92 \\
F2MJ082502.0$+$471652\tablenotemark{b} &  61.12&  63.24&  51.85&  52.71 &  50.5 &   15.68& 15.93 & $-$0.67 \\
F2MJ083011.1$+$375951                  &   6.42&   6.50&   6.79&   4.64 &   5.4 &    1.02&  0.97 & $-$1.06 \\
F2MJ083407.0$+$350601                  &   1.22&   0.55&   0.72& \nodata&   1.4 &    0.12&\nodata& $-$1.00 \\
F2MJ084104.9$+$360450                  &   6.49&   6.58&   6.60&   6.39 &   7.4 &    3.92&  4.06 & $-$0.29 \\
F2MJ090450.5$-$014524                  &   2.23&   2.75&   2.03&   4.08 &   2.8 &    0.44&  0.39 & $-$0.86 \\
F2MJ090651.5$+$495236                  &  67.87&  75.11&  67.15&  67.75 &  57.1 &   78.76& 80.00 &\phs0.09 \\
F2MJ091501.7$+$241812                  &   9.80&  10.12&   5.84&   5.67 &   9.0 &   24.00& 24.50 &\phs0.79 \\
F2MJ091608.5$-$031939\tablenotemark{b} & 117.25& 125.77& 118.40& 118.64 & 127.4 &   54.27& 55.90 & $-$0.44 \\
F2MJ092145.6$+$191812                  &   4.40&   4.71&   3.84&   4.19 &   4.7 &    1.03&  1.10 & $-$0.74 \\
F2MJ100424.8$+$122922                  &  11.42&  12.32&  11.36&  11.62 &  11.8 &   11.92& 12.19 &\phs0.03 \\
F2MJ101230.4$+$282527                  &   9.23&   8.76&   5.71&  11.40 &  10.0 &    8.30&  8.52 &\phs0.21 \\
F2MJ103633.5$+$282821                  &   4.25&   4.37&   3.97&   3.67 &   5.7 &    4.68&  4.75 &\phs0.09 \\
F2MJ110607.2$+$281247                  & 211.47& 214.79& 224.94& 226.26 & 224.7 &  380.66&386.28 &\phs0.29 \\
F2MJ111354.6$+$124438\tablenotemark{b} &   2.99&   4.23&\nodata& \nodata&   4.6 &    2.09&  2.06 & \nodata \\
F2MJ111733.8$-$023600\tablenotemark{b} & 608.04& 961.26& 905.55& 930.15 & 996.1 &  233.99&272.48 & $-$0.76 \\
F2MJ111811.0$-$003341                  &   1.30&   1.81&   0.78& \nodata&   2.8 &    0.28&  0.22 & $-$0.57 \\
F2MJ115124.0$+$535957\tablenotemark{c} &   3.52&   3.60&   3.29& \nodata&   3.5 &    0.55&  0.71 & $-$1.00 \\
F2MJ115931.8$+$291443                  &1855.80&1952.59& 2023.5& 2027.4 &2030.8 &  1743.2& 1766.3 & $-$0.08 \\
F2MJ120921.2$-$010717                  &   1.42&   1.25&   1.77& \nodata&   2.8 &    1.11&\nodata& $-$0.26 \\
F2MJ122703.2$+$505356                  &   3.87&   4.27&   3.66&   3.94 &   3.9 &    0.99&  0.94 & $-$0.73 \\
F2MJ130404.0$+$461253                  &   2.14&   2.63&   2.38&   2.59 &   3.2 &     .75&\nodata& $-$0.65 \\
F2MJ130700.6$+$233805                  &   2.99&   2.63&   2.68&   2.38 &   3.0 &    0.87&  0.79 & $-$0.63 \\
F2MJ134108.1$+$330110                  &  69.41&  69.81&  63.12&  63.28 &  71.6 &   23.06& 23.50 & $-$0.56 \\
F2MJ134408.3$+$283932                  &   9.76&  10.41&   7.95&  10.41 &  10.8 &    1.82&  1.80 & $-$0.83 \\
F2MJ135308.6$+$365751                  &   3.71&   3.32&   3.75&   3.24 &   3.5 &    0.76&  0.69 & $-$0.89 \\
F2MJ140811.6$+$324350                  &   1.09&   2.26&   0.63& \nodata&   1.4 &    0.69&\nodata&\phs0.05 \\
F2MJ141522.8$+$333306\tablenotemark{b} &   5.21&   6.21&   4.12&   5.51 &   8.9 &    1.32&\nodata& $-$0.64 \\
F2MJ142744.3$+$372337                  &   1.71&   1.52&   0.99& \nodata&   1.4 &    0.96&\nodata& $-$0.02 \\
F2MJ153150.4$+$242317                  &   2.24&   2.00&   6.66& \nodata&   1.4 &    0.28&\nodata& $-$1.78 \\
F2MJ153233.1$+$241526\tablenotemark{b,c}&  7.43&   9.37&  11.69& \nodata&   5.3 &    7.39&  7.54 & $-$0.26 \\
F2MJ154938.8$+$124507                  &   1.80&   1.79&   1.94&   2.07 &   2.0 &    0.53&  0.49 & $-$0.73 \\
F2MJ160034.5$+$352227                  &   3.17&   3.30&   1.17& \nodata&   3.4 &    0.60&  0.67 & $-$0.37 \\
F2MJ161809.7$+$350208                  &  13.73&  16.22&  34.25&  37.32 &  35.8 &    7.02&  8.50 & $-$0.89 \\
F2MJ165647.1$+$382136                  &   4.12&   3.89&   4.09&   4.34 &   4.5 &    0.71&  0.72 & $-$0.98 \\
F2MJ170802.5$+$222725                  &   1.31&   1.74&   1.05& \nodata&   2.8 &    0.37&  0.37 & $-$0.58 \\
F2MJ172428.0$+$333119\tablenotemark{b} &   1.58&   1.35&$<$1.20&$<$1.20 &   2.8 & \nodata&\nodata&\nodata  \\
F2MJ221633.7$-$005451                  &   1.31&   0.80&   1.44& \nodata&   2.0 &    0.21&  0.32 & $-$1.08 \\
F2MJ222252.7$-$020257                  &   2.43&   2.66&\nodata& \nodata&   2.5 &    0.16&  0.30 & \nodata \\
F2MJ232549.8$-$105243                  &   1.46&   1.10&\nodata& \nodata&   2.8 &    0.25&  0.20 & \nodata \\
F2MJ233903.8$-$091221                  &   4.34&   4.06&   2.70& \nodata&   3.9 &    0.35&  0.39 & $-$1.14 \\
\enddata
\tablenotetext{a}{Observations were conducted using the CnD configuration of the VLA, unless otherwise noted.}
\tablenotetext{b}{Observations were conducted using the D configuration of the VLA}
\tablenotetext{c}{Gaussian fitting was forced to a point source, with the FWHM fixed at the synthesized beam size.}

\end{deluxetable}

\begin{deluxetable}{cccrrrrrrrrrcc}
\tabletypesize{\scriptsize}
\rotate
\tablewidth{0pt}
\tablehead{
\colhead{} & \colhead{} & \colhead{} & \multicolumn{3}{c}{GSC II} & 
\multicolumn{3}{c}{APM} &  \multicolumn{3}{c}{SDSS} & 
\colhead{} & \colhead{} \\
\cline{4-6} \cline{7-9} \cline{10-12}  \\ 
\colhead{RA} & \colhead{Dec} & \colhead{$z$} & 
\colhead{$F$} & \colhead{$J$} & \colhead{$J-F$} & 
\colhead{$E$} & \colhead{$O$} & \colhead{$O-E$} & 
\colhead{$r$} & \colhead{$g$} & \colhead{$g-r$} & 
\colhead{$K$} & \colhead{$E(B-V)$} \\
\colhead{(1)} & \colhead{(2)} & \colhead{(3)} & 
\colhead{(4)} & \colhead{(5)} & \colhead{(6)} &
\colhead{(7)} & \colhead{(8)} & \colhead{(9)} &
\colhead{(10)} & \colhead{(11)} &\colhead{(12)} & 
\colhead{(13)} & \colhead{(14)} 
}
\tablecaption{Optical Properties of Blue F2M Quasars\label{table:blue_optical}}
\startdata
08 10 58.98&$+$41 34 02.6&0.506&   18.33&18.27&$-$0.06&   18.27&   18.67&\phn0.40& 18.64 & 18.53 & $-$0.11&14.199&\phn0.06\\
09 06 51.52&$+$49 52 36.0&1.635&$>$20.80&20.82&$<$0.02&$>$20.00&$>$21.50& \nodata& 21.23 & 21.50 &\phn0.27&15.141&\phn0.10\\
11 46 58.31&$+$39 58 34.2&1.088&   18.17&18.47&   0.30&   18.03&   19.02&\phn0.99& 18.65 & 18.30 &\phn0.35&13.530& $-$0.04\\
11 59 31.84&$+$29 14 43.9&0.730&   17.59&17.67&   0.08&   16.58&   17.53&\phn0.95&\nodata&\nodata& \nodata&11.474& $-$0.06\\
16 42 58.80&$+$39 48 37.2&0.593&   16.68&17.42&   0.74&   14.70&   15.48&\phn0.78& 16.47 & 16.57 &\phn0.10&12.398&\phn0.03\\
\enddata
\end{deluxetable}

\begin{deluxetable}{ccccccrrcll}
\tabletypesize{\scriptsize}
\rotate
\tablewidth{0pt}
\tablehead{
\colhead{} & \colhead{} & \multicolumn{2}{c}{FIRST} &
\colhead{} &\colhead{} & \multicolumn{2}{c}{Spectral Index} & \colhead{} & \colhead{} & \colhead{}\\
\colhead{RA} & \colhead{Dec} & 
\colhead{$F_{pk}$} & \colhead{$F_{int}$} &\colhead{$NVSS$} & 
\colhead{$6$ cm} &\colhead{$\alpha^{FIRST}_{6\mathrm{cm}}$} & \colhead{$\alpha^{NVSS}_{6\mathrm{cm}}$} &
\colhead{$\log R$\tablenotemark{a}} &\colhead{NED} & \colhead{Comment} \\
\colhead{(J2000)} & \colhead{(J2000)} & \colhead{(mJy)} & \colhead{(mJy)} & \colhead{(mJy)} & \colhead{(mJy)} &
\colhead{} & \colhead{} & \colhead{} & \colhead{} & \colhead{}\\
\colhead{(1)} & \colhead{(2)} & \colhead{(3)} & 
\colhead{(4)} & \colhead{(5)} & \colhead{(6)} &
\colhead{(7)} & \colhead{(8)} & \colhead{(9)} &
\colhead{(10)} & \colhead{(11)} 
}
\tablecaption{Radio Properties of Blue F2M Quasars\label{table:blue_radio}}
\startdata
08 10 58.98&$+$41 34 02.6&  170.0&  189.3&   219.7&     168&$-$0.10&$-$0.22&2.98 &y & RASS X-ray source \citet{Brinkmann00} \\
09 06 51.52&$+$49 52 36.0&   67.9&   75.1&    57.1&      96&   0.20&   0.42&3.51 &y & \\
11 46 58.31&$+$39 58 34.2&  335.6&  338.5&   330.9&     733&   0.62&   0.64&3.55 &y & \\
11 59 31.84&$+$29 14 43.9& 1855.8& 1952.6&  2030.8&    1542&$-$0.19&$-$0.22&3.64 &y & Blazar \citet{Ghosh00} \\
16 42 58.80&$+$39 48 37.2& 6050.1& 6598.6&  7098.6&    6560&$-$0.01&$-$0.06&4.20 &y & Blazar \citet{Webb94,Raiteri98} \\
\enddata
\tablenotetext{a}{Radio loudness parameter, defined as in \citet{Stocke92} $\log R = \log(f(5 \mathrm{GHz})/f(B))$}
\end{deluxetable}

\begin{figure}
\epsscale{1}
\plotone{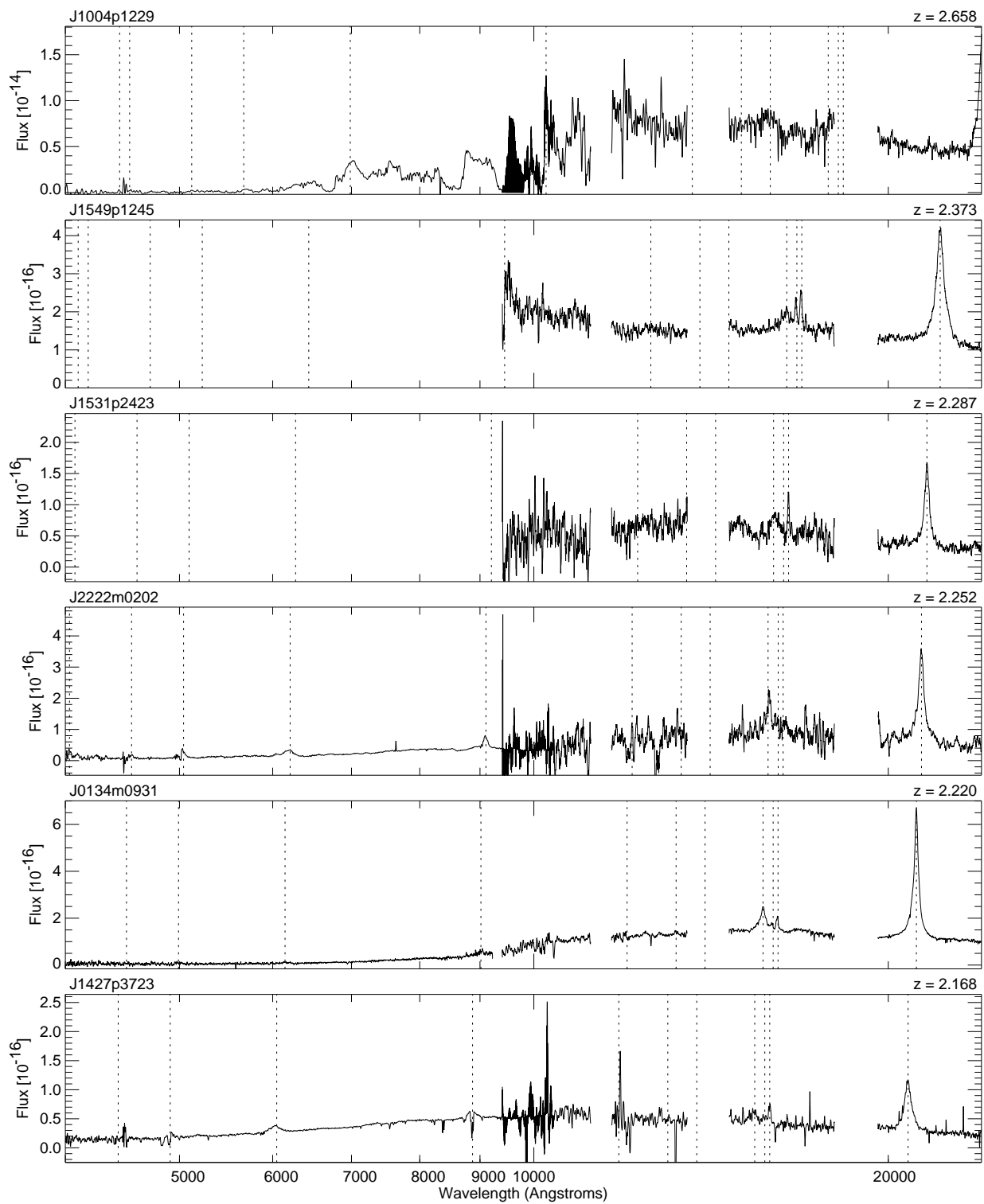}
\caption{
All available spectra of F2M candidates identified as quasars, ordered by
redshift.  The dotted lines show expected positions of
prominent emission lines in the optical and near-infrared:
Ly$\alpha$~1216,
N~V~1240,
Si~IV~1400,
C~IV~1550,
C~III]~1909,
Mg~II~2800,
[O~II]~3727,
H$\delta$~4102,
H$\gamma$~4341,
H$\beta$~4862,
[O~III]~4959,
[O~III]~5007,
H$\alpha$~6563,
Pa$\gamma$~10941,
Pa$\beta$~12822.
}\label{fig:spectra}
\end{figure}

\begin{figure}
\figurenum{3b}
\plotone{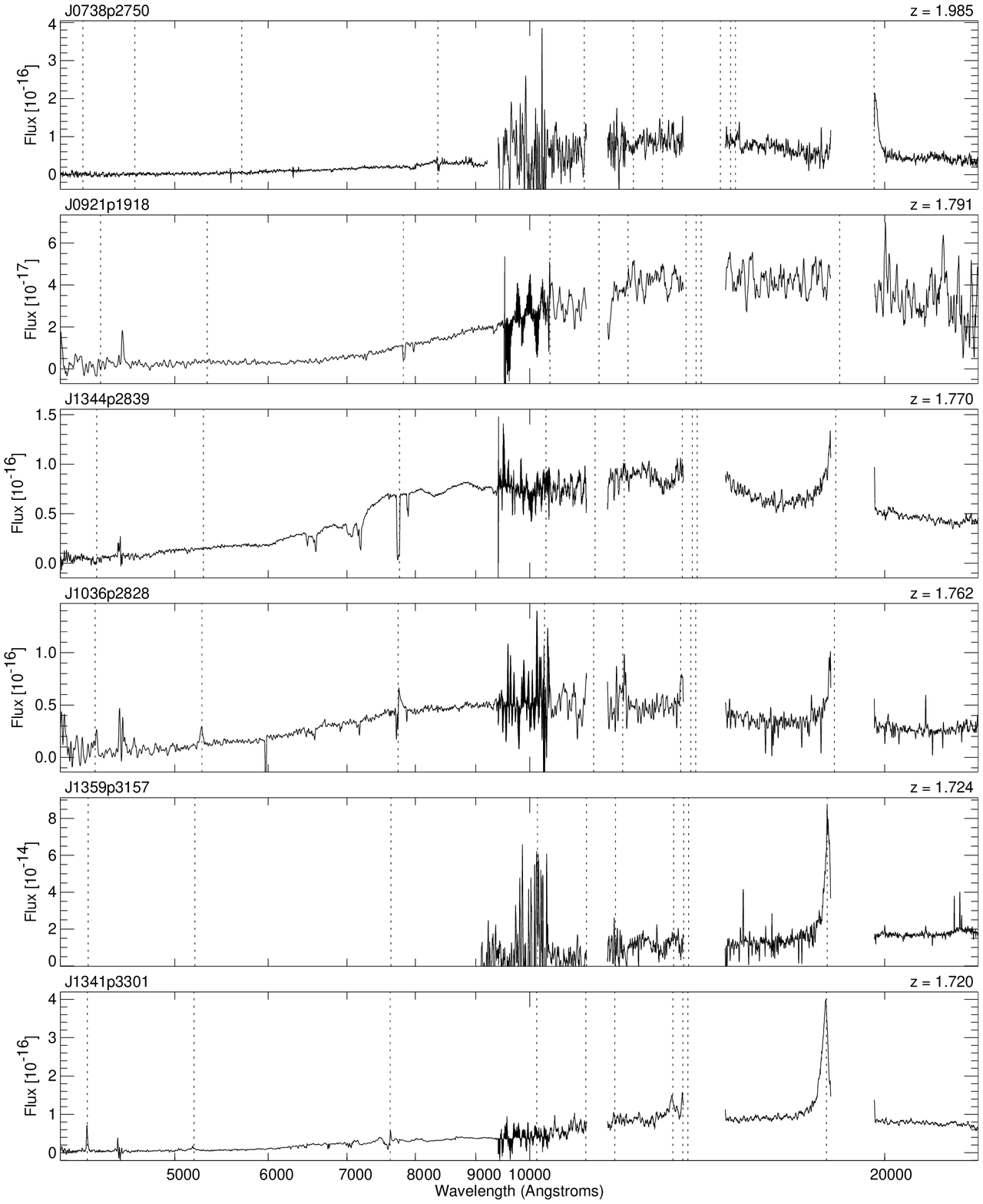}
\caption{{\it Continued.} Spectra of F2M quasars.}
\end{figure}

\begin{figure}
\figurenum{3c}
\plotone{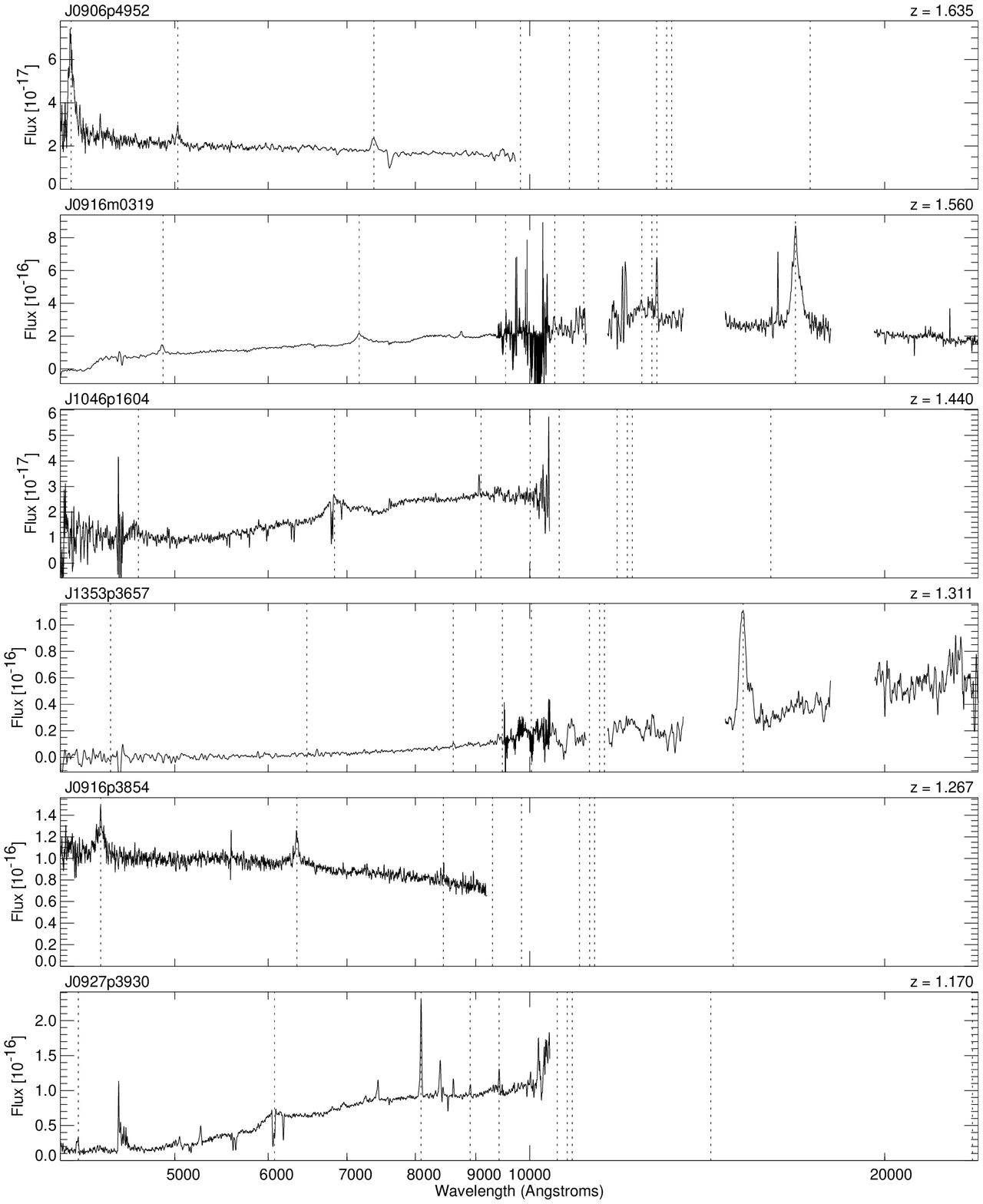}
\caption{{\it Continued.} Spectra of F2M quasars.}
\end{figure}

\begin{figure}
\figurenum{3d}
\plotone{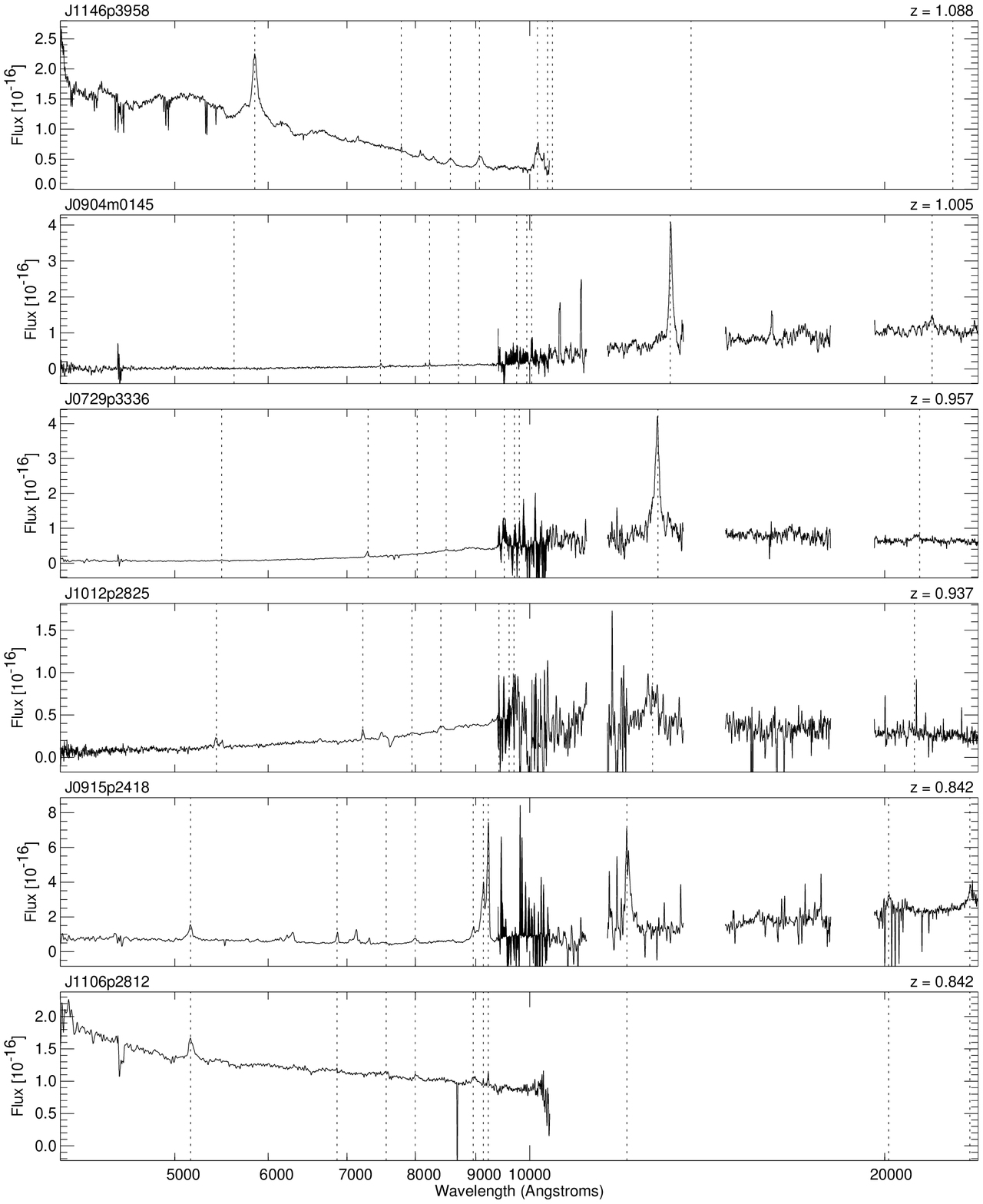}
\caption{{\it Continued.} Spectra of F2M quasars.}
\end{figure}

\begin{figure}
\figurenum{3e}
\plotone{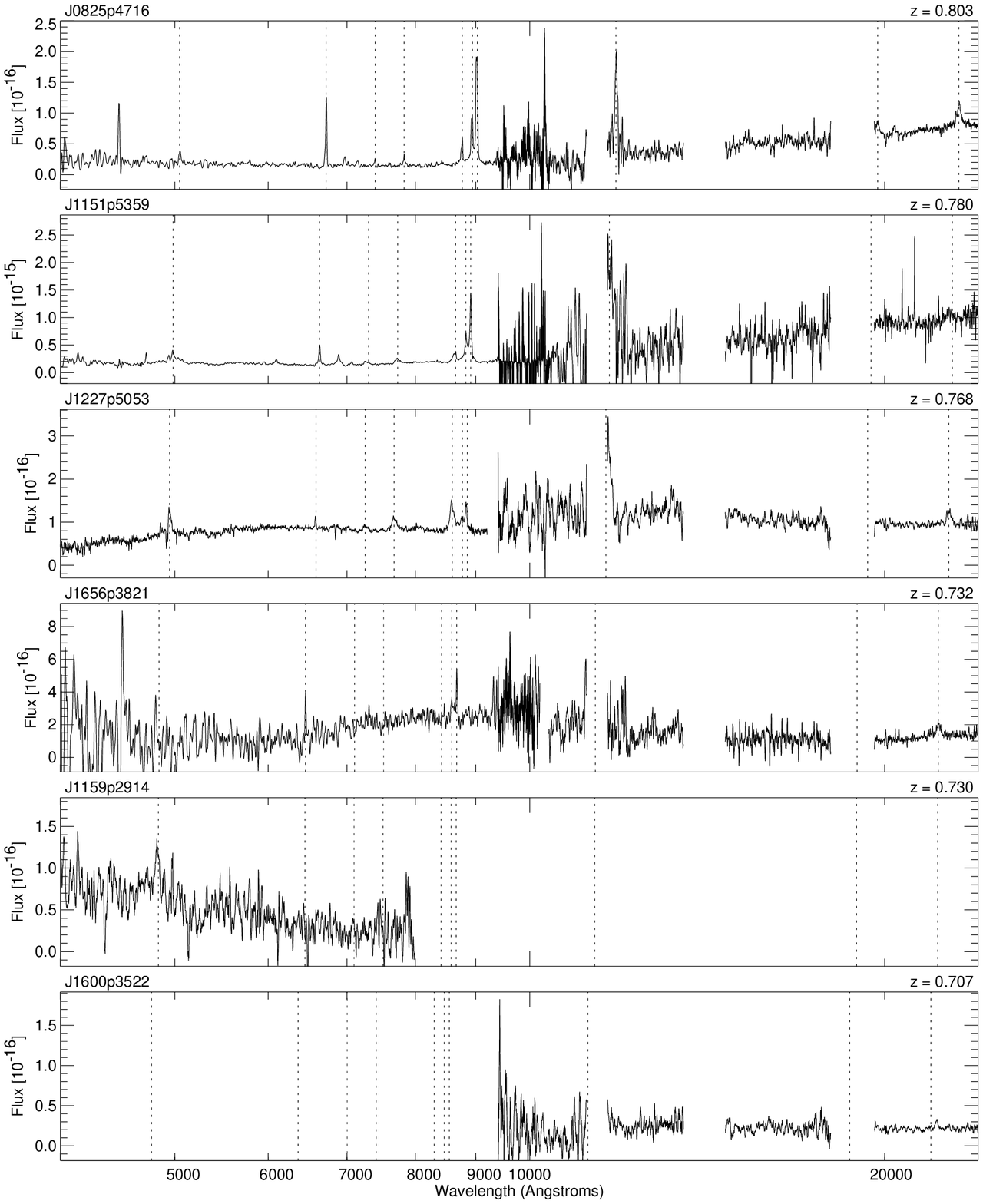}
\caption{{\it Continued.} Spectra of F2M quasars.}
\end{figure}

\begin{figure}
\figurenum{3f}
\plotone{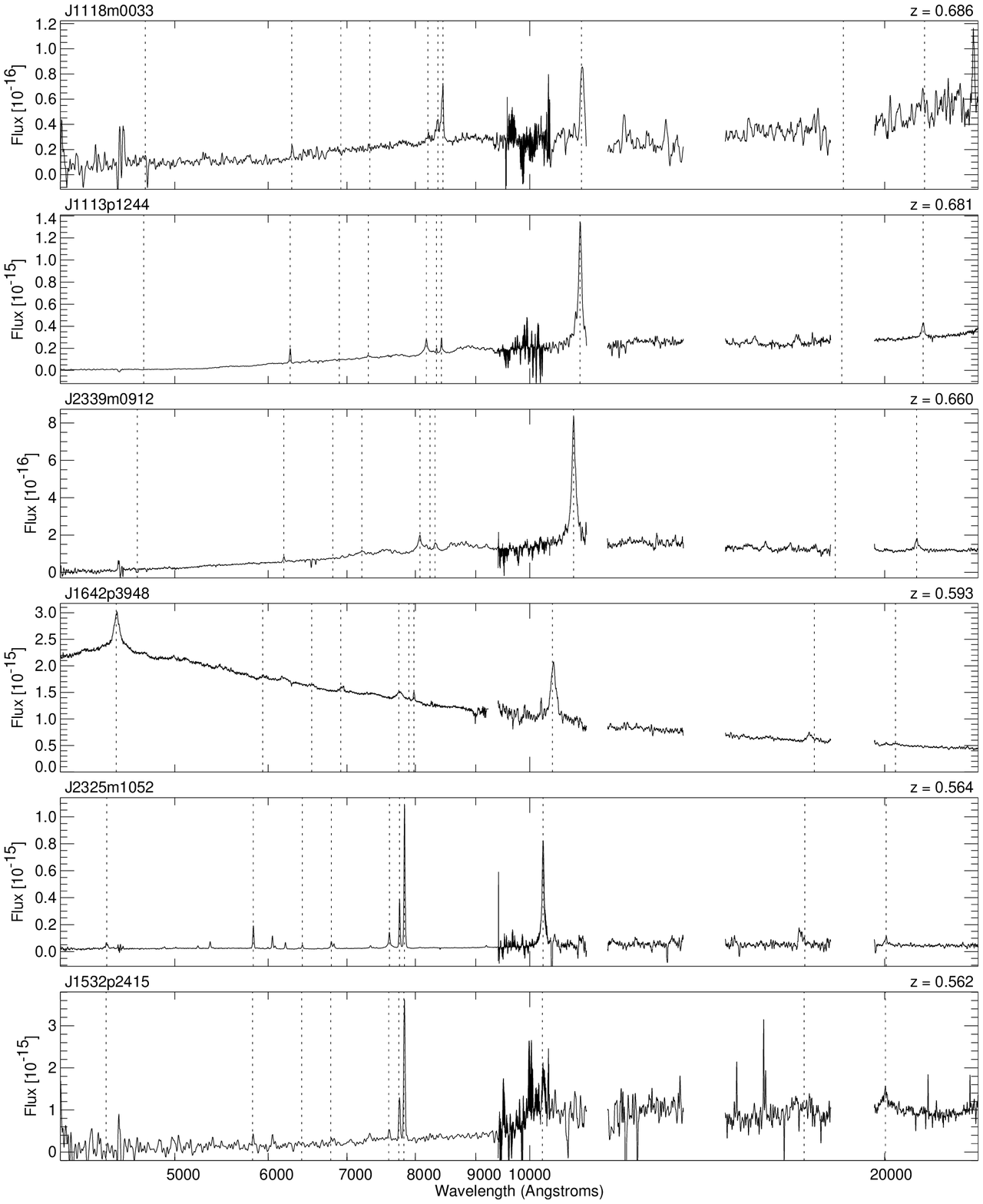}
\caption{{\it Continued.} Spectra of F2M quasars.}
\end{figure}

\clearpage

\begin{figure}
\figurenum{3g}
\plotone{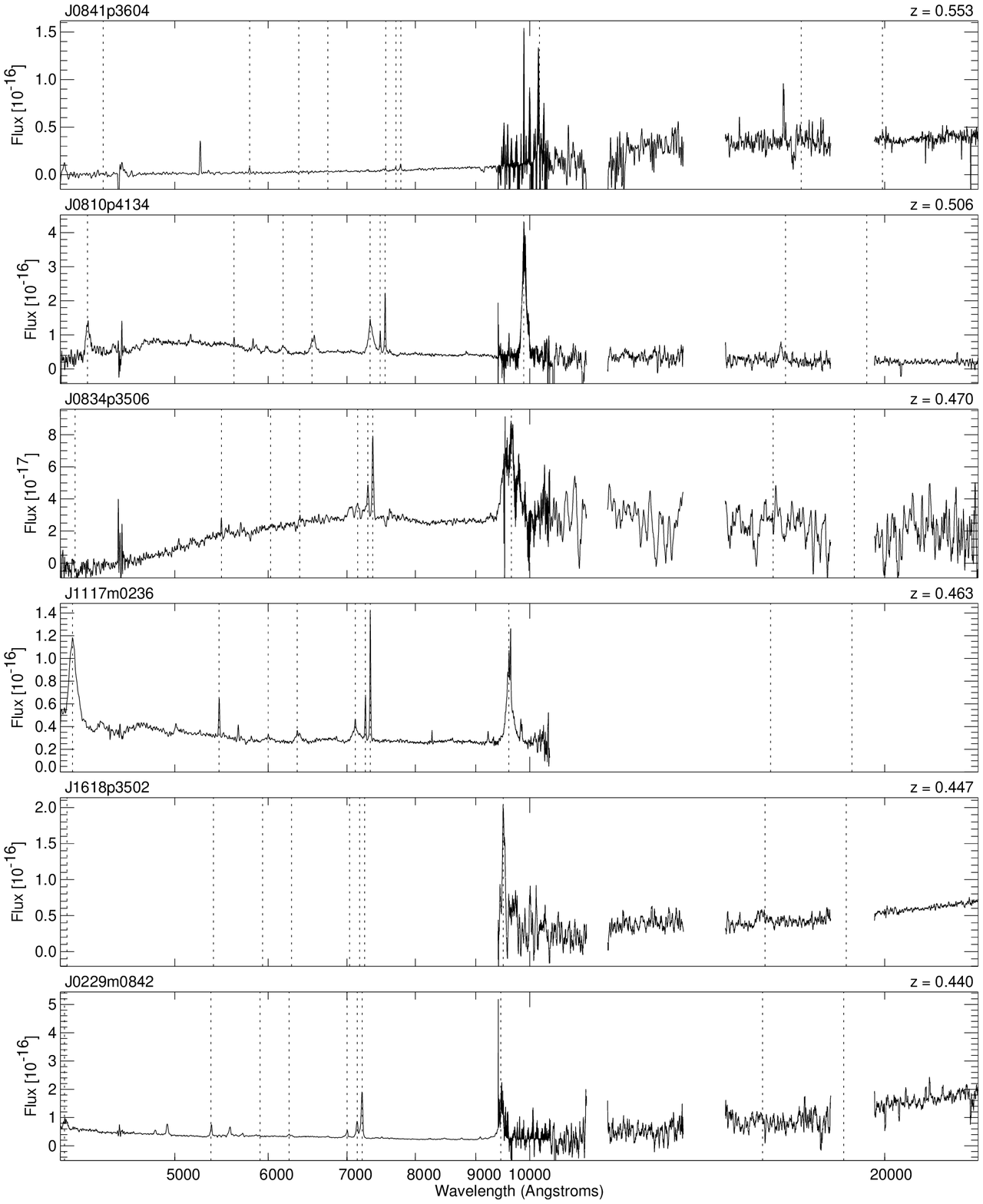}
\caption{{\it Continued.} Spectra of F2M quasars.}
\end{figure}

\begin{figure}
\figurenum{3h}
\plotone{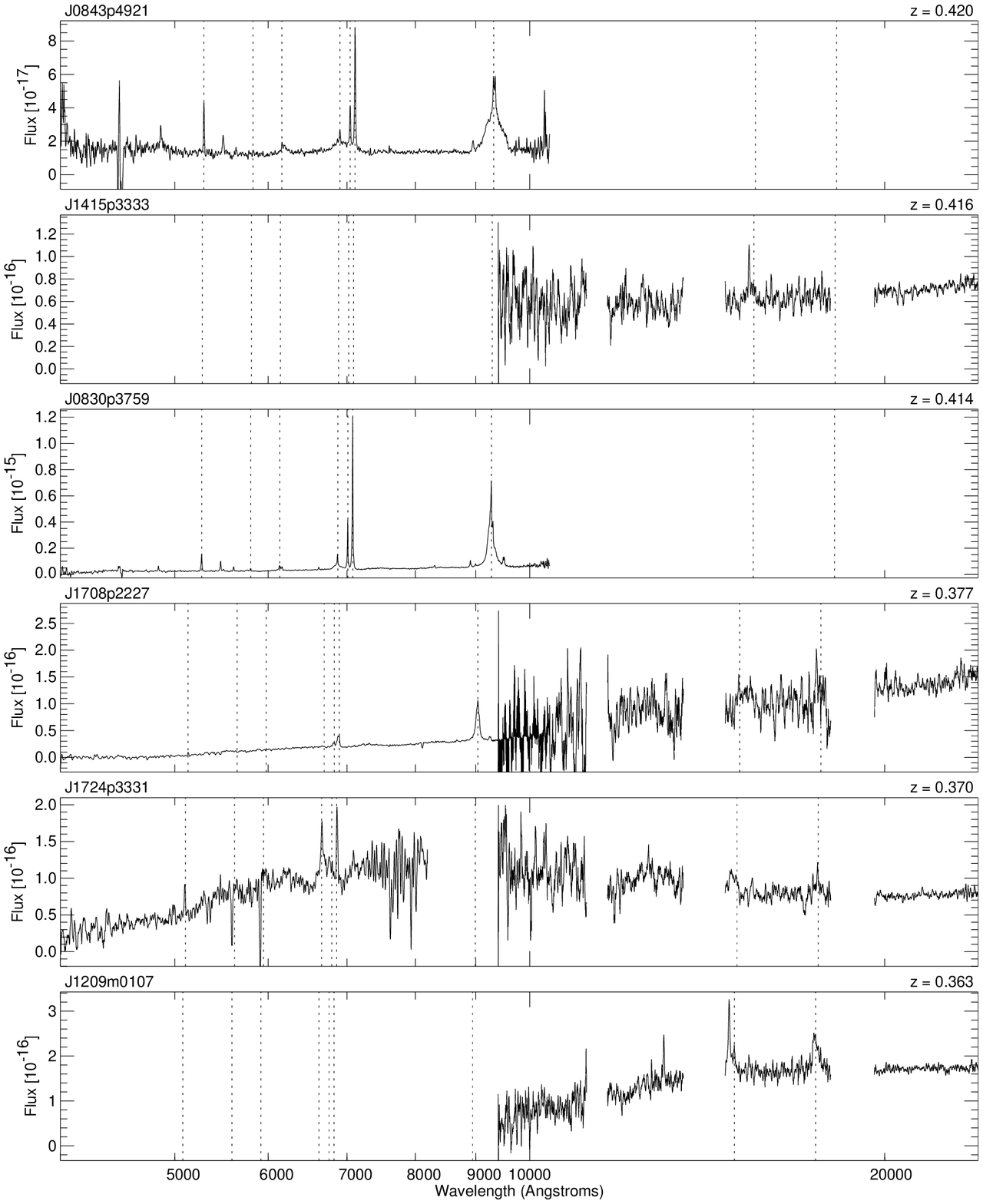}
\caption{{\it Continued.} Spectra of F2M quasars.}
\end{figure}

\begin{figure}
\figurenum{3i}
\plotone{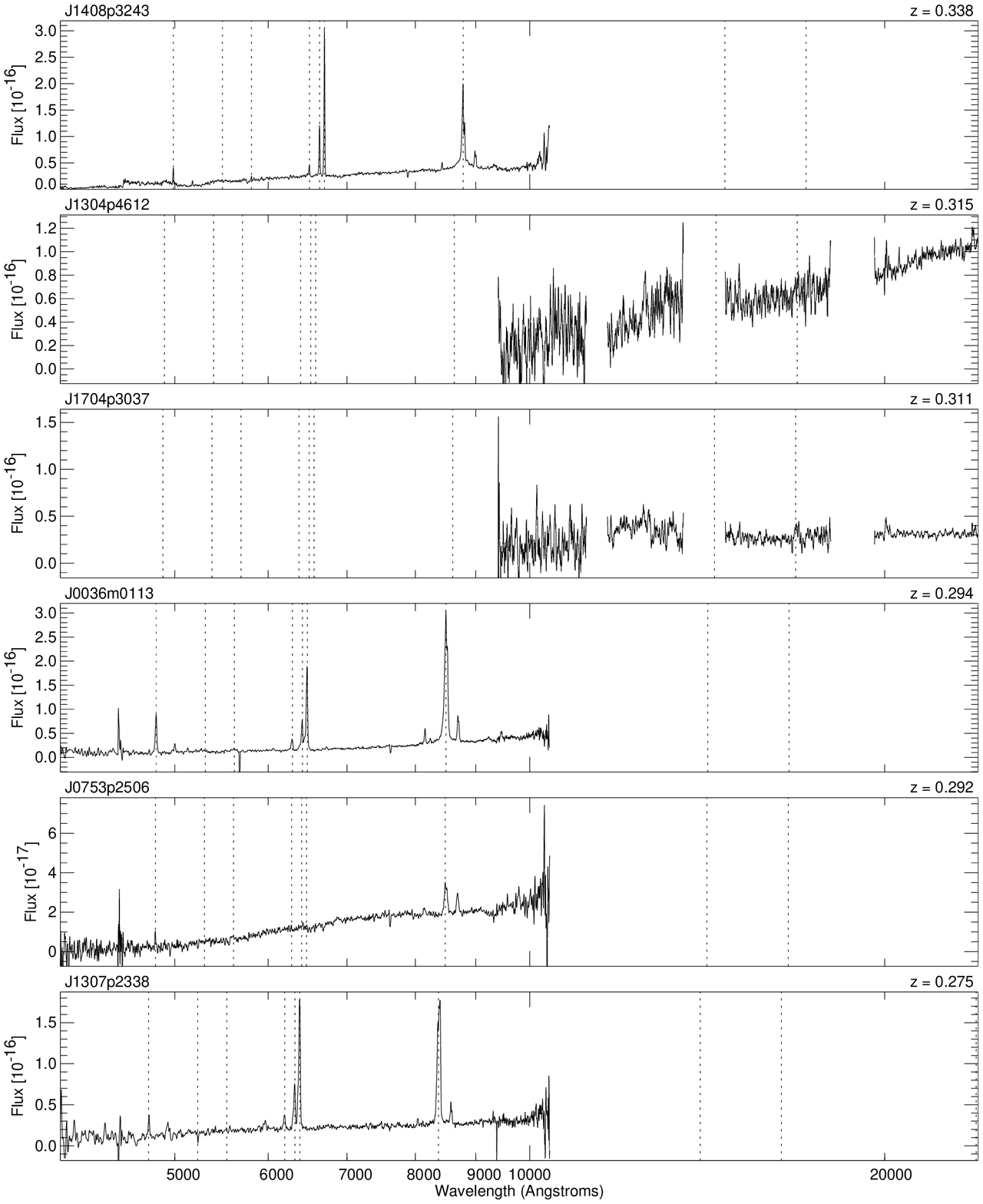}
\caption{{\it Continued.} Spectra of F2M quasars.}
\end{figure}

\begin{figure}
\figurenum{3j}
\plotone{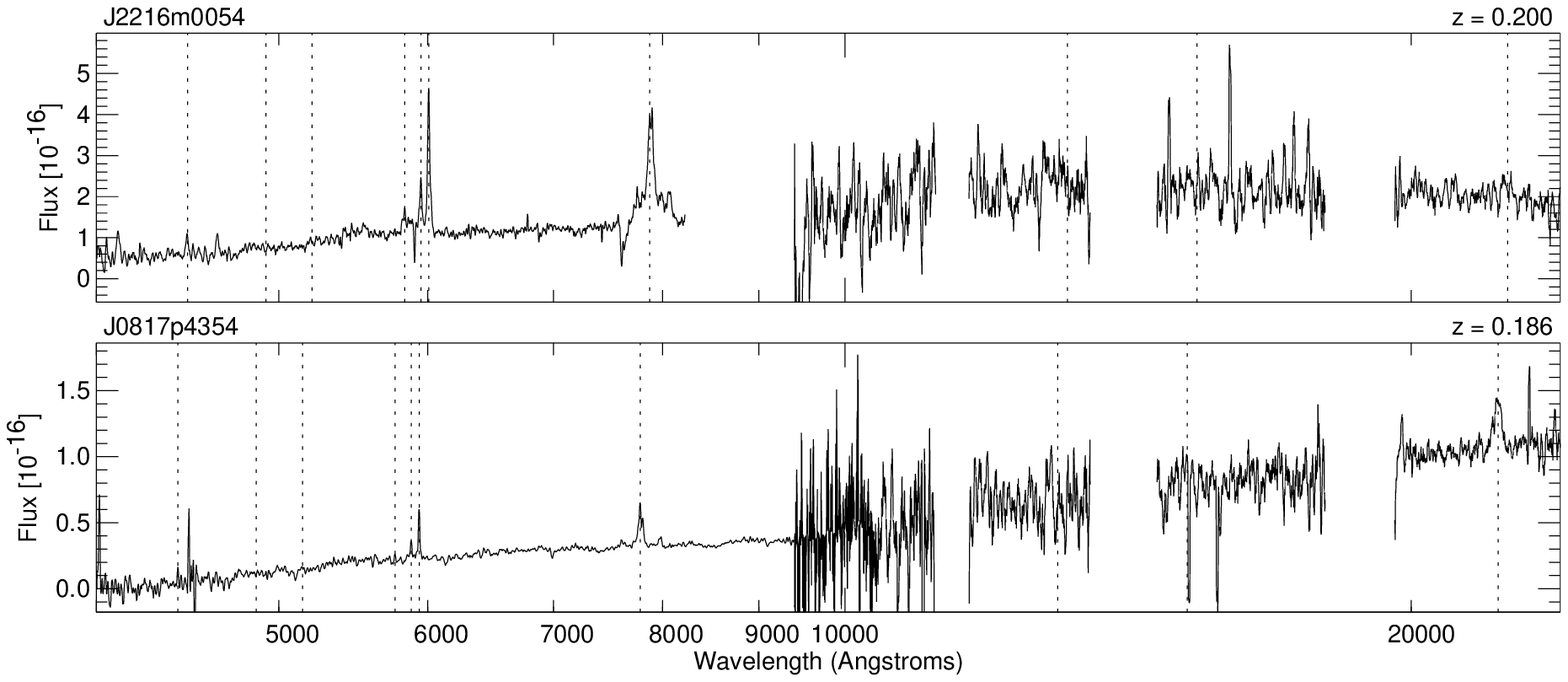}
\caption{{\it Continued.} Spectra of F2M quasars.}
\end{figure}

\begin{figure}
\plotone{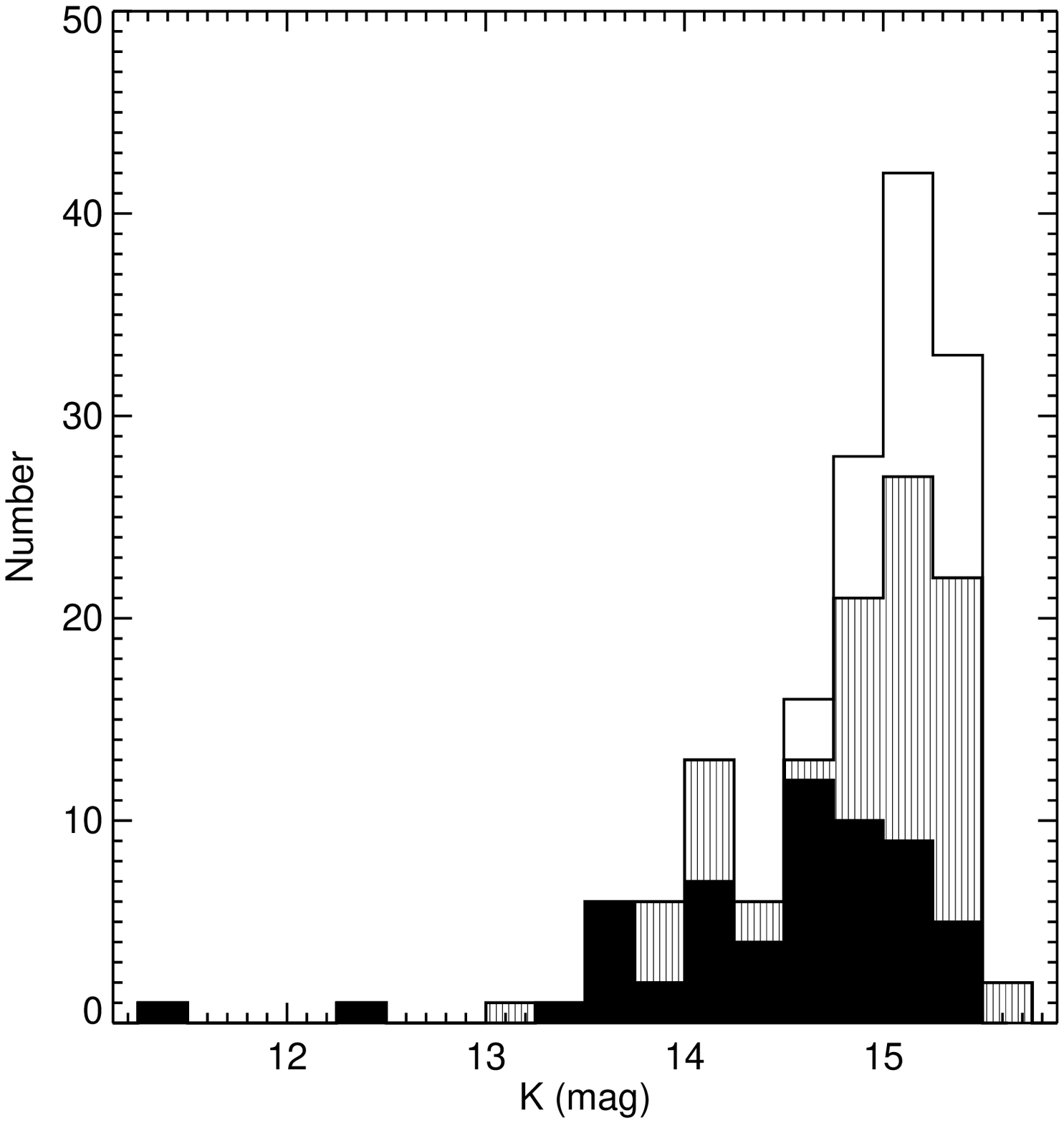}
\caption{Histogram of the $K$ magnitudes for the F2M candidates ($unshaded$).  All spectroscopically identified objects are overplotted in the shaded histogram, and the quasars are overplotted in black. }\label{fig:khist}
\end{figure}

\clearpage

\begin{figure}
\plotone{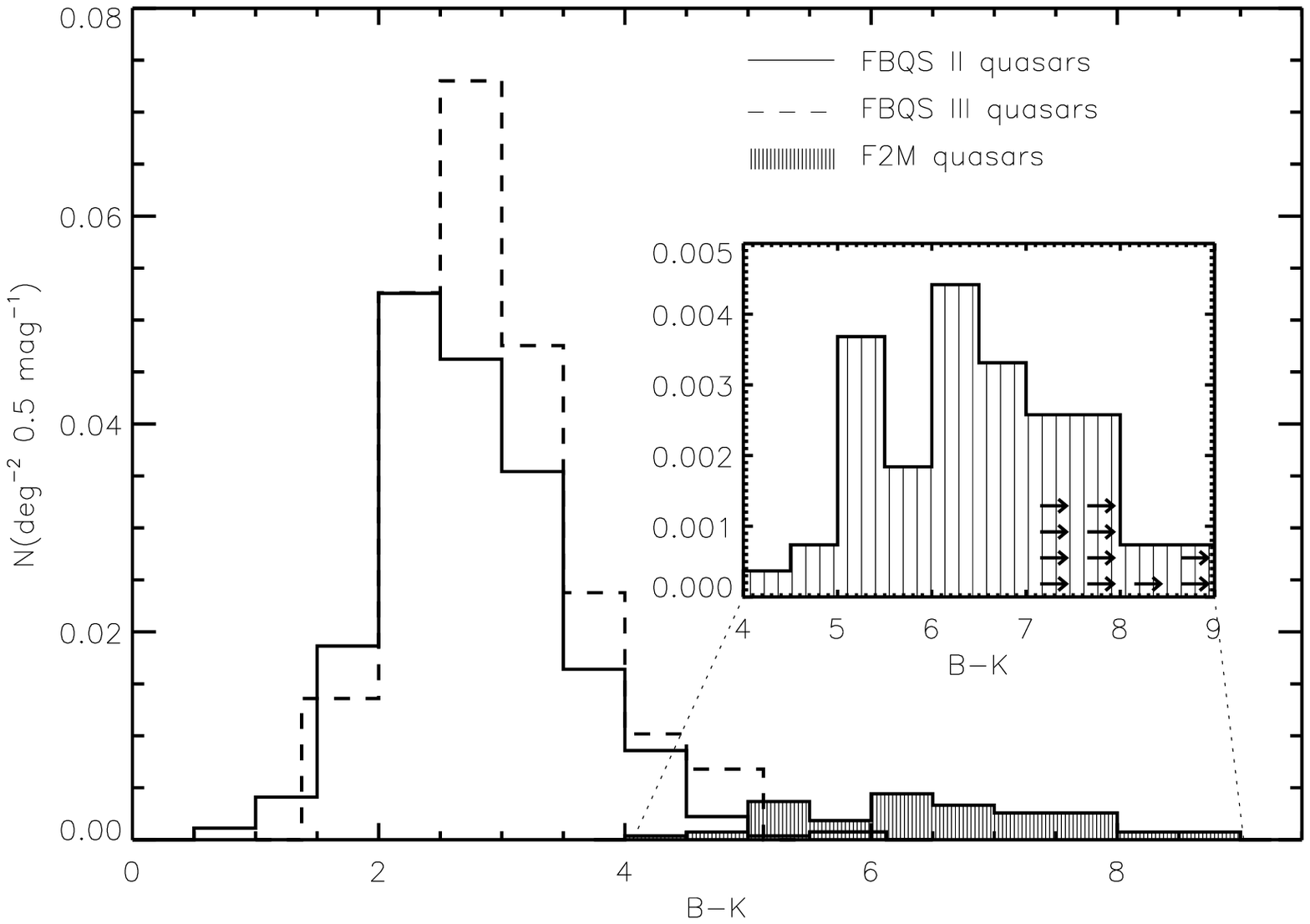}
\caption{$B-K$ distribution of F2M quasars compared with FBQS II and III; arrows denote bins which are lower limits.}\label{fig:bkhist}
\end{figure}

\begin{figure}
\plotone{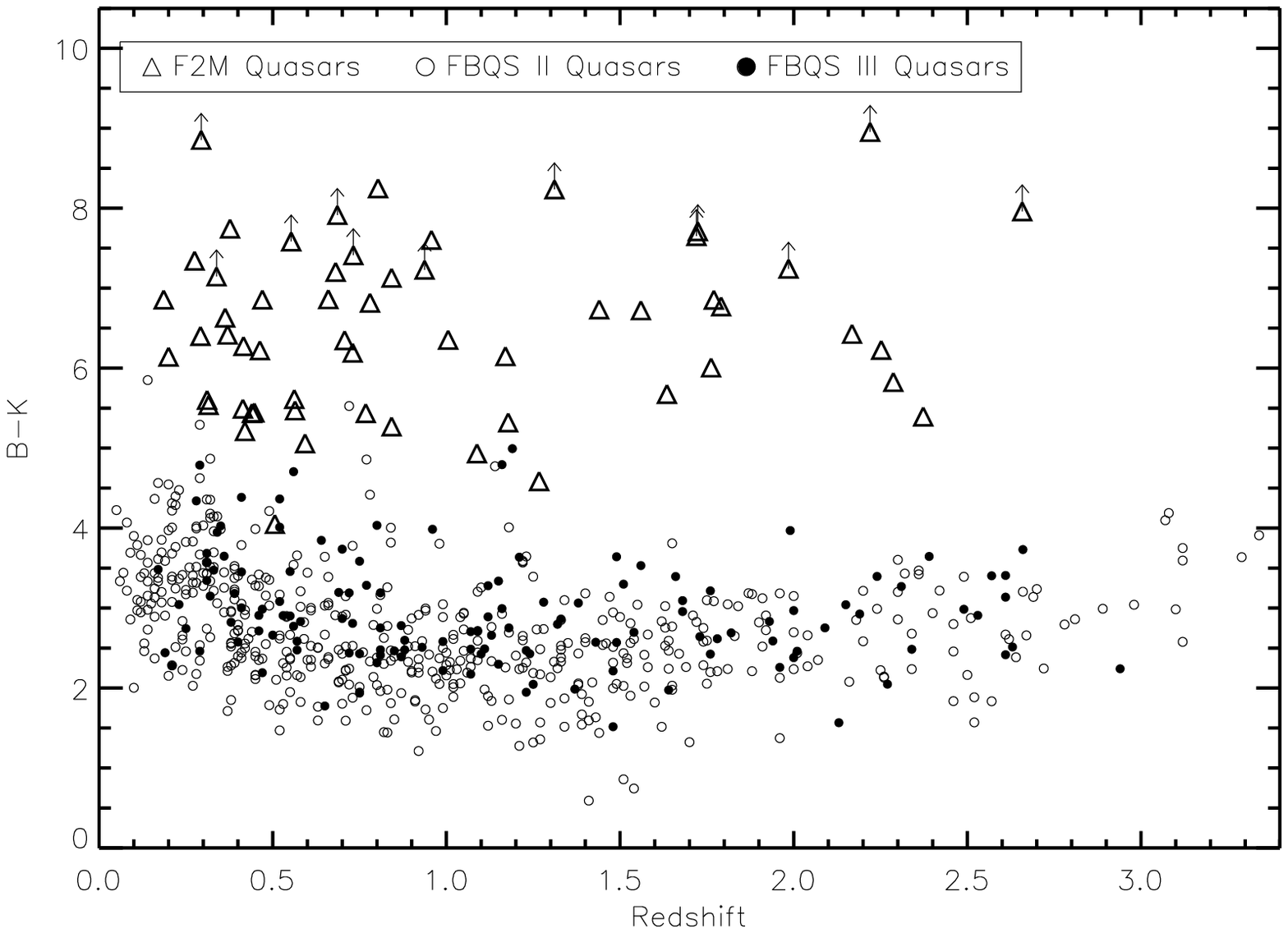}
\caption{Redshift distribution of as a function of color of F2M quasars ({\em triangles}) compared with FBQS II ({\em open circles}) and III ({\em filled circles}).}\label{fig:z_dist}
\end{figure}

\begin{figure}
\plotone{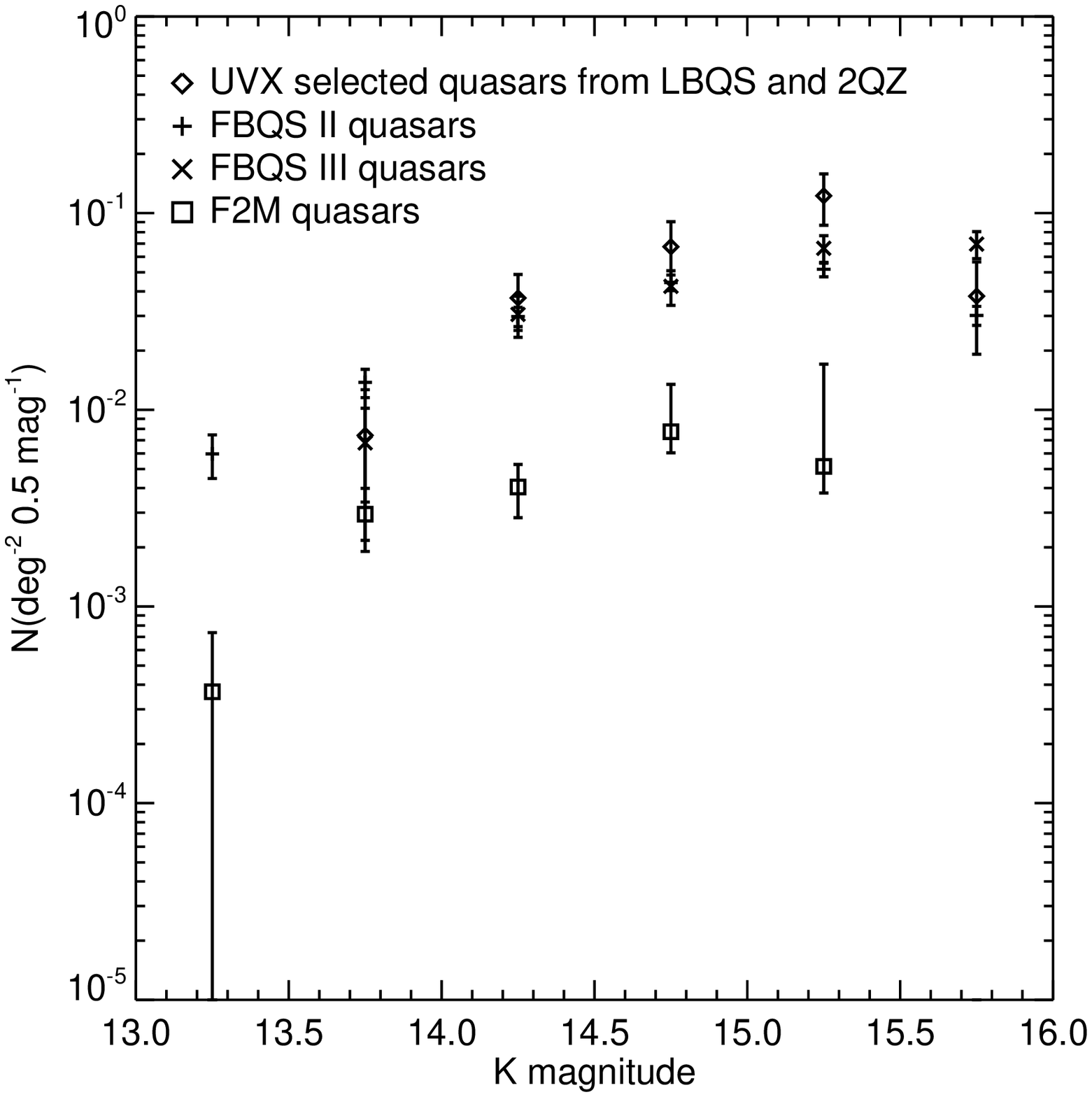}
\caption{Spatial density of quasars on the sky of F2M red quasars compared with FBQS II and III as well as radio-detected UVX-selected quasars from the LBQS and 2QZ quasar surveys.}\label{fig:space_dist}
\end{figure}

\clearpage
\begin{figure}
\plotone{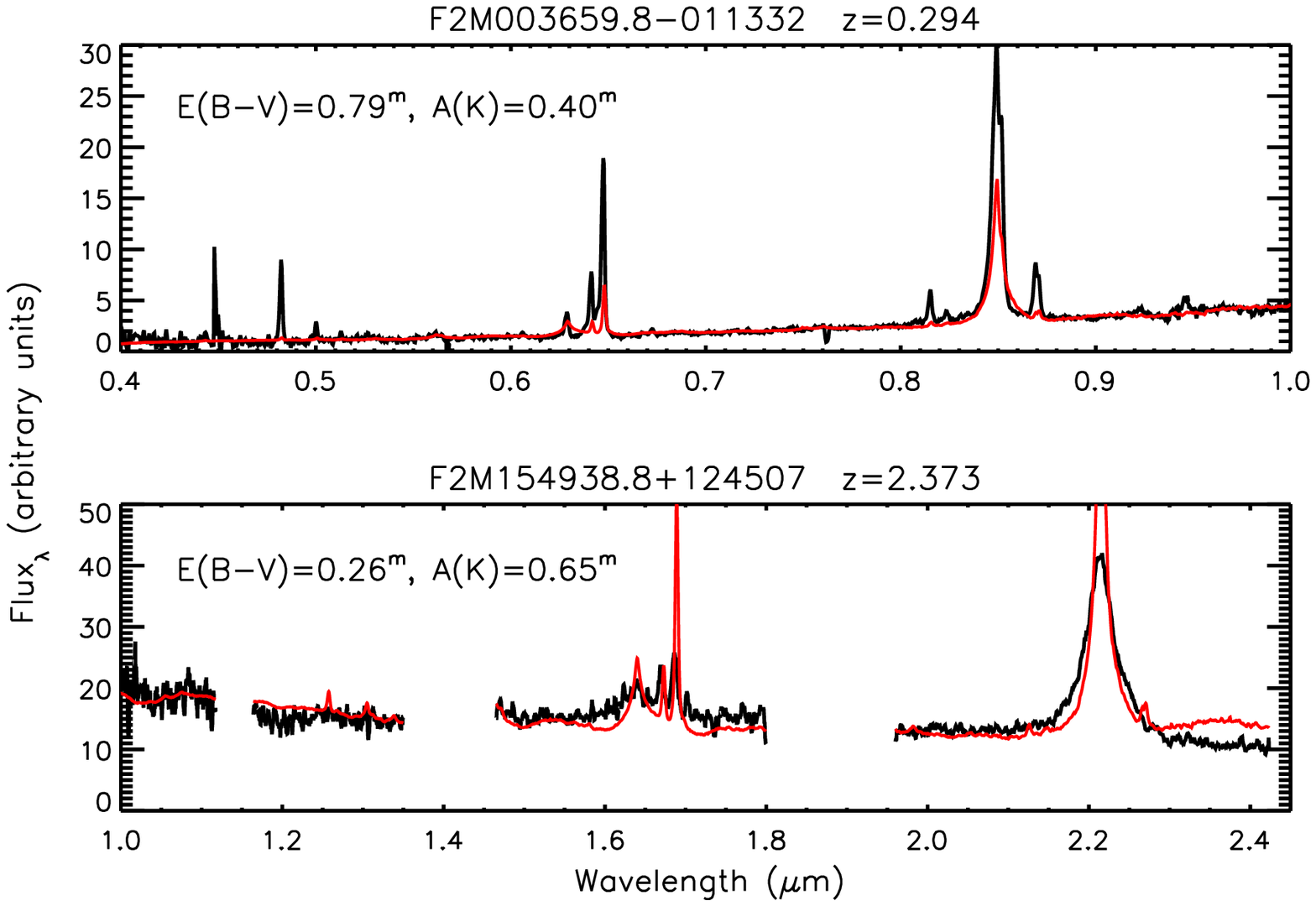}
\caption{Examples of a reddened composite spectrum fitted to two spectra.  Top -- F2M003659.8$-$011332 has an optical spectrum and is best fit by a composite spectrum reddened with $E(B-V)=0.79$, which is overplotted in {\em red}.  Bottom -- F2M154938.8$+$124507 has only a near-infrared spectrum and is best fit by the same composite reddened with $E(B-V)=0.26$, overplotted in {\em red}.}\label{fig:single_fit}
\end{figure}

\begin{figure}
\plotone{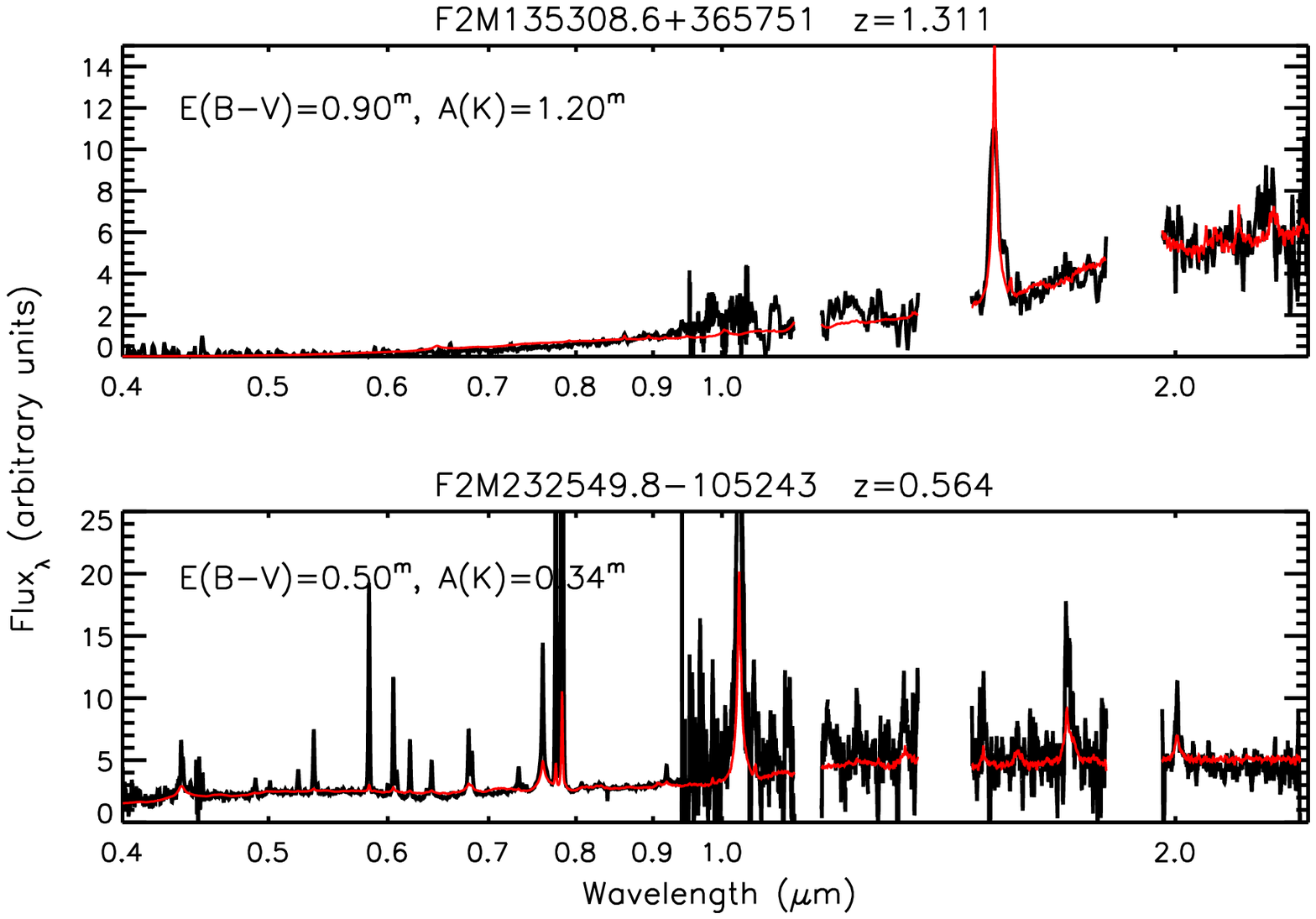}
\caption{Two examples of combined optical and near-infrared spectra with the best-fit reddened composite overplotted in {\em red}.}\label{fig:combined_fit}
\end{figure}

\begin{figure}
\plotone{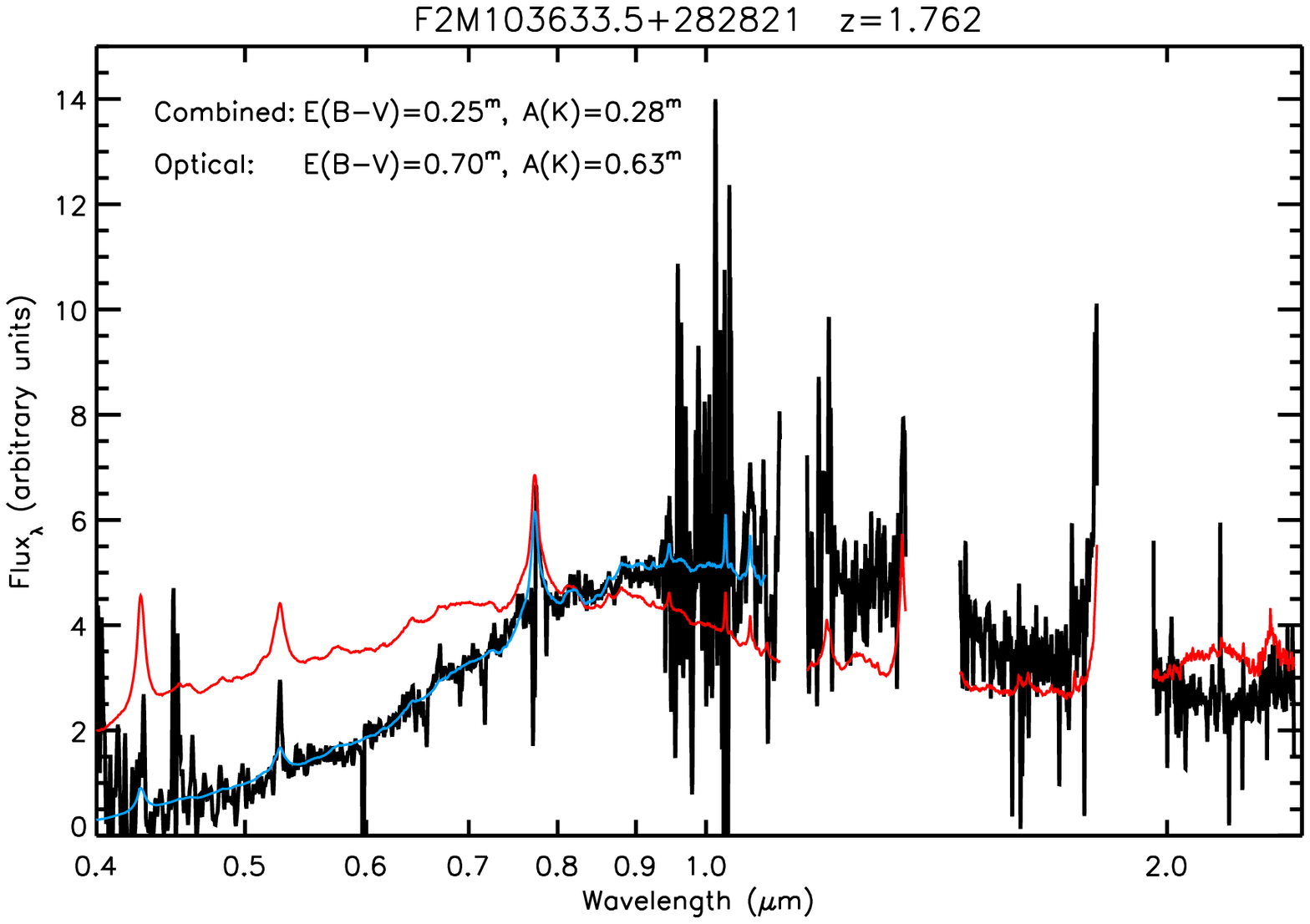}
\caption{An example of a spectrum whose combined optical plus near-infrared spectrum was poorly fit by a reddened composite, overplotted in {\em red}.  The optical spectrum alone provides an excellent fit to the reddened composite, overplotted in {\em blue}. }\label{fig:compare_fits}
\end{figure}

\begin{figure}
\plotone{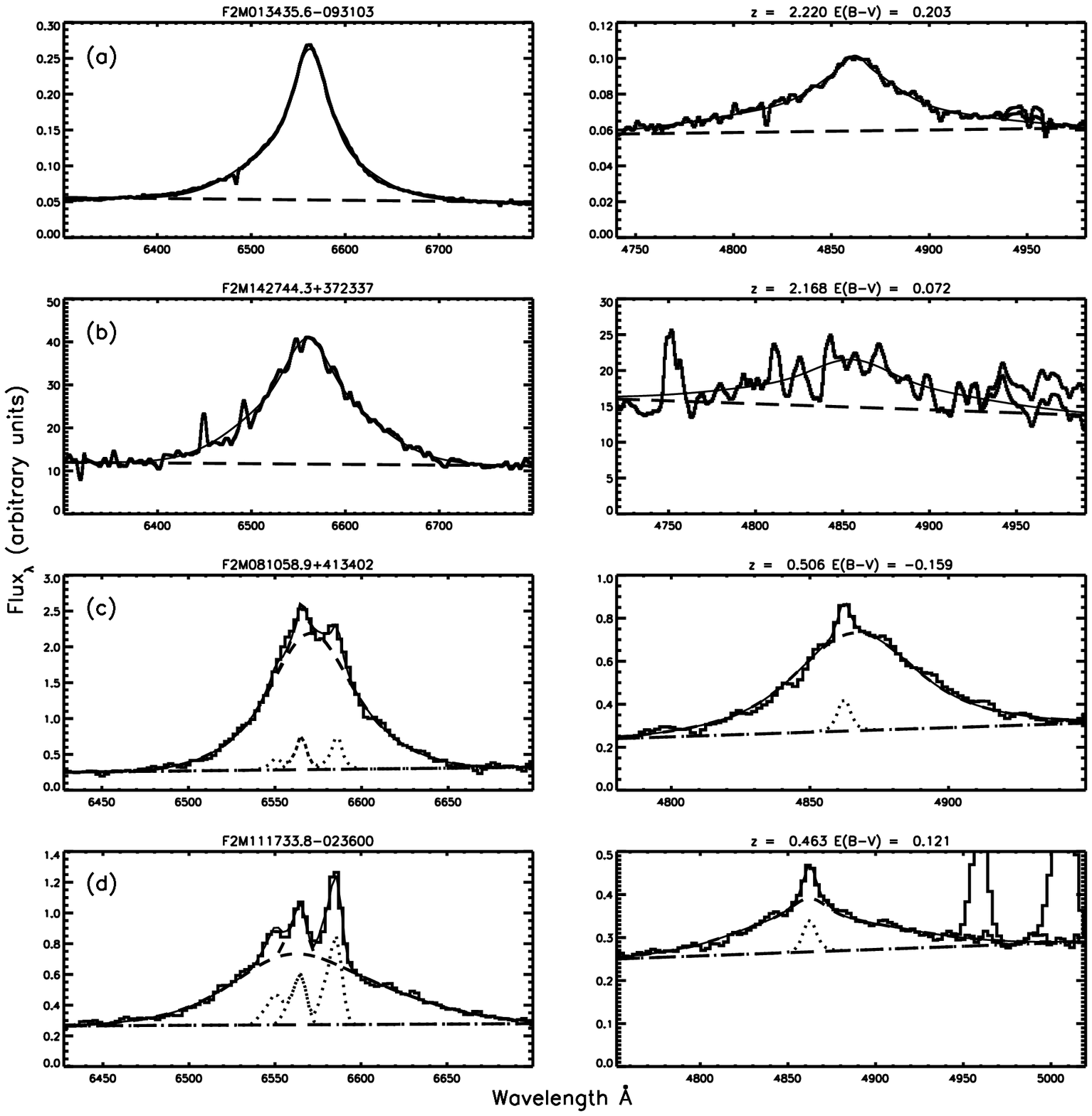}
\caption{Examples of H$\alpha$ (left) and H$\beta$ (right) fits for computing Balmer decrements using the different modeling techniques described in the text. The object's name is labeled on the left panel; the redshift and $E(B-V)$ value is given above the right panel for each object.  The total fit is plotted with a {\em solid line}.  When applicable, the narrow-line components are overplotted with a {\em dotted line} and the broad-line component is marked with a {\em dashed line}.  The linear continuum is overplotted with {\em long-dashed line}.  (a) High signal-to-noise near-infrared spectrum with no narrow-line model.  (b) Lower signal-to-noise near-infrared spectrum with no narrow-line model. (c) High-signal-to-noise optical spectrum with [\ion{O}{3}] used for the narrow-line model.  (d) High signal-to-noise spectrum with [\ion{S}{2}] used for the narrow-line model.}\label{fig:bd_example}
\end{figure}

\begin{figure}
\plotone{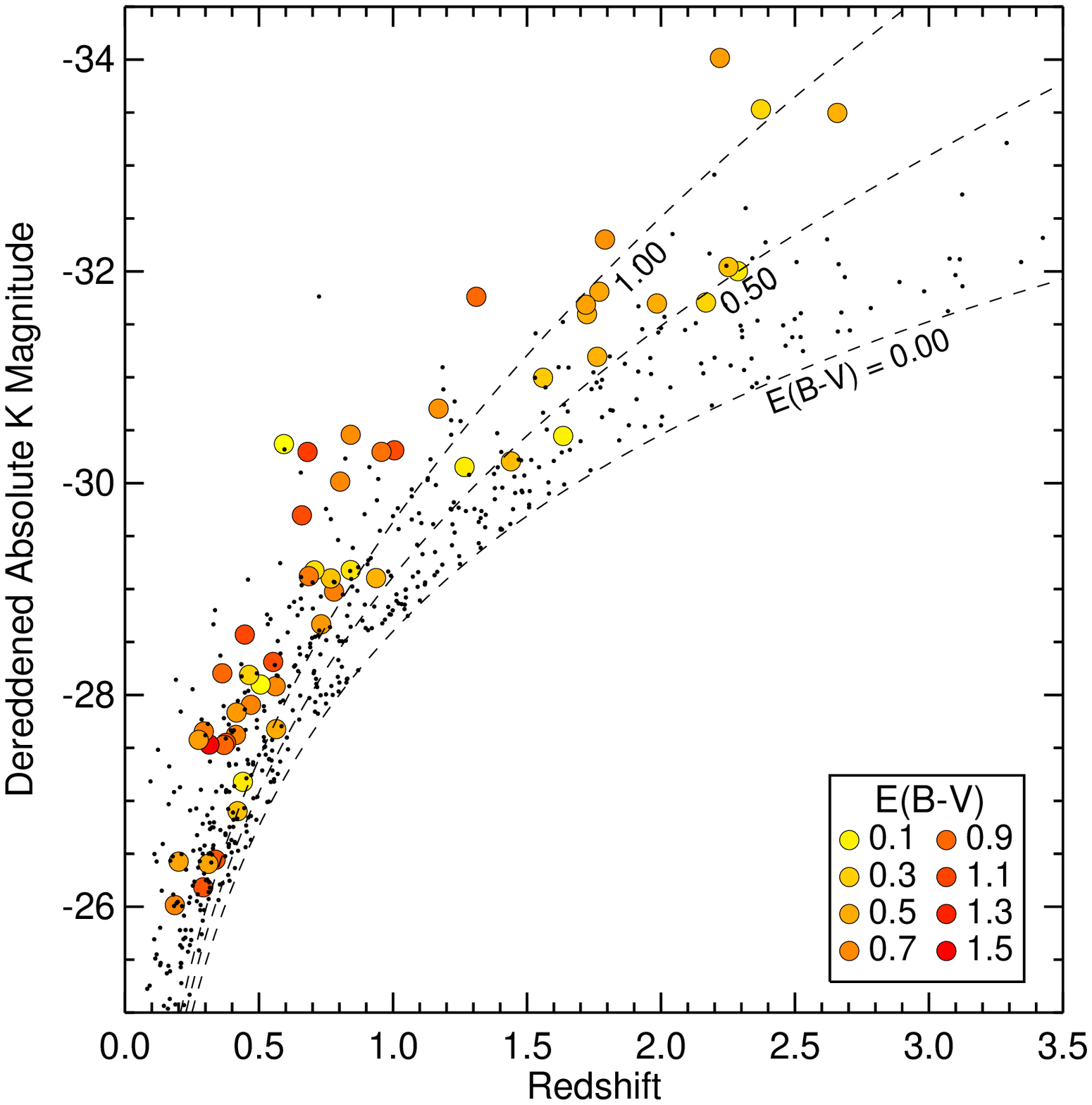}
\caption{Dereddened $K$-band absolute magnitude as a function of redshift.  The colors of the circles correspond to the amount of extinction, ranging from no extinction represented by light shading (bright yellow) to heavily reddened (dark red).  The dotted lines indicate the survey limit ($K<15.5$) for increasing amounts of extinction.  The small dots are FBQSII and FBQSIII quasars, which we assume are largely unabsorbed.}\label{fig:eb_min_v_color}
\end{figure}

\begin{figure}
\plotone{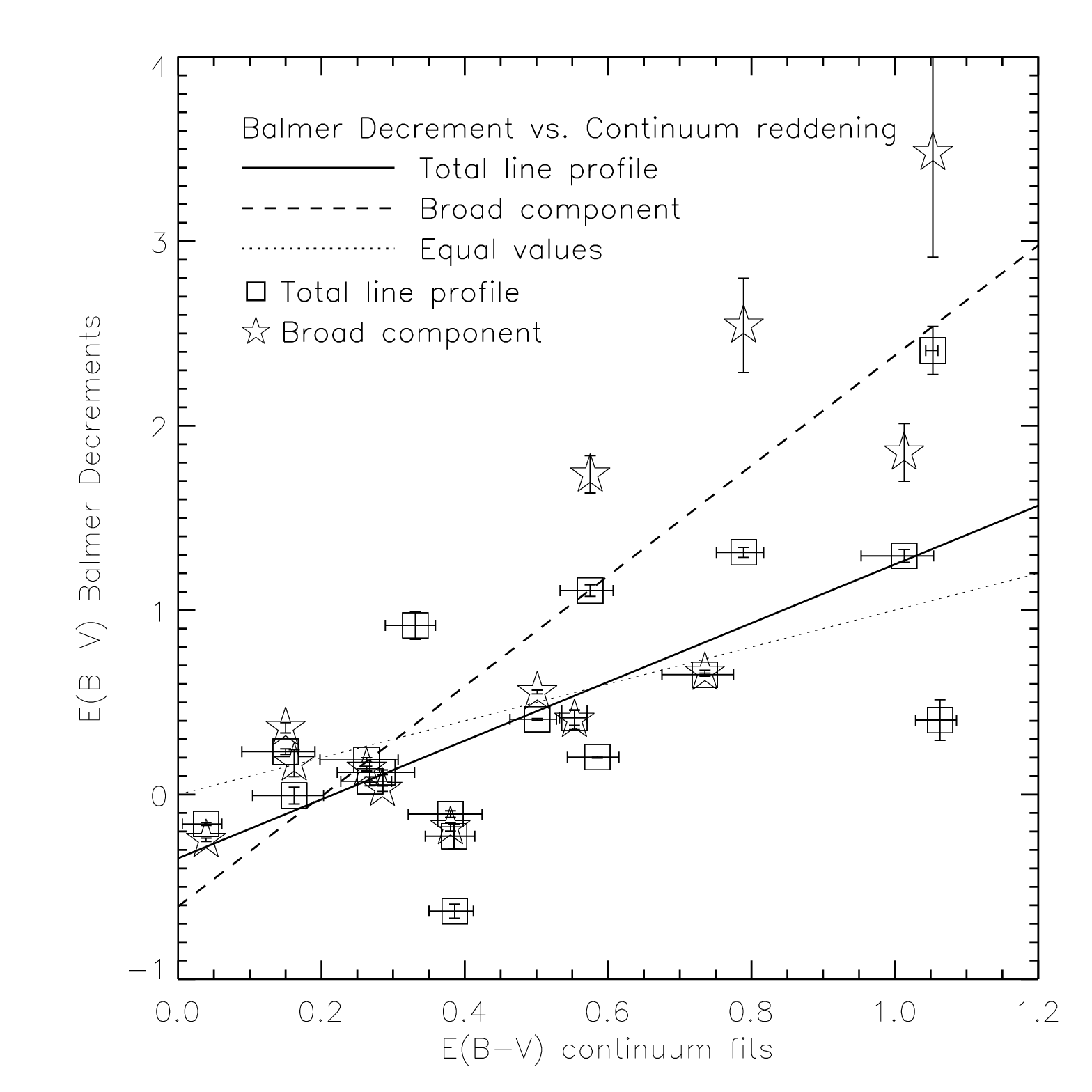}
\caption{Color excess derived from Balmer Decrement measurements versus color excess derived from reddened composite fits to spectra.  {\em Square} points are from Balmer decrements measured from the total line profile, while Balmer decrements derived from the broad component of the lines are plotted with {\em stars}.   The Balmer decrement in the total hydrogen line profile versus the reddening in the continuum is best fit by the {\em solid line}.  The {\em dashed line} shows the best fit to the extinctions derived from the Balmer decrement in only the broad component of the hydrogen line. The {\em dotted line} indicates equivalent measurements between the color-excess derived from the Balmer decrement versus the reddening in the continuum. }\label{fig:ebv_compare}
\end{figure}

\clearpage
\begin{figure}
\plotone{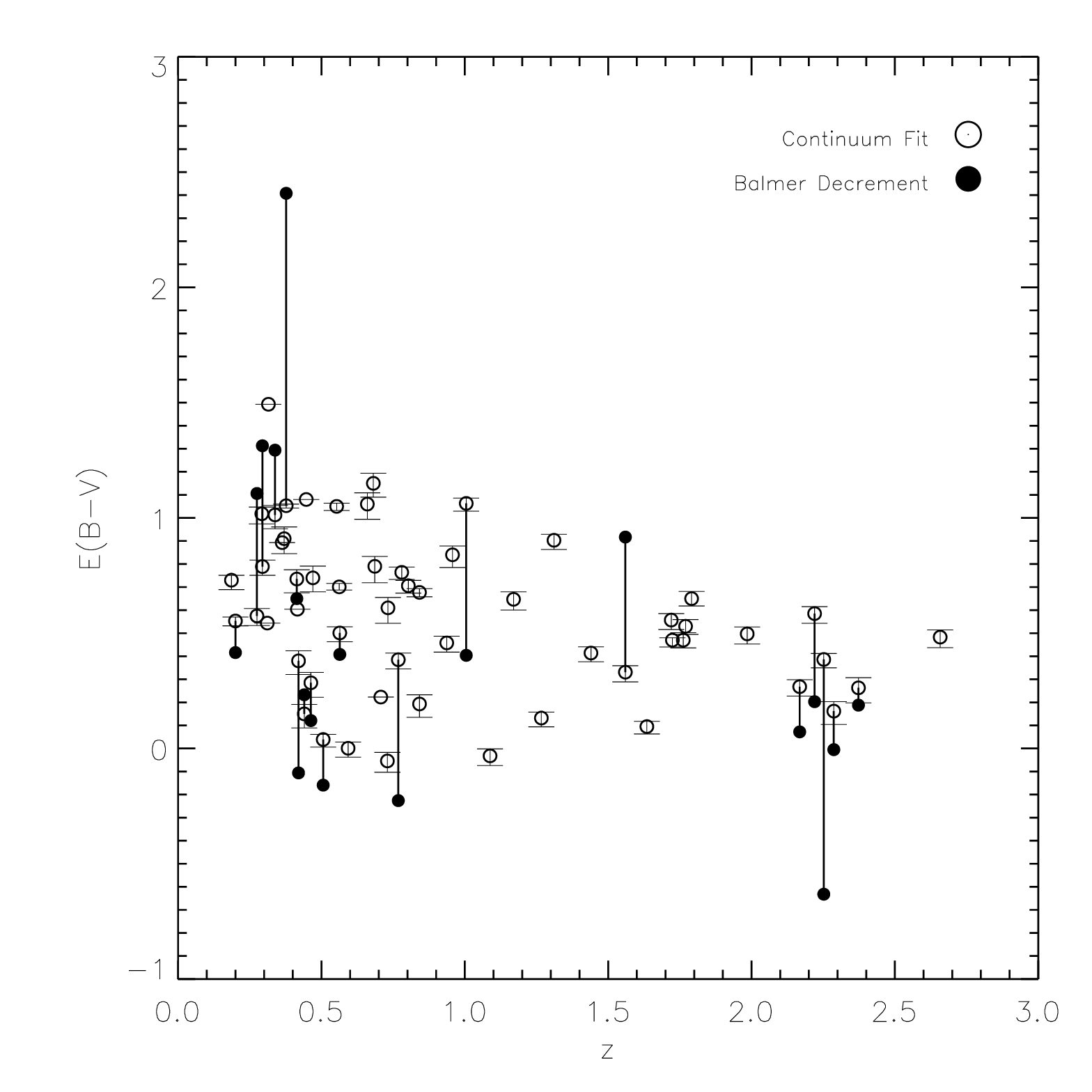}
\caption{Color excess derived from composite fitting ({\it open circles}) and from Balmer decrements using the total line profile ({\it closed circles}) as a function of redshift.  Fits from the same object are connected with a line.  The error bars on the open circles indicate the range of values obtained from the steep and flat template spectra (spectral slopes $\alpha_\nu = -0.25$ and $\alpha_\nu = -0.76$, respectively). }\label{fig:z_ebv}
\end{figure}

\begin{figure}
\plotone{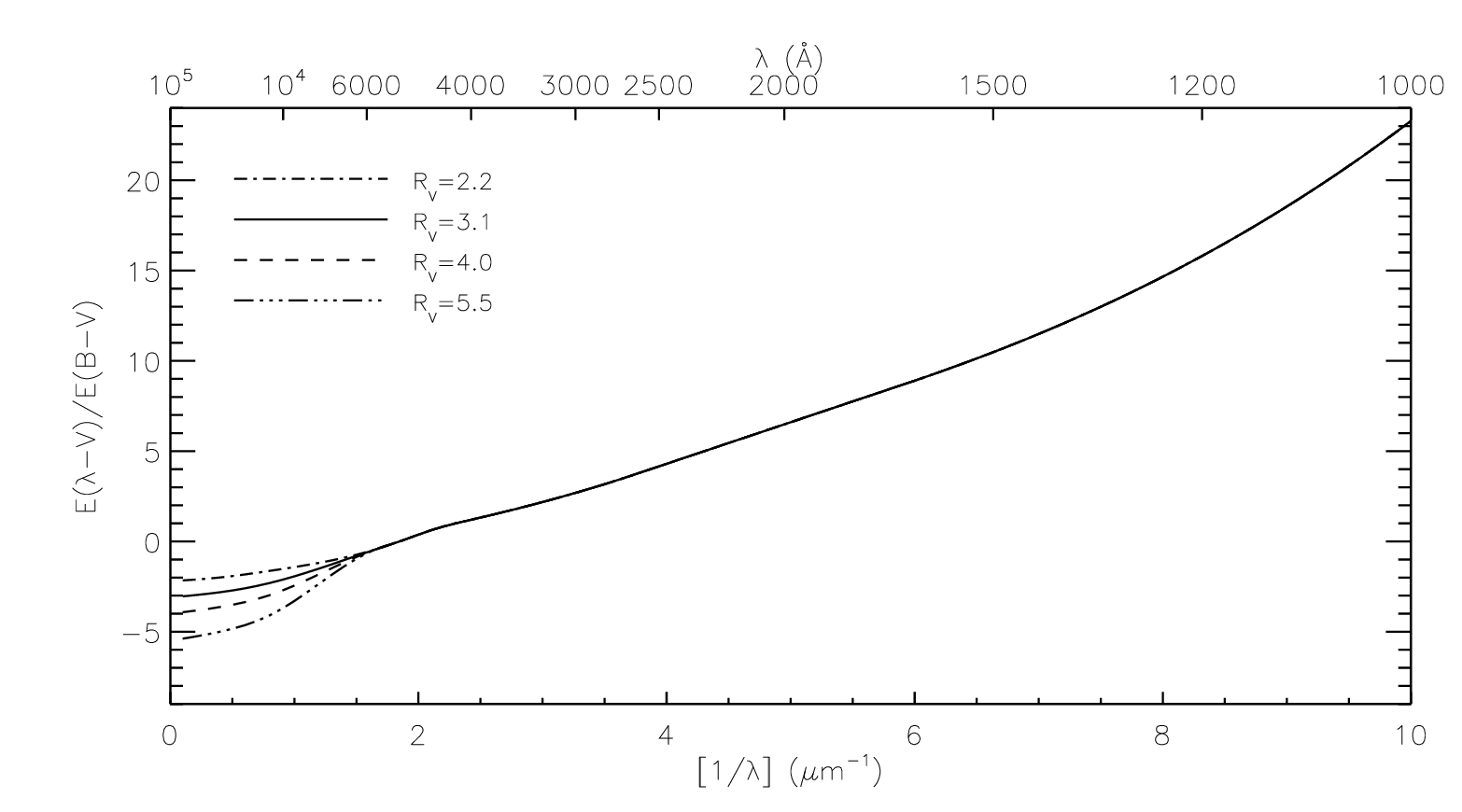}
\caption{Extinction curves from \citet{Gordon98} with varying $R_V$ values spanning the Infrared-though-UV wavelength range.  The extinction as a function of wavelength is identical in the UV and optical for the different dust laws.  They diverge around $7000$\AA\ where larger dust grains characterized by larger $R_V$ values lead to less extinction in the infrared.}\label{fig:extinction}
\end{figure}

\begin{figure}
\plotone{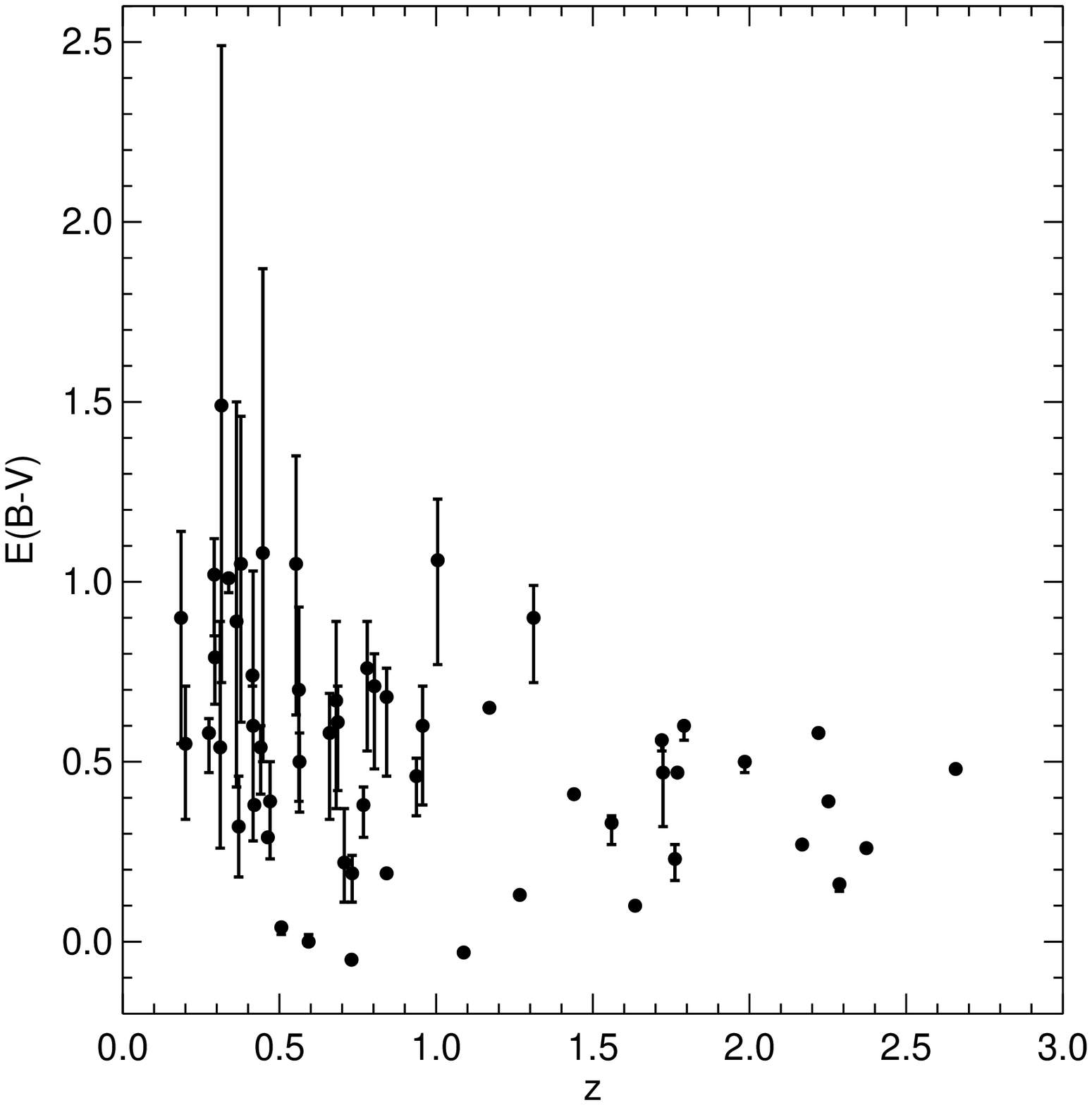}
\caption{Color excess as a function of redshift for $R_V = 3.1$ ({\em filled circles}).  The error bars represent the extreme extinctions derived from the $R_V=2.2$ and 5.5 dust laws.  Higher extinction is derived from the $R_V=2.2$ dust law, since it is more sensitive in the near-infrared, while the lower value is computed with the $R_V=5.5$ dust law as it is less sensitive to dust at these wavelengths (see Figure \ref{fig:extinction}). }\label{fig:rv_compare}
\end{figure}

\clearpage
\begin{figure}
\plotone{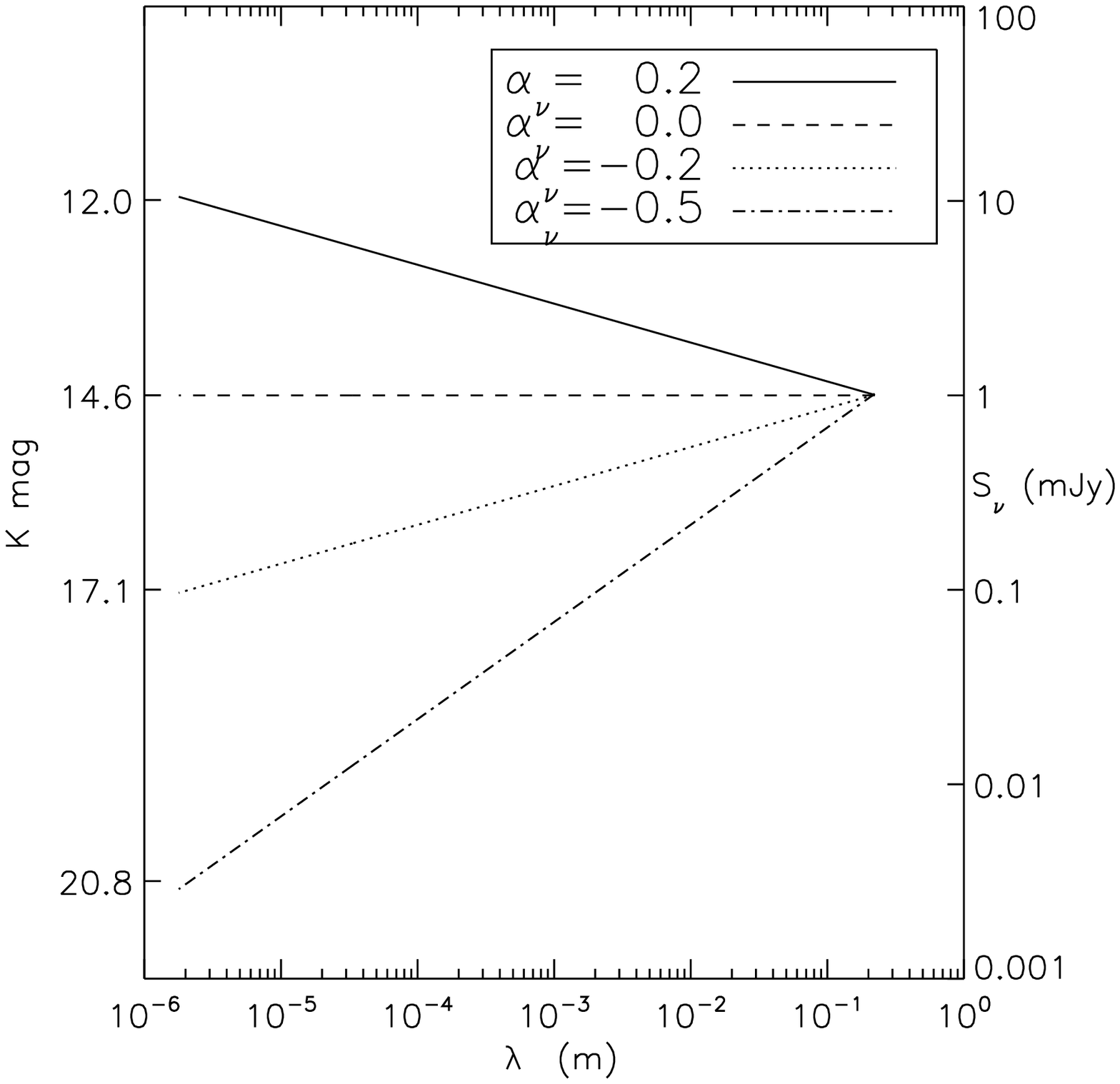}
\caption{Flux-density as a function of wavelength for power-law spectra with flat spectral indices ($F_\nu \propto \nu^\alpha$ with $\alpha = -0.5,-0.2,0.0,0.2$). The left-hand y-axis presents the 2.2 \micron\ brightness in magnitudes, while the right-hand y-axis presents the flux density in milliJansky.  This model makes the simple assumption that there is no change in the spectral shape over six dex in wavelength between 20 cm and 2 \micron\ and determines the amount of flux contributed in the $K$-band if the 1.4 GHz flux density is 1 mJy.}\label{fig:synch_rad}
\end{figure}

\begin{figure}
\plotone{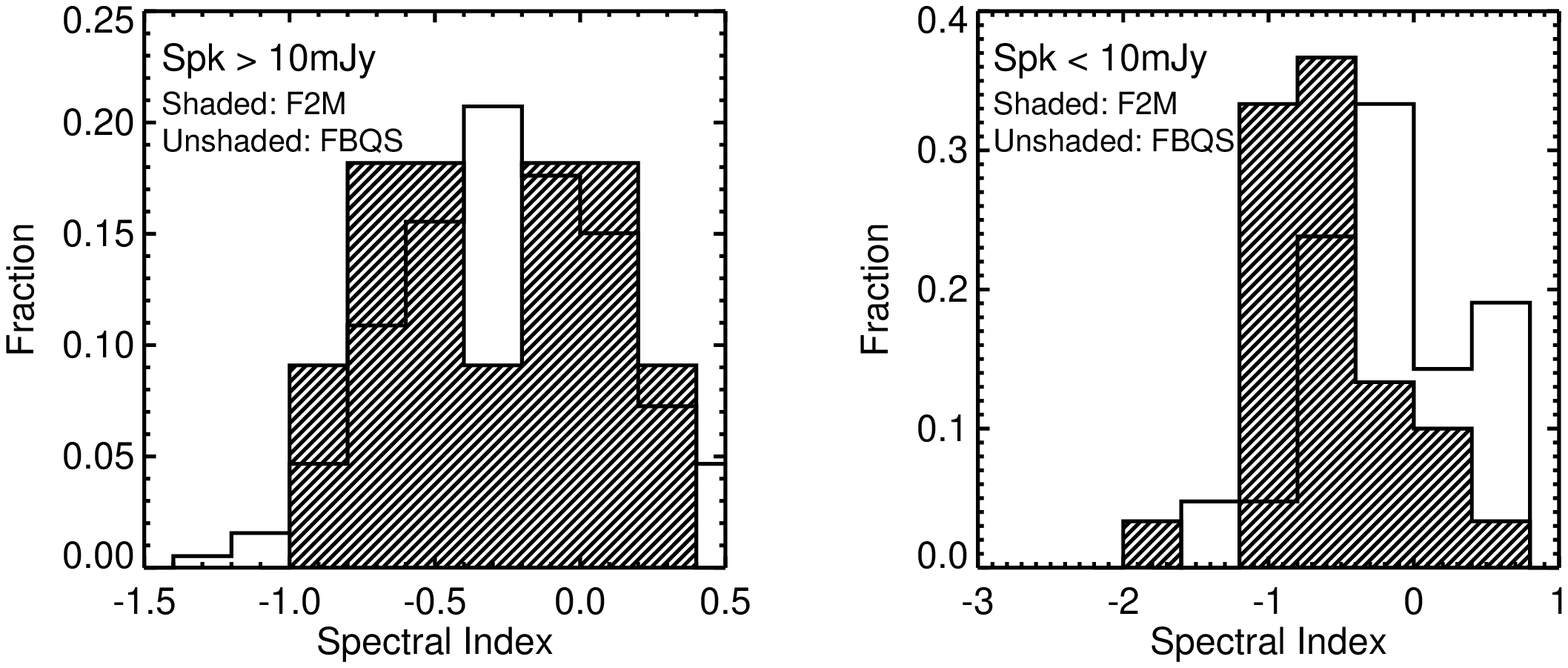}
\caption{Spectral index histogram for 41 F2M and 214 FBQS quasars, normalized by the number of quasars in each sample. Left -- 11 F2M and 193 FBQS sources brighter than 10 mJy at 20 cm.  Right -- 30 F2M and 21 FBQS sources fainter than 10 mJy at 20 cm.}\label{fig:vla}
\end{figure}

\begin{figure}
\plotone{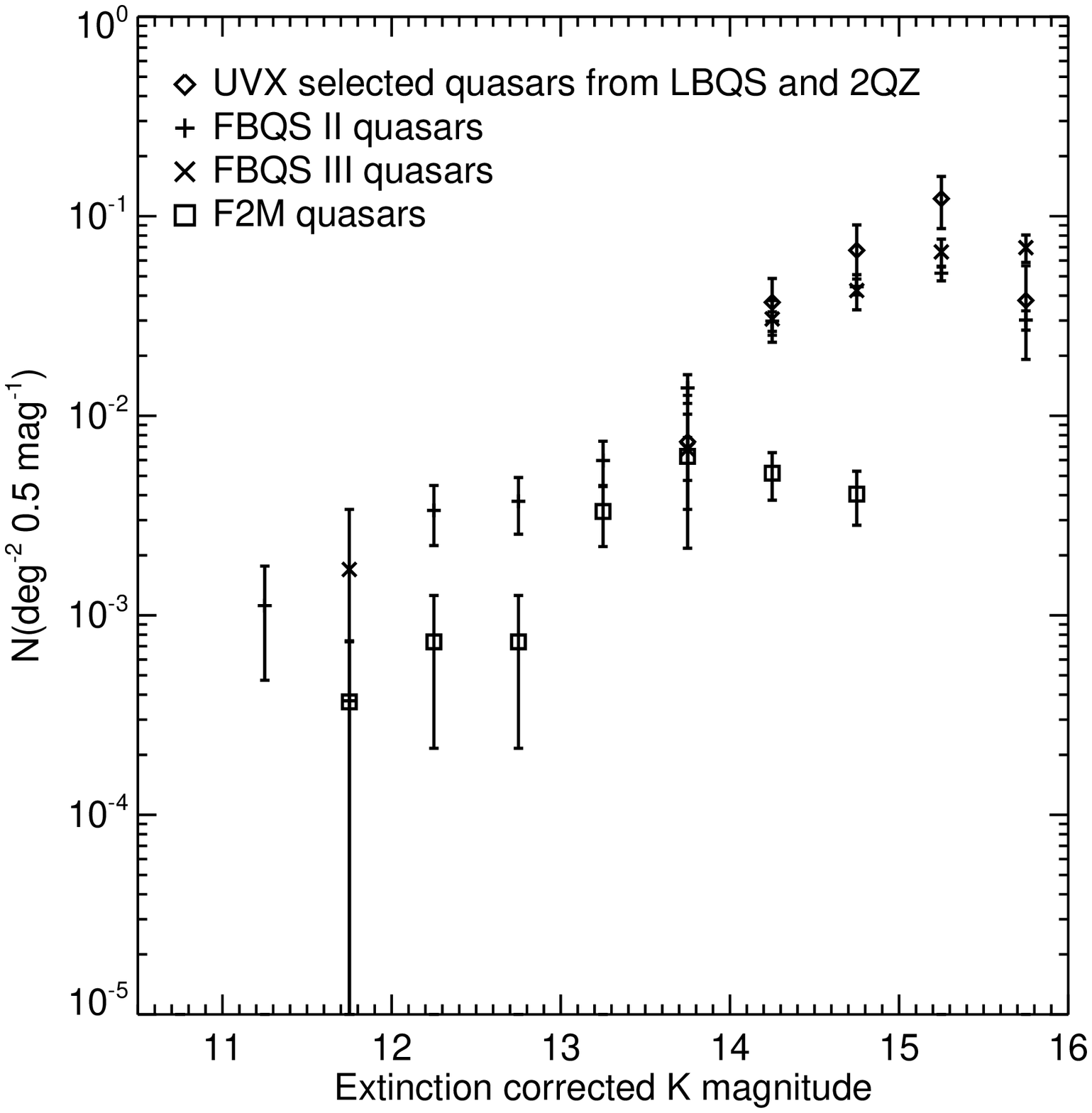}
\caption{Spatial density of quasars on the sky of F2M red quasars corrected for $K$-band absorption and compared with FBQS II and III (assuming no absorption for those quasars).}\label{fig:acor_sd}
\end{figure}

\begin{figure}
\plotone{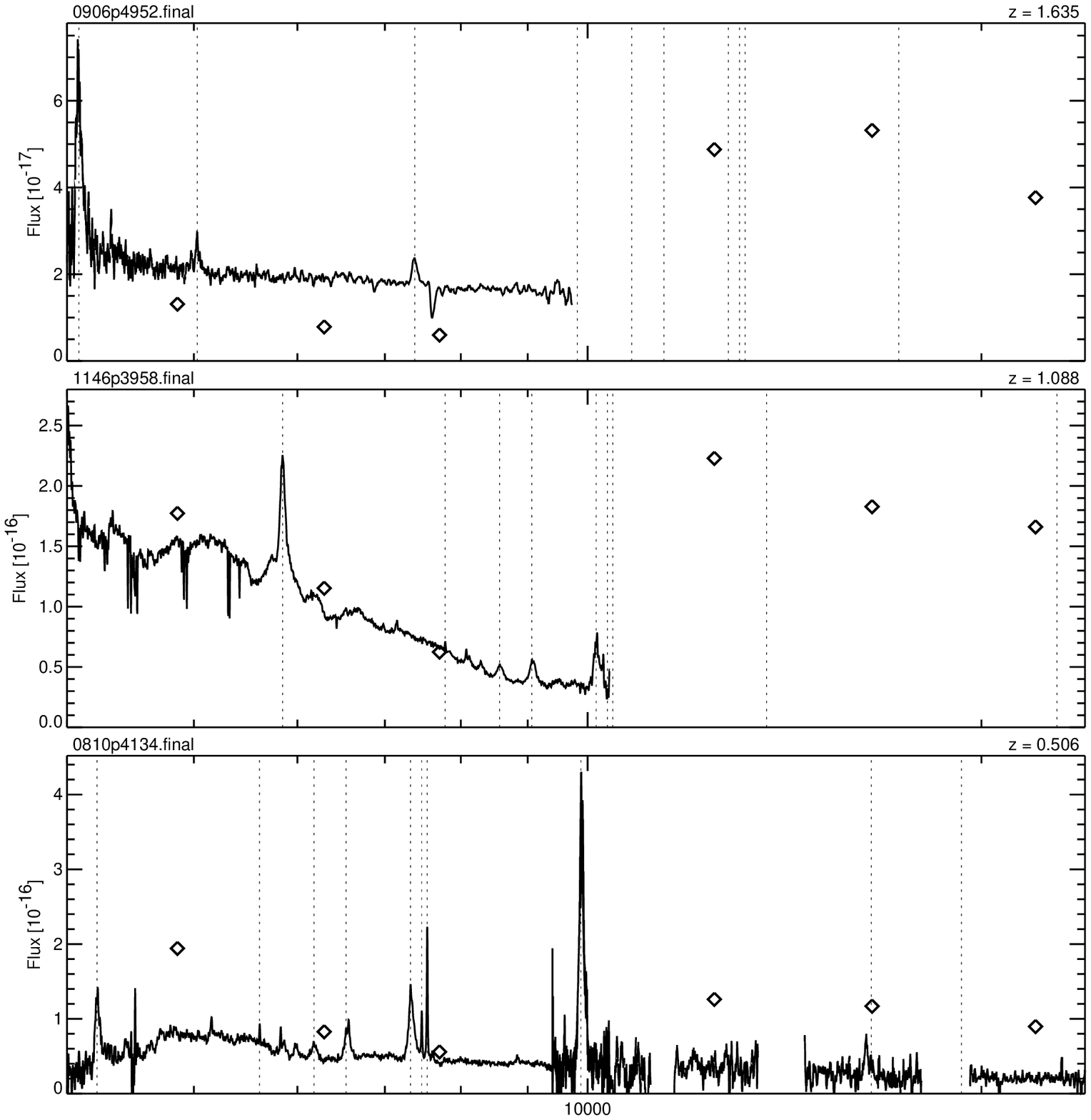}
\caption{Blue quasars detected in the F2M sample, displayed as in Figure \ref{fig:spectra}.  The {\em diamond} symbols represent three SDSS bands, $g$, $r$, and $i$, followed by the three 2MASS bands, $J$, $H$, and $K_s$.}\label{fig:blue}
\end{figure}

\end{document}